\documentclass[numberedappendix]{emulateapj}
\usepackage{amsmath}
\usepackage{graphicx}

\begin{document}

\title{On the use of shot noise for photon counting}
\author{Jonas Zmuidzinas}
\affil{Division of Physics, Mathematics, and Astronomy,
California Institute Institute of Technology,
Pasadena, CA 91125}
\email{jonas@caltech.edu}

%
%
\newcommand{\tmpbf}{}
\newcommand{\tmpcom}[1]{}

\begin{abstract}
\citet{Lieu15} have recently claimed that it is possible to substantially 
improve the sensitivity of radio-astronomical observations.
In essence, their proposal is to make use 
of the intensity of the photon shot noise as a measure of the 
photon arrival rate.
\citet{Lieu15} provide a detailed quantum-mechanical calculation of a
proposed measurement scheme that uses two detectors
and conclude that this scheme avoids the sensitivity degradation
that is associated with photon bunching.
If correct, this result could have a profound impact on radio astronomy.
Here I present a detailed  analysis of the sensitivity attainable using
shot-noise measurement schemes that use either one or two detectors,
and demonstrate that neither scheme can avoid the photon bunching penalty.
I perform both semiclassical and fully quantum calculations of the sensitivity,
obtaining consistent results, and provide a formal proof of the
equivalence of these two approaches. These direct calculations are
furthermore  shown to be consistent with an indirect argument based 
on a correlation method that establishes an independent limit to
the sensitivity of shot-noise measurement schemes.
\tmpbf{Furthermore, these calculations are directly applicable to the regime of
interest identified by Lieu et al.}
Collectively, these results conclusively demonstrate that
the photon-bunching sensitivity penalty applies to shot-noise measurement
schemes just as it does to ordinary photon counting, in contradiction to the
fundamental claim made by \citet{Lieu15}. The source of this contradiction
is traced to a logical fallacy in their argument.

\end{abstract}


\maketitle

\newcommand{\manual}{rm}        
\newcommand{\bs}{\char '134 }     
\newcommand{\oop}{\mathcal{O}}
\newcommand{\ffn}{F(g \left| \alpha \right.)}
\newcommand{\ket}[1]{\left| {#1} \right>}
\newcommand{\bra}[1]{\left< {#1} \right|}
\newcommand{\braket}[2]{\left< {#1} {|} {#2} \right>}
\newcommand{\mat}[3]{\left< {#1} \right. |{#2}|\left. {#3} \right>}
\newcommand{\ave}[1]{\left< {#1} \right>}

\newcommand{\re}{{\rm Re}}
\newcommand{\im}{{\rm Im}}
\newcommand{\tr}{{\rm Tr}}
\newcommand{\sinc}{{\rm sinc}}
\newcommand{\rth}{\rho_{\rm th}}
\newcommand{\Gb}{\bar{\Gamma}}
\newcommand{\dG}{\delta \Gamma}
\newcommand{\hI}{\hat{I}}
\newcommand{\IW}{I^{(W)}}

\newcommand{\hdG}{\delta \hat{\Gamma}}
\newcommand{\SG}{S_\Gamma}
\newcommand{\CG}{C^{(\Gamma)}}

\newcommand{\AItwo}{A^d_{\Delta t}}
\newcommand{\IdT}{I^d_{\Delta t}}

\newcommand{\Gsemi}{G^\mathrm{(sc)}}
\newcommand{\Gquant}{G^\mathrm{(qm)}}

\newcommand{\be}{\begin{equation}}
\newcommand{\ee}{\end{equation}}

\section{Introduction}
\label{sec:Introduction}

In the infrared, optical, or x-ray bands,
detection sensitivities are ultimately limited by the Poisson statistics of photon counting,
with r.m.s. count fluctuations given by $\sqrt{N}$ 
where $N$ is the mean number of photons collected \citep{Geh86}.
Thus the Poisson uncertainty in the flux measured for an astronomical source
is proportional to the square root of the 
total intensity of the radiation falling on the detector.
Meanwhile, sensitivities for radio-astronomical observations 
are calculated using the radiometer equation \citep{Dic46},
which states that the measurement uncertainty
is proportional to the the total radiation intensity rather than its square root.

\begin{figure}[htb]
\begin{center}
\includegraphics[width=3in]{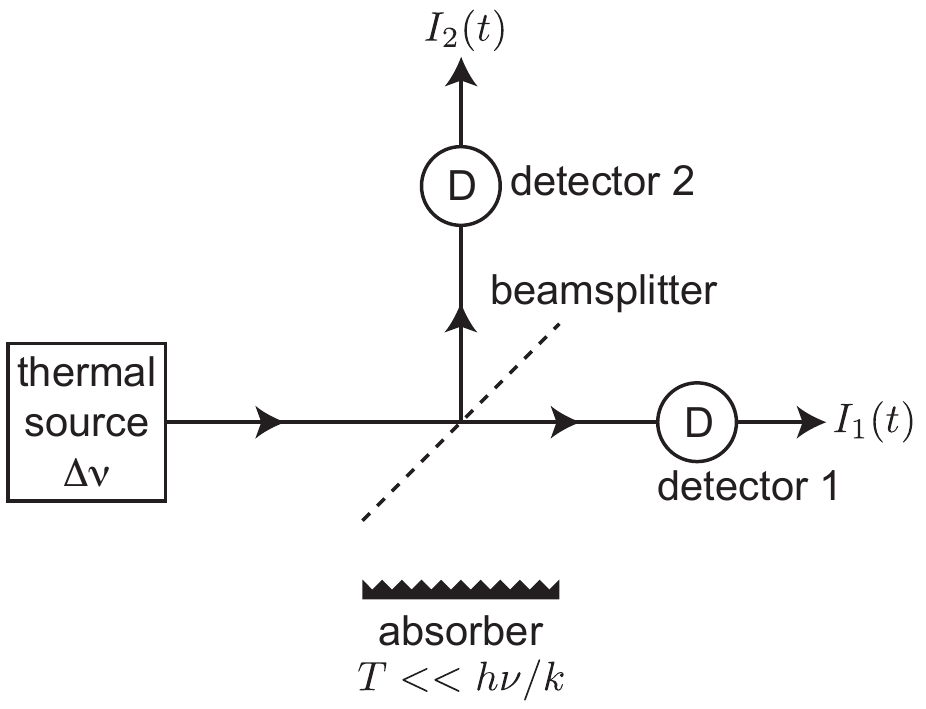}
\caption{A two-detector experiment similar to that used 
by \citet{Han56} and others (e.g., Harwit 1960)
to demonstrate photon correlations. A bright thermal light source
with narrow spectral bandwidth $\Delta \nu$ illuminates two
photon detectors via a 50/50 beamsplitter, producing photocurrents
$I_1(t)$ and $I_2(t)$. 
Photon bunching results in a nonzero correlation between
the photocurrents: $\ave{I_1(t) I_2(t)} \ne 0$.
The unused input port of the beamsplitter is
terminated with a cold (dark) absorber to prevent stray light from entering.
\label{fig:TwoDetectors}}
\end{center}
\end{figure}

A transition between these two regimes -- radio and optical -- is therefore inevitable,
and corresponds to a shift from a classical description involving fields and waves
to a quantum description involving photons,
sometimes referred to as the radio-optical dichotomy \citep{Nit94}
and ultimately stemming from the wave-particle duality of quantum mechanics.
The nature of this transition was clarified through the demonstration 
of correlated photon arrivals at two independent detectors
by  Hanbury Brown \& Twiss (1956; HBT),
using a setup similar to that illustrated in Figure~\ref{fig:TwoDetectors}.
The HBT correlations are a manifestation of photon bunching,
which causes the photon arrivals to be clustered in time rather
than being purely random.
As described in more detail in section~\ref{sec:PhotocurrentSpectrumSingle},
bunching causes the photon count fluctuations for a single detector
to increase to $\sqrt{N(1+n)}$ rather than the usual $\sqrt{N}$ for Poisson statistics.
Here $n$ represents the photon mode occupation number for a detector
with unit efficiency \citep{Zmu03},
given by the Bose-Einstein formula $n = 1/(e^{h\nu/kT}-1)$
for thermal blackbody radiation at a temperature $T$.
Bunching is usually ignorable for astronomical observations made in the infrared to x-ray bands
because  $n <<1$;\footnote{Thermal radiation at optical wavelengths
with high occupation number $n$
may readily be generated in the laboratory using stochastically modulated
coherent laser radiation, 
e.g. produced by scattering from rotating ground glass \citep{Rou71}.}
in contrast,  photon bunching is a large effect in the radio band since $n \sim kT / h\nu >> 1$.
Furthermore, at radio wavelengths both $N$ and $n$ scale with the intensity of the radiation
being detected; therefore, $\sqrt{N(1+n)}$
is also proportional to the intensity,
in agreement with the Dicke equation.

\begin{figure}[htb]
\begin{center}
\includegraphics[width=3in]{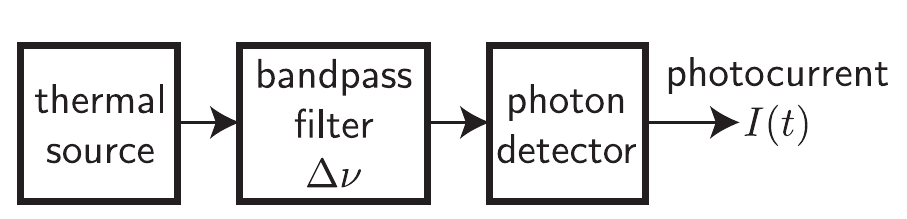}
\caption{A simple setup consisting of a thermal radiation source,
an optical bandpass filter with transmission bandwidth $\Delta \nu$,
and an ideal photon detector. The detector output is represented
by the photocurrent $I(t)$.
The thermal source and filter comprise the light source 
in Figure~\ref{fig:TwoDetectors}.
\label{fig:SourceFilterDetector}}
\end{center}
\end{figure}

\begin{figure}[htb]
\begin{center}
\includegraphics[width=3in]{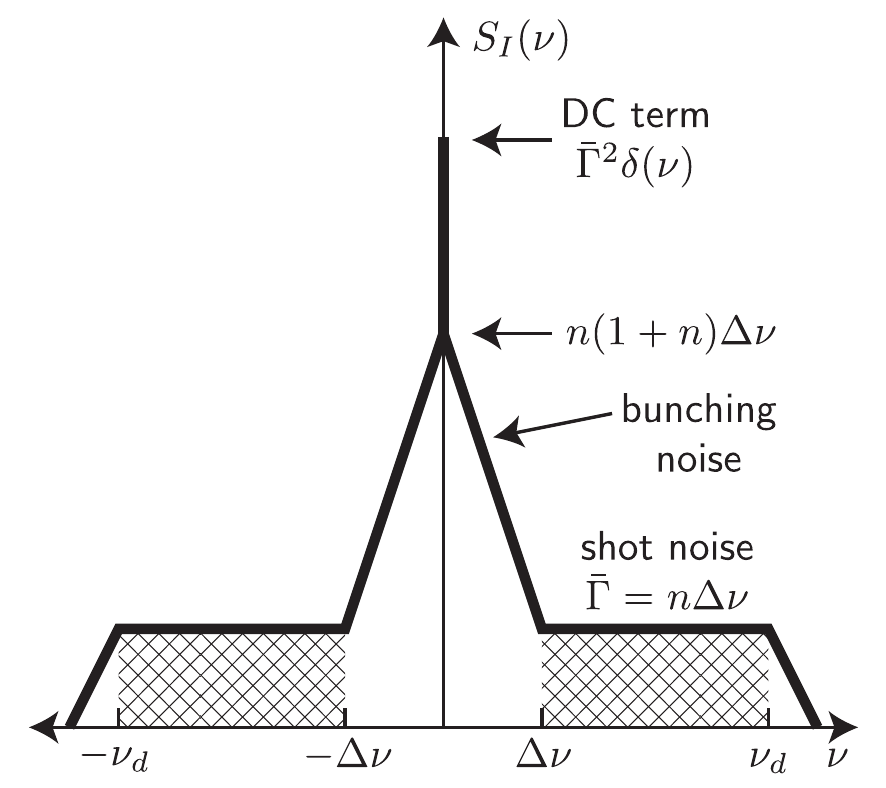}
\caption{The photocurrent noise spectrum $S_I(\nu)$ for an 
ideal photon detector
illuminated with filtered thermal radiation consists of 
three components:
(1) a DC term contributed by the mean photocurrent; 
(2) a photon bunching component that extends to $\pm \Delta \nu$,
where $\Delta \nu$ is the bandwidth of the radiation;
(3) a white shot noise term that rolls off at $\pm \nu_d$,
the detector bandwidth.
Both the DC photocurrent and the shot noise intensity are proportional
to the average photon arrival rate $\Gb = n \Delta \nu$.
The shaded region shows the portion of the spectrum that is
available for measurement of the shot noise intensity
without interference from the bunching noise component.
See section~\ref{sec:PhotocurrentSpectrumSingle}
and equation~(\ref{eqn:PhotocurrentSpectrumSingle}) for details.
\label{fig:SingleDetectorNoiseSpectrum}}
\end{center}
\end{figure}

In a recent paper, \citet{Lieu15}
claim to have found a method for avoiding the extra noise associated with
photon bunching and thereby potentially increasing the sensitivity
of radio telescopes by $\sqrt{1+n}$,
which is a large factor, 
e.g., over an order of magnitude for the example of Arecibo described in their paper.
Such a possibility is of obvious interest given the large sums spent
on the construction of radio telescopes and the associated receiving equipment.
The essence of the \citet{Lieu15} proposal is to use the wide-band shot noise 
at the output of a fast photon-counting detector as a measure of the radiation
intensity.\footnote{A fast photon detector operating at radio frequencies may
represent a serious technical challenge, but not one of fundamental principle:
tunnel junction detectors offer one possible method of 
implementation \citep{Tuc78, Sch99}.}
Consider the simple single-detector setup illustrated in Figure~\ref{fig:SourceFilterDetector}.
It is helpful to visualize the noise spectrum (the power spectral density, or PSD)
at the output of the 
detector as illustrated in Figure~\ref{fig:SingleDetectorNoiseSpectrum},
which graphically summarizes the quantum-mechanical calculations 
presented later in section~\ref{sec:PhotocurrentSpectrumSingle}.
The shot-noise spectrum \citep{Sch18}
is white and featureless within the output bandwidth of the detector,
and has an intensity that is proportional to the mean photon arrival rate $\Gb$.
Meanwhile,  the bunching noise component is confined to lower frequencies, 
determined by the bandwidth $\Delta \nu$ of the radiation being detected.
In principle, the radiation bandwidth $\Delta \nu$ may be made arbitrarily
small using narrow-band filters preceding the detector, 
so the use of a  fast detector with an output bandwidth 
$\nu_d >> \Delta \nu$
allows the region of the photocurrent noise spectrum where the white shot noise dominates
to be accessed and measured with appropriate signal processing techniques.
Clearly, it is advantageous to use a large measurement bandwidth $B = \nu_d - \Delta \nu$,
since the fractional precision with which the shot noise intensity
may be determined cannot be better than $1/\sqrt{B T}$,
where $T$ is the measurement time \citep{Dic46}.

\begin{figure}[htb]
\begin{center}
\includegraphics[width=3in]{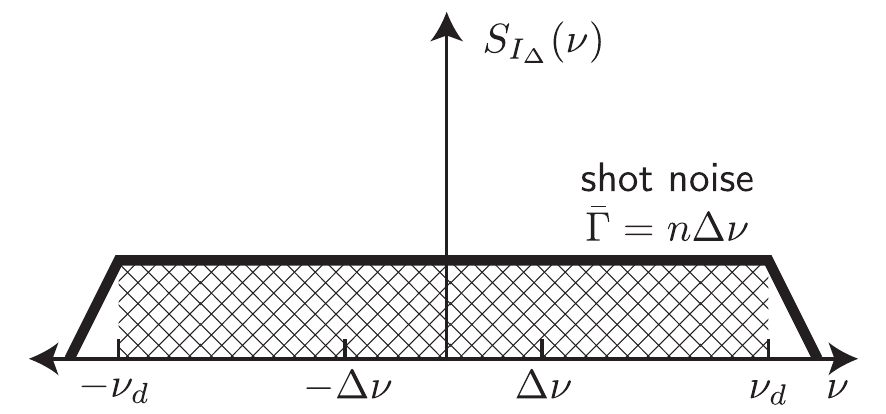}
\caption{The noise spectrum $S_{I_\Delta}(\nu)$ for
the output $I_\Delta$ of the two-detector scheme proposed by \citet{Lieu15} 
and illustrated in Fig.~\ref{fig:TwoDetectors}. 
Here $I_\Delta = I_1 - I_2$ represents the difference in the photocurrents
for the two detectors.
Taking this difference eliminates the DC component as well as the bunching
component in the spectrum, leaving only the white shot noise component
that is proportional to the mean photon rate $\Gb$.
The full spectrum is available for measurement of the shot noise
intensity, as illustrated by the shaded region.
See section~\ref{sec:PhotocurrentSpectrumMultipleDetectors}
and equations~(\ref{eqn:PhotocurrentSpectrumBeamsplitter}), (\ref{eqn:Cdelta}), 
and (\ref{eqn:PhotocurrentSpectrumDifference}) for details.
\label{fig:TwoDetectorNoiseSpectrum}}
\end{center}
\end{figure}

Alternatively, as specifically proposed by \citet{Lieu15}
and shown in Figure~\ref{fig:TwoDetectors},
a 50/50 beamsplitter or its
radio equivalent may be used to feed two detectors.
Each detector individually has an output noise spectrum similar
to that shown in Figure~\ref{fig:SingleDetectorNoiseSpectrum},
although with half the total photon rate ($\Gb/2$) per detector.
The DC term may be eliminated
by taking the difference of the two photocurrents.
Differencing also eliminates
the bunching noise lying in the frequency interval
$[-\Delta \nu, \Delta \nu]$,
because this component is fully correlated at the
two detectors, as is demonstrated through a quantum-mechanical
calculation in section~\ref{sec:PhotocurrentSpectrumMultipleDetectors}.
Indeed, this component is responsible for the HBT correlations.
Thus, only the white shot-noise spectrum survives
after taking the difference,
as shown in Figure~\ref{fig:TwoDetectorNoiseSpectrum};
the shot noise intensity may then be measured 
using relatively simple signal processing techniques.
Although the measurement bandwidth may now be increased to 
$B = \nu_d$ instead of $\nu_d - \Delta \nu$, the resulting improvement 
is modest when $\nu_d >> \Delta \nu$.
\citet{Lieu15} present a full quantum-mechanical calculation
of the sensitivity of such a shot noise measurement scheme
using two detectors,
which is a nontrivial task involving computation of eighth-order moments of photon operators,
and conclude that the $\sqrt{N}$ Poisson uncertainty may be
achieved instead of the usual bunching-degraded $\sqrt{N (1+n)}$ uncertainty
as expressed by the Dicke equation.
This result is quite surprising, and if correct and amenable
to practical implementation,
would represent a significant discovery with the potential to stimulate
large advances in radio astronomy.
\tmpbf{However, as demonstrated in section~\ref{sec:Resolution},
the fundamental conclusion of the \citet{Lieu15} paper 
rests on a logical fallacy and is therefore not valid.
Section~\ref{sec:Resolution} also contains a simple intuitive
argument that demonstrates why the measurement
scheme proposed by Lieu et al. is in fact subject to
the photon bunching penalty; those uninterested in
the detailed calculations in the following sections may wish
to jump straight to section~\ref{sec:Resolution}.}

\begin{figure}[htb]
\begin{center}
\includegraphics[width=3in]{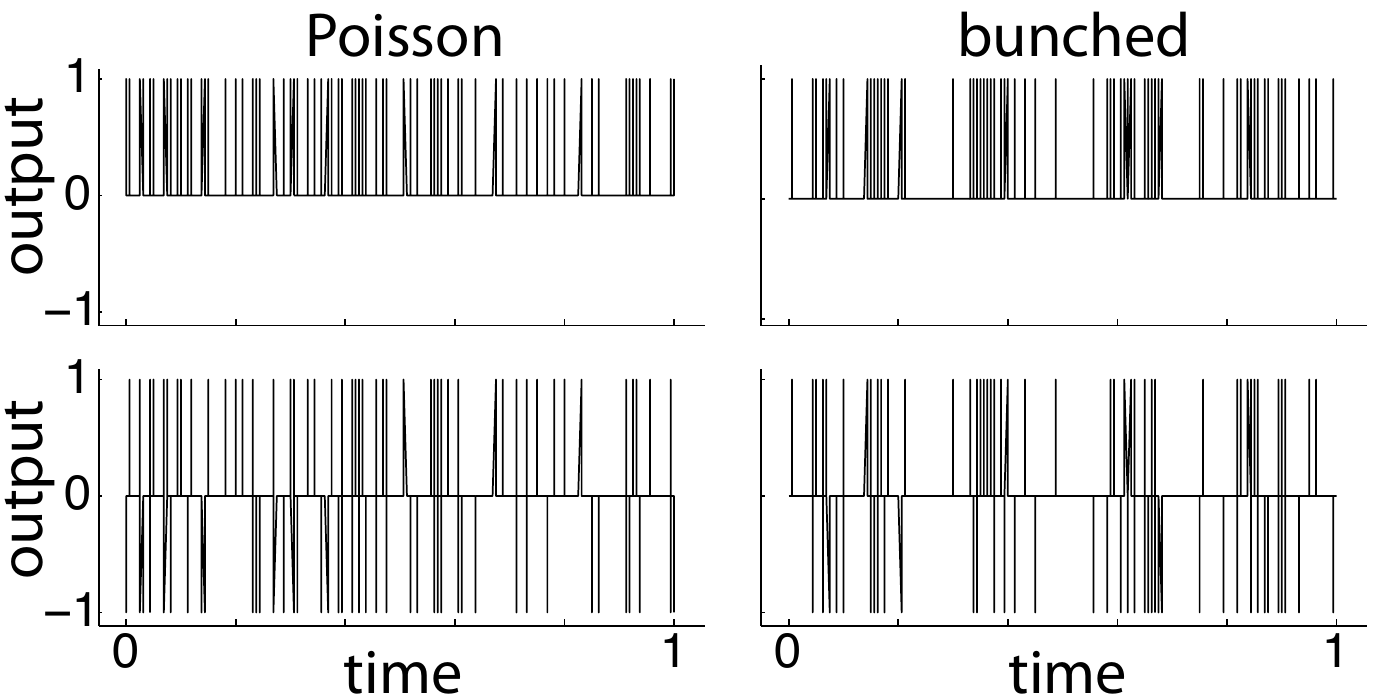}
\caption{
Top left: Simulated output time stream for a single detector 
(see diagram in Figure~\ref{fig:SourceFilterDetector})
illuminated by a coherent source or a thermal source
with low occupation number.
The photon arrival rate is constant with time,
which is the case of Poisson statistics. 
Bottom left: Differenced output of a beamsplitter-fed
pair of detectors (Figure~\ref{fig:TwoDetectors})
for the same Poisson case.
Top right: Output time stream for a single detector
illuminated with strongly bunched thermal radiation 
(high occupation number). 
Bottom right: Differenced output of a detector pair
for the bunched case.
Horizontal and vertical scales are in arbitrary units.
\label{fig:PulsePlot}}
\end{center}
\end{figure}

The work of \citet{Lieu15}, and the quantum calculations presented here,
may pose a challenge to those
who are more familiar with classical concepts such as fields
and voltages than with photon operators and quantum mechanics.
Nonetheless, the essence of the problem is quite easy to understand by use
of a familiar analogy.
The analogy relies on the fact that thermal photon
bunching can be correctly described by a photon arrival rate that varies with
time in a random way, as will be discussed below.
Imagine listening to the sound of rain landing on a roof:
this is the shot noise produced by the random arrivals of a large number
of individual raindrops. The intensity of the sound depends on
how hard it is raining, i.e. the arrival rate of the 
raindrops.\footnote{The raindrop size should be kept fixed for the analogy to hold.}
If the raindrop arrival rate changes with time, 
as often occurs over timescales of seconds to minutes,
the intensity of the sound will vary accordingly.
Thus, while a measurement of the total precipitation may be made
by integrating the acoustic shot noise intensity over time,
this shot noise measurement will reflect the fluctuations of the 
raindrop arrival rate in the same way as would a direct measurement,
e.g. observations of the water level in a standard rain gauge.
Note that the spectral character or the ``sound'' of the acoustic rain noise remains
constant as the intensity changes.
Furthermore, note that the connection between the two measurement methods --
acoustic noise vs. rain gauge -- 
is purely classical and has nothing to do with quantum mechanics.
This statement is also true for the photon detection problem.
Indeed, the output of a photodetector is an entirely classical quantity -- 
a train of electrical pulses -- whose
properties are fully specified by the statistics of the photon arrival times.
While we may need to turn to quantum mechanics to calculate
the arrival time statistics, once the arrival time statistics are known,
in principle we can generate a simulated classical pulse train numerically,
as illustrated in Figure~\ref{fig:PulsePlot},
and use this time stream to calculate any other quantity of interest, 
e.g. the mean and variance of the photon counts, 
or the mean and variance of the shot noise intensity,
or the correlation between the photon counts and shot noise intensity, etc.
These quantities are all related to various moments of the same classical time stream.
It is therefore not surprising that bunching affects
standard photon counting measurements and photon shot noise measurements
in similar ways, and therefore the sensitivity degradation
due to bunching cannot be avoided.
Indeed, in section~\ref{sec:Correlation}
I present a calculation
that demonstrates that the shot noise intensity has a nonzero
correlation with the photon counts,
and then use this correlation
to establish a rigorous sensitivity bound for the shot noise measurement.
This bound shows that the shot noise measurement is subject to 
the same $\sqrt{1+n}$ sensitivity degradation due to photon bunching
as for ordinary photon counting.

Let us continue to accept the claim that photon bunching can
be correctly described by a photon arrival rate that varies randomly
with time. Would we expect to see the noise spectra
illustrated in Figures~\ref{fig:SingleDetectorNoiseSpectrum} and
\ref{fig:TwoDetectorNoiseSpectrum},
which were derived from the quantum calculations presented
in sections~\ref{sec:PhotocurrentSpectrumSingle}
and \ref{sec:PhotocurrentSpectrumMultipleDetectors} ?
It is helpful to visualize the 
detector outputs as a function of
time as shown in Figure~\ref{fig:PulsePlot}.
The detector bandwidth is assumed to be larger than the
photon arrival rate, $\nu_d > \Gb$, 
so the detected photons are visible as sharp, well-separated output pulses.
For the case of two detectors, taking the difference of the two
outputs means that the pulses may be positive or
negative depending on which detector receives the photon.
The two subplots on the left correspond to the case that
the photon arrival rate is kept constant, 
which produces Poisson statistics;
in contrast, the two subplots on the right were generated using a
time-variable photon arrival rate in order to simulate photon bunching.
Imagine that these output time streams are averaged over a timescale $\tau$ that is
long compared to $1 / \Gb$. 
For the single detector case,
Poisson arrivals would produce a DC component
along with small fractional fluctuations of order $1/\sqrt{\Gb \tau}$.
Meanwhile, a time-variable arrival rate would lead to a DC component
along with significantly larger fluctuations,
in accordance with the noise spectrum shown in 
Figure~\ref{fig:SingleDetectorNoiseSpectrum}.
For two detectors,
the positive and negative pulses largely cancel when performing
the time average. This cancellation occurs for both the Poisson and bunched cases,
in agreement with the noise spectrum shown in 
Figure~\ref{fig:TwoDetectorNoiseSpectrum}.
However, both positive and negative pulses deliver the same
high-frequency energy, on average, to the subsequent circuitry and thus contribute
equally to the shot noise intensity. Therefore, the shot noise intensity
for the single-detector and two-detector cases should
be the same.
Thus, our conclusion is that a description of photon bunching
in which the photon arrival rate
varies randomly with time could indeed reproduce the
noise spectra in Figures~\ref{fig:SingleDetectorNoiseSpectrum}
and \ref{fig:TwoDetectorNoiseSpectrum}.

Is it in fact correct to view photon bunching as resulting from a time-varying photon arrival rate?
Is Figure~\ref{fig:PulsePlot} a faithful depiction of 
the photon bunching?
Indeed, this was how \citet{Han57} viewed the phenomenon in their original work.
In their words:
``... we are dealing essentially with an interference phenomenon which can
be interpreted, on the classical wave picture, as a correlation between 
intensity fluctuations due to beats between waves of different frequency;
the concept of a photon need only be introduced at the stage where energy
is extracted from the light beam in the process of photoemission.''
This physical picture is why the excess noise due to photon bunching
continues to be referred to as ``wave noise''.
 \citet{Han57} computed the effect using exactly this semiclassical
 approach, in which the light is first treated as a classical wave, consisting of a random
 superposition of components at different frequencies, resulting in an intensity that
has fractional variations of order unity that occur on a ``coherence'' 
timescale $\tau \sim \Delta \nu^{-1}$
that is set by the fastest beat frequency that can be produced if the spectrum is
restricted to an optical bandwidth $\Delta \nu$.
The photoemission rate is assumed to be proportional to the light intensity,
and therefore the output of each photon detector may be described by a compound
Poisson process in which the photon arrival rate varies stochastically with time.
The classical light intensities calculated
for the two detectors shown in Figure~\ref{fig:TwoDetectors} would be identical;
\citet{Han57} therefore conclude that although the photoemission rates
for the two detectors both fluctuate,  the fluctuations of these rates are perfectly correlated,
and this leads to a nonzero correlation of the detector outputs.
A similar semiclassical approach involving a compound Poisson process
was described by \citet{Man59}.
Section~\ref{sec:SemiclassicalSingleDetector} presents a semiclassical
analysis of the sensitivity of a shot-noise measurement scheme
using a single detector; the case of multiple detectors in 
treated in section~\ref{sec:SemiclassicalMultipleDetectors}.
The conclusion of the semiclassical analysis
for both cases is that the 
shot noise schemes cannot improve on the $\sqrt{N(1+n)}$ 
bunching-limited sensitivity for standard photon counting.

In addition to this historical basis,
the interpretation of photon bunching in terms of a time-variable
photon arrival rate is both supported by experiment
and fully consistent with the predictions of quantum mechanics.
The fact that the photon arrival rate for thermal radiation
does indeed vary with time was directly demonstrated in the laboratory
by \citet{Mor66} through measurement of the correlation of the arrival times
of individual photons at a single detector.
Indeed, \citet{Mor66} give a simple, concise description of the photon bunching effect:
``In time intervals of order or less than the coherence time of the light,
the probability of counting two pulses is greater than that expected
for random events'',
just as depicted in Figure~\ref{fig:PulsePlot}.
Furthermore, the results of the semiclassical analysis
which invokes a stochastic, time-variable photon arrival rate
are reproduced by a full quantum calculation.
\cite{Kel64} describe
a quantum-mechanical theory of photon detection 
that uses a density matrix to describe the state of the electromagnetic field;
the HBT photon bunching effect may be studied through use of a
density matrix appropriate for thermal radiation.
A fully quantum analysis for shot-noise measurements
using a single detector is described in section~\ref{sec:ShotNoiseQuantum},
and agrees with the semiclassical results in section~\ref{sec:SemiclassicalSingleDetector}.
As discussed in section~\ref{sec:ShotNoiseQuantum},
the extension of the fully quantum analysis to the case of two detectors is
straightforward and agrees with the corresponding semiclassical analysis in
section~\ref{sec:SemiclassicalMultipleDetectors}.
Thus, both the semiclassical and fully quantum
calculations show that the bunching noise
cannot be evaded through use of a shot noise measurement scheme,
whether one uses one or two detectors.
This agreement is a reflection of
the equivalence of the quantum and semiclassical
descriptions of light as shown by \citet{Sud63}, who made use of the coherent state
representation introduced by \cite{Gla63}.
In Appendix~\ref{app:QuantumVsSemiclassical},
the equivalence of the quantum and semiclassical (i.e., compound Poisson)
descriptions of the photocurrent statistics is demonstrated explicitly.
Thus, the interpretation of photon bunching in terms of
a time-varying photon arrival rate
as illustrated in Figure~\ref{fig:PulsePlot}
is in fact predicted by the full quantum theory and also
supported by laboratory measurements.

A potentially confusing aspect of the discussion
is the fact that the shot noise spectrum is 
white -- indeed, the output spectrum for the two-detector case 
(Figure~\ref{fig:TwoDetectorNoiseSpectrum})
is flat and featureless.
Where is the bunching noise hiding?
The answer is simple: the shot noise spectrum is
white regardless of whether the photon arrival rate
is constant or if it varies randomly with time due to bunching,
as can easily be understood.
The variation of shot noise intensity
due to a randomly-varying event rate is similar
to that of steady shot noise subjected to a random amplitude modulation.
The effect of amplitude modulation (AM) of a sinusoidal carrier
is very well known to radio engineers:
modulation sidebands are produced below and above the carrier frequency.
Mathematically, a carrier at frequency $\nu_c$ that is AM-modulated
at frequency $\nu_m$ develops sidebands at $\nu_\pm = \nu_c \pm \nu_m$:
\begin{align}
\cos(2 \pi \nu_c t) & \left[ 1 + a \cos(2 \pi \nu_m t) \right] = \cos(2 \pi \nu_c t) 
\nonumber \\
&  + \frac{a}{2} \cos (2 \pi \nu_+ t) + \frac{a}{2} \cos (2 \pi \nu_- t)\ .
\end{align}
This result is easily generalized to a Fourier superposition of modulation frequencies.
Thus, the bunching noise component illustrated in Figure~\ref{fig:SingleDetectorNoiseSpectrum}
may be interpreted as the sidebands on a DC 
carrier -- the mean photocurrent -- that are produced by the random
modulation of the photon arrival rate. 
Indeed, these sidebands extend out to
$\pm \Delta \nu$ which corresponds to the bandwidth of the 
arrival rate (or light intensity) fluctuations.
Similarly, a Fourier component of shot noise at some frequency $\nu_1$ will develop sidebands
extending over $\nu_1 \pm \Delta \nu$ as a result of the arrival rate fluctuations.
Because all Fourier components of shot noise develop these sidebands in the same way,
it is clear that the net result must be a white spectrum.
However, the sideband generation process introduces the possibility that the
Fourier components of shot noise
at different frequencies are correlated.
Shot noise with a variable event rate arises in other contexts and is well studied,
e.g. in the theory of diode mixers \citep{Hel78}
or in the detection of fast optical pulse trains \citep{Qui13}.
For these examples, the event rate modulation is deterministic and periodic
and the correlations between different Fourier components
of shot noise play an essential role.
However, these correlations vanish for the present case of photon bunching
because the event rate varies randomly rather than deterministically.
Indeed,  the photon shot noise must be a stationary process,
in the sense that all statistics such as the autocorrelation function $\ave{I(t) I(t+\tau)}$
are invariant under time translation $t \rightarrow t + t_1$,
because the time-varying photon arrival rate is also a stationary random
process.
A translation in time changes the phase of a product of two
different Fourier components, 
$\hat{I}^*(\nu) \hat{I}(\nu') \rightarrow \hat{I}^*(\nu) \hat{I}(\nu') e^{-i 2 \pi (\nu - \nu') t_1}$,
and therefore time-translation symmetry
requires that the correlation of different Fourier components vanish.
A more detailed mathematical proof of these statements is given in 
section~\ref{sec:ShotNoise},
culminating with equation~(\ref{eqn:ShotNoiseSpectrumRandomRate}),
which agrees with previous work \citep{Pic70}.
Thus, the fact that the shot noise spectrum is white
tells us nothing about possible time-dependent variations of the shot noise intensity.
So where is the  bunching noise hiding ?
To find it, we must go beyond the noise spectrum,
which relates to the second order statistics of the photocurrent,
and look at the fourth-order photocurrent statistics that are needed to describe
the fluctuations of the shot noise intensity.
This calculation is presented in
section~\ref{sec:SemiclassicalSingleDetector},
leading to equation~(\ref{eqn:ShotNoiseUncertaintyC}),
which reveals the presence of the bunching noise in the shot noise intensity.

\section{Photocurrent spectrum for a single detector}
\label{sec:PhotocurrentSpectrumSingle}

I now turn to a straightforward quantum-mechanical calculation of
the output noise spectrum of a detector that is illuminated with
filtered thermal radiation with bandwidth $\Delta \nu$ and
occupation number $n$,
as in the setup shown in Figure~\ref{fig:SourceFilterDetector}.
The treatment uses a conventional quantum formalism 
described in \citet{Zmu03};
\tmpbf{the calculations presented in this section are fairly standard and
mainly serve to introduce the formalism and notation.}
The principal result, stated below in equation~(\ref{eqn:PhotocurrentSpectrumSingle})
and illustrated in Figure~\ref{fig:SingleDetectorNoiseSpectrum},
shows that the spectrum consists of three components:
(1) a DC term corresponding
to the average output; (2) a component due to photon bunching
that is confined to a bandwidth equal to the optical bandwidth $\Delta \nu$;
and (3) a white-noise component due to photon shot noise
that is limited only by the detector output bandwidth $\nu_d$.
For conventional photon counting, the observable quantity is the
time integral of the photocurrent,\footnote{The time integral of the photocurrent
is a useful and analytically tractable quantity for quantifying the performance of ideal detectors
but is not the optimal statistic for real detectors that have additional
non-ideal sources of noise, e.g. amplifier noise.
For example, one might use
Wiener filtering followed by peak detection to locate 
and count the individual photon pulses in the output timestream;
this approach rejects most of the noise emanating from
the detection system during the time intervals between photon events.}
which makes use of the fact
that the DC photocurrent is proportional  
to the mean photon rate $\Gb = n \Delta \nu$.
For the alternative shot noise measurement technique,
the observable is the time integral of the noise intensity
in the white-noise region $\Delta \nu \le |\nu| \le \nu_d$,
and makes use of the proportionality of the shot noise intensity
to the mean photon rate $\Gb$.

Consider an ideal photon detector illuminated by a single mode of the
radiation field. The radiation field is described by photon creation and destruction
operators,
\begin{align}
b^\dagger(t) &= \int_0^\infty d\nu \, e^{-i 2 \pi \nu t}\, b^\dagger(\nu)
\nonumber \\
b(t) &= \int_0^\infty d\nu \, e^{+i 2 \pi \nu t} \, b(\nu)
\nonumber
\end{align}
which are defined only for positive frequencies
and obey Bosonic commutation relations
\be
[b(\nu), b^\dagger(\nu')] = \delta(\nu-\nu')\ .
\ee
I assume that the radiation field is in a thermal state described by
the density matrix
\be
\rho = \exp\left({\int_0^\infty  d\nu\, \left\{-x(\nu)\, b^\dagger(\nu) b(\nu) 
+ \ln \left[ 1 - e^{-x(\nu)} \right] \right\} }\right)\ .
\label{eqn:ThermalDensityMatrix}
\ee
This is of the standard form for thermal equilibrium, $\rho \propto \exp(-H/kT)$, 
where the Hamiltonian  consists of a sum of harmonic oscillators,
\be
H = \int_0^\infty d\nu h \nu \, b^\dagger(\nu) b(\nu)\ ,
\ee
and where $x(\nu) = h \nu / kT$ is the normalized inverse temperature.
The $\ln[1 - e^{-x(\nu)}]$ term provides the required normalization $\tr \rho = 1$.
It is readily shown \citep{Zmu03}
that this density matrix gives expectation values of
\be
\ave{b^\dagger(\nu) b(\nu')} = \tr[\rho\, b^\dagger(\nu) b(\nu')] = n(\nu) \delta(\nu - \nu')
\ee
where the occupation number is given by the Bose-Einstein formula
\be
n(\nu) = \frac{1}{e^{x(\nu)} - 1} = \frac{1}{e^{h \nu / kT} - 1}\ .
\ee
Note that the excitation temperature need not be the same for all frequencies;
we may easily generalize to $x(\nu) = h \nu / kT(\nu)$.

An ideal photodetector produces one pulse at its output for every photon absorbed.
For example, in a superconductor-insulator-superconductor (SIS) tunnel junction
detector \citep{Tuc78}, 
each absorbed photon causes one electron to tunnel across the junction.
Such a detector may be described by a Hermitian photocurrent operator
\be
I_F(t) = \int_{-\infty}^t dt' \, F(t-t') b^\dagger(t') b(t')
\ee
where $F(t)$ describes the shape of the current pulse produced by one photon.
The detector output need not be an electrical current.
More generally, we can consider $I_F(t)$ to be the output signal of the detector
when illuminated by the radiation field, and $F(t)$ to be the output signal
produced when a single photon is absorbed;
however I will continue to call $I_F(t)$ the photocurrent operator.
Note that I am assuming that the
detector is operating in a linear regime: doubling the photon absorption
rate doubles the output signal.
If the detector response is fast relative to the timescales of interest,
we may approximate 
\be
F(t) \approx \delta(t)\ ,
\ee
and therefore consider the Hermitian operator
\be
I(t) = b^\dagger(t) b(t)\ 
\label{eqn:PhotocurrentOperator}
\ee
where now $I(t)$ has units of $s^{-1}$.

The impact of photon bunching on the sensitivity of measurements
performed using conventional photon counting
may be demonstrated by considering the operator
\be
N_T = \int_{-T/2}^{T/2} dt\, I(t)\ ,
\label{eqn:NTdef}
\ee
which represents the number of photons detected in a measurement 
time interval $[-T/2, T/2]$.
The mean and variance of this operator are given
in equations 41 and 42 of \citet{Zmu03}: 
\begin{align}
\ave{N_T} & = \tr \left( \rho N_T \right)  = T \int_0^\infty d\nu \, n(\nu)
\label{eqn:NTave}
\\
\sigma_{N_T}^2  & = \ave{N_T^2} - \ave{N_T}^2 = T \int_0^\infty d\nu \, n(\nu) \left(1 + n(\nu) \right)\ .
\label{eqn:sigmaNT}
\end{align}
Thus, the fractional measurement uncertainty is
\be
\frac{\sigma_{N_T}}{\ave{N_T}} = \frac{\sqrt{1 + \tilde{n}}}{\sqrt{\ave{N_T}}}
\label{eqn:BunchingFractionalUncertainty}
\ee
and is degraded by $\sqrt{1 + \tilde{n}}$ due to photon bunching
as compared to the 
fractional Poisson uncertainty of $1 / \sqrt{\ave{N_T}}$.
Here the effective occupation number $\tilde{n}$ is defined by
\be
\tilde{n} = \frac{\int_0^\infty d\nu \, \left[ n(\nu)\right]^2}{\int_0^\infty d\nu \, n(\nu)}\ .
\ee
For the simple case that $n(\nu) = n$ inside an optical bandwidth $\Delta \nu$ and zero outside,
one readily finds $\tilde{n} = n$.

The calculation of the output noise spectrum of the detector
makes use of the Fourier transform of the photocurrent operator:
\begin{align}
\hat{I}(\nu) & = \hat{I}^\dagger(-\nu) 
\nonumber \\
&= \int_{-\infty}^{+\infty} dt \, e^{- i 2 \pi \nu t} I(t)
\nonumber \\
&= \int_{-\infty}^{+\infty} dt \, e^{- i 2 \pi \nu t} \, b^\dagger(t) b(t)\
\nonumber \\
&= \int_0^\infty d\nu_1 \int_0^\infty d\nu_2 
b^\dagger(\nu_1) b(\nu_2)
\nonumber \\
& \times
\int_{-\infty}^{+\infty} dt  \, e^{+i 2 \pi \nu_1 t} \, e^{- i 2 \pi \nu_2 t} \, e^{- i 2 \pi \nu t}
\nonumber \\
&= \int_0^\infty d\nu_1  \, b^\dagger(\nu_1+\nu) b(\nu_1) \theta(\nu_1 + \nu)\ .
\label{eqn:PhotocurrentSpectrumOperator}
\end{align}
The photon operators $b$ and $b^\dagger$ are
only defined for positive frequencies;
the unit step function $\theta(\nu_1 + \nu)$ is needed to guarantee
$\nu_1 + \nu \ge 0$ even when $\nu <0 $.
For brevity of notation, I will not write the step function 
explicitly but instead rely on the interpretation
$b(\nu_1) \rightarrow b(\nu_1) \theta(\nu_1)$ and similarly for $b^\dagger(\nu_1)$.
Clearly, the effect of a finite pulse width $F(t)$ will be multiplication by
the corresponding frequency-domain filter $\hat{F}(\nu)$,
\be
\hat{I_F}(\nu)  = \int_{-\infty}^{+\infty} dt \, e^{- i 2 \pi \nu t} I_F(t) = \hat{F}(\nu) \hat{I}(\nu)\ .
\ee
The power spectrum $S_I(\nu) = S_I(-\nu)$ of the photocurrent is defined by
\be
\ave{\hat{I}^\dagger(\nu) \hat{I}(\nu')} = S_I(\nu) \, \delta(\nu - \nu')\ ;
\ee
including the pulse shape $F(t)$ would lead to a filtered power spectrum
\be
S_{I_F}(\nu) = \left| F(\nu) \right|^2 S_I(\nu)\ .
\ee
Because the thermal density matrix is gaussian,
the required expectation value may be found by 
combining the photon operators pairwise,
\begin{align}
\ave{\hat{I}^\dagger(\nu) \hat{I}(\nu')} 
&= 
\int_0^\infty d\nu_1   d\nu_2  \, 
\nonumber \\ &
\ave{ b^\dagger(\nu_1) b(\nu_1+\nu) 
b^\dagger(\nu_2+\nu') b(\nu_2)}
\nonumber \\
&=
\int_0^\infty d\nu_1  d\nu_2  \, 
\nonumber \\ &
\ave{ b^\dagger(\nu_1) b(\nu_1+\nu)} 
\ave{b^\dagger(\nu_2+\nu') b(\nu_2)}
\nonumber \\
& + 
\ave{ b^\dagger(\nu_1)  b(\nu_2)}
\ave{b(\nu_1+\nu) b^\dagger(\nu_2+\nu')}
\nonumber \\
&= 
 \delta(\nu - \nu')
\int_0^\infty d\nu_1  d\nu_2  \, 
\left\{ n(\nu_1) n(\nu_2) \delta(\nu)    \right.
\nonumber \\
& + \left.
n(\nu_1) \left[ n(\nu_1+\nu) + 1 \right] \delta(\nu_1 - \nu_2) \right\}  \ .
\label{eqn:PhotocurrentPowerSpectrumCalc}
\end{align}
The photocurrent power spectrum is therefore given by a sum of three terms,
\be
S_I(\nu) = 
{\Gb}^2 \, \delta(\nu)  + \Gb
 + \int_0^\infty d\nu_1 \, n(\nu_1) n(\nu_1+\nu)  \ ,
\label{eqn:PhotocurrentSpectrumSingle}
\ee
where, according to eqn.~(\ref{eqn:NTave}), the mean photon arrival rate is
\be
\Gb =  \int_0^\infty d\nu_1 \, n(\nu_1)\ .
\ee
When $n(\nu)$ is constant within a bandpass $\Delta \nu$ and zero outside,
we obtain $\Gb = n \Delta \nu$, 
so the occupation number $n$ may be interpreted as the
number of photons per second per Hertz of optical spectrum.

The first term in $S_I(\nu)$, proportional to $\delta(\nu)$, represents the contribution to the
power spectrum from the  DC value of the photocurrent.
At nonzero frequencies, only the second and third term contribute.
The second term, $\Gb$,  is white noise independent of frequency $\nu$,
and represents photon shot noise.
The third term is due to photon bunching, and is not white.
Indeed, for a rectangular optical bandpass of width $\Delta \nu$,
the spectrum of the bunching term has a triangular shape that
is symmetric with respect to $\nu = 0$ and
extends over $- \Delta \nu \le \nu \le + \Delta \nu$:
\be
S_I^\mathrm{(bunching)}(\nu) = 
\left\{
\begin{array}{cl}
n \Gb \left[1 - \frac{|\nu|}{\Delta \nu} \right] & |\nu| \le \Delta \nu \\
0 & \mathrm{otherwise} 
\end{array}
\right.
\ee
The sum of these three terms is plotted in Figure~\ref{fig:SingleDetectorNoiseSpectrum}.
At high frequencies $|\nu| > \Delta \nu$, only the white photon shot
noise term contributes. This suggests the following idea: 
place a high-pass filter $\hat{F}(\nu)$ at the detector output
that transmits only at frequencies $|\nu| > \Delta \nu$.
The spectral density of the shot noise is $\Gb$;
we can therefore measure the photon rate by measuring the
noise intensity at $|\nu| > \Delta \nu$.
However, as we shall see, this method does not avoid
the sensitivity degradation due to photon bunching.

\section{Photocurrent cross-spectrum for multiple detectors}
\label{sec:PhotocurrentSpectrumMultipleDetectors}

\begin{figure}[htb]
\begin{center}
\includegraphics[width=3.25in]{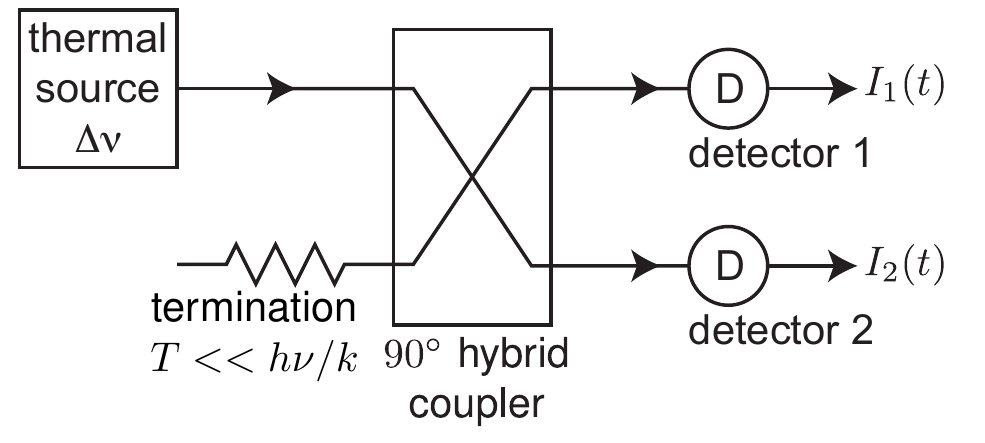}
\caption{The radio-frequency equivalent of the two-detector
plus beamsplitter setup shown in Figure~\ref{fig:TwoDetectors}.
Free-space propagation is replaced by
guided-wave propagation in transmission lines or waveguides,
the function of the beamsplitter is performed by a $90^\circ$ 3~dB hybrid 
coupler \citep{Poz12},
and the unused port of the coupler is connected to a cold termination.
As shown by equation~(\ref{eqn:PhotocurrentSpectrumDifference}),
the spectrum of the difference of the photocurrents
$I_\Delta(t) = I_1(t) - I_2(t)$ does not contain the DC or bunching
noise components, and is therefore white across the full 
detector output bandwidth.
\label{fig:CouplerDiagram}}
\end{center}
\end{figure}

I now generalize the discussion of section~\ref{sec:PhotocurrentSpectrumSingle}
to the case of multiple detectors, 
in order to
analyze the two-detector scheme illustrated in Figure~\ref{fig:TwoDetectors},
or its radio-frequency equivalent shown in Figure~\ref{fig:CouplerDiagram}.
The detection scheme proposed by \citet{Lieu15} uses an identical 
two-detector setup.
I start with a more general case in which an arbitrary passive linear optical
system is used to illuminate a set of detectors,
and calculate the cross-spectral density of the output currents.
\tmpbf{The principal result (equation~\ref{eqn:PhotocurrentSpectrumMultiple})
is a straightforward generalization of equation~(\ref{eqn:PhotocurrentSpectrumSingle})
for a single detector; to our knowledge this is a new result.}
Following \citet{Zmu03},
the optical system is represented by a
passive linear $N$-port network with a scattering matrix $S$;
a 50/50 beamsplitter is an example of a four-port network.
The network is illuminated by incoming radiation described by 
the photon operators $a_i(\nu)$, 
and produces outgoing radiation according to the scattering equation
\be
b_i(\nu) = \sum_j S_{ij}(\nu) a_i(\nu) + c_i(\nu)\ ,
\ee
where the $c_i(\nu)$ are operators representing noise added by the network.
Here the indices $1 \le i, j \le N$ label the ports of the network.
The noise operators satisfy commutation relations
\be
[c_i(\nu), c_j(\nu')] = [1 - S(\nu) S^\dagger(\nu)]_{ji} \delta(\nu-\nu')\ ,
\ee
as required to preserve the Bosonic commutation relations for 
the output operators,
\be
[b_i(\nu), b^\dagger_j(\nu')] = \delta_{ij} \delta(\nu-\nu')\ ,
\ee
given that the input operators also satisfy the same commutation relations,
\be
[a_i(\nu), a^\dagger_j(\nu')] = \delta_{ij} \delta(\nu-\nu')\ .
\ee

If the input radiation is thermal and the $N$-port is passive,
the output radiation is also thermal and may be fully described
by a mode occupation matrix $B_{ij}(\nu)$, defined through
\be
\ave{b_i(\nu) b_j^\dagger(\nu')} = B_{ij}(\nu) \delta(\nu - \nu')\ ,
\label{eqn:BmatrixDef}
\ee
which is a generalization of the mode occupation number $n(\nu)$.
Photocurrent operators and their Fourier transforms
may be introduced for each port:
\begin{align}
I_i(t) &= b_i^\dagger(t) b_i(t)\ .
\\
\hat{I}_i(\nu) &= \int_0^\infty d\nu_1  \, b_i^\dagger(\nu_1+\nu) b_i(\nu_1)\ .
\end{align}
The photocurrent power cross-spectrum $C_{ij}(\nu)$ is defined through the expression
\be
\ave{\hat{I}^\dagger_i(\nu) \hat{I}_j(\nu')} = C_{ij}(\nu)\, \delta(\nu - \nu')\ ,
\ee
which I calculate
using pairwise evaluation of the resulting fourth-order moments of the photon operators,
as for eqn.~(\ref{eqn:PhotocurrentPowerSpectrumCalc}).
The power cross-spectrum is found to be
\begin{align}
C_{ij}(\nu) & = 
\Gb_i \Gb_j \, \delta(\nu)  + \Gb_i \, \delta_{ij}
\nonumber \\
&
 + \int_0^\infty d\nu_1 \, B_{ij}(\nu_1) B_{ji}(\nu_1+\nu)  \ ,
\label{eqn:PhotocurrentSpectrumMultiple}
\end{align}
where the mean photon rates at the detectors are
\be
\Gb_i =  \int_0^\infty d\nu_1 \, B_{ii}(\nu_1)\ .
\ee
As was found for the single detector case and illustrated in
Figure~\ref{fig:SingleDetectorNoiseSpectrum},
we see that the power cross-spectrum 
$C_{ij}(\nu)$
for multiple detectors consists of three terms:
a DC term $\left( \Gb_i \Gb_j \, \delta(\nu) \right)$, 
a white spectrum from photon shot noise that is uncorrelated
between detectors  $\left( \Gb_i \, \delta_{ij} \right)$,
and a  bunching spectrum 
\be
C^\mathrm{(bunching)}_{ij} (\nu) = \int_0^\infty d\nu_1 \, B_{ij}(\nu_1) B_{ji}(\nu_1+\nu)
\ee
that exhibits correlations between detectors but is limited to frequencies
$\nu \le \Delta \nu$.

We now specialize to a four-port network 
appropriate for a beamsplitter or $90^\circ$ 3~dB coupler \citep{Poz12},
as illustrated in Figures~\ref{fig:TwoDetectors} and \ref{fig:CouplerDiagram},
with a scattering matrix given by
\be
S = \frac{1}{\sqrt{2}}
\left[
\begin{array}{cccc}
0 & 0 & i & 1 \\
0 & 0 & 1 & i \\
i & 1 & 0 & 0 \\  
1 & i  & 0 & 0
\end{array}
\right]\ .
\ee
This matrix is reciprocal, $S^T = S$, as required by time-reversal symmetry,
and $S$ is also unitary, $S S^\dagger = 1$, and therefore the network is lossless.
We will call ports $1$ and $2$ the output ports and place detectors on them,
and ports $3$ and $4$ will serve as the input ports.
The incoming fields $a_i(\nu)$ are assumed to be in independent
thermal states described by occupation numbers $n_i(\nu)$.
Port $4$ will be illuminated with occupation number $n_4(\nu)$.
Port $3$ will be terminated in a vacuum state with zero occupation
number; furthermore,
the detectors are assumed to be cold and therefore do not
radiate toward the beamsplitter, so $n_1(\nu) = n_2(\nu) = n_3(\nu) = 0$.
We may now calculate
\begin{align}
B_{ij}(\nu) &= \sum_k S_{ik}(\nu) S_{jk}^*(\nu) n_k(\nu)\ .
\nonumber \\
&= S_{i4}(\nu) S_{j4}^*(\nu) n_4(\nu)\ .
\end{align}
Therefore,
\be
B_{11}(\nu) = B_{22}(\nu) = \frac{n_4(\nu)}{2}
\label{eqn:BbeamsplitterDiag}
\ee
while
\be
B_{12}(\nu) = B_{21}^*(\nu) = -i \frac{n_4(\nu)}{2}\ .
\label{eqn:BbeamsplitterOffDiag}
\ee
Thus, the bunching power cross-spectrum for the detector ports $1$, $2$ is
\be
C^\mathrm{(bunching)} = 
\frac{1}{4} \left[
\begin{array}{cc}
1 & 1 \\
1 & 1 \\
\end{array}
\right]
\,
\int_0^\infty d\nu_1 \, n_4(\nu_1) n_4(\nu_1+\nu) \ ;
\label{eqn:PhotocurrentSpectrumBeamsplitterBunching}
\ee
the bunching noise is fully correlated between the detectors.
The matrix in this expression has eigenvalues of $2$ and $0$ for 
the symmetric and antisymetric eigenvectors $(1, 1)$ and $(1, -1)$.
We therefore see that the bunching noise term will be absent 
for the difference of the two detector photocurrents, 
$I_\Delta(t) = I_1(t) - I_2(t)$.
Neglecting the DC term, the noise matrix is
\begin{align}
C(\nu) & = 
\frac{1}{2} 
\left[
\begin{array}{cc}
1 & 0 \\
0 & 1 \\
\end{array}
\right] 
\,
\int_0^\infty d\nu_1 \, n_4(\nu_1)
\nonumber \\
& +
\frac{1}{4} 
\left[
\begin{array}{cc}
1 & 1 \\
1 & 1 \\
\end{array}
\right]
\,
\int_0^\infty d\nu_1 \, n_4(\nu_1) n_4(\nu_1+\nu) 
\label{eqn:PhotocurrentSpectrumBeamsplitter}
\end{align}
where the first term represents the shot noise,
which is uncorrelated between detectors.
This is the only term that remains when we
calculate the spectral density of $I_\Delta$,
\begin{align}
C_\Delta(\nu) & = 
\left[
\begin{array}{cc}
1 & -1
\end{array}
\right] 
\, C(\nu) \,
\left[
\begin{array}{c}
1 \\
-1
\end{array}
\right]
\nonumber \\
&  =  \int_0^\infty d\nu_1 \, n_4(\nu_1) = \Gb\ ,
\label{eqn:Cdelta}
\end{align}
which is the same as the shot noise intensity for a single detector without the beamsplitter.
To summarize, the difference of the two detector currents 
$I_\Delta(t) = I_1(t) - I_2(t)$ has a white
spectrum, given by
\be
\ave{\hat{I}^\dagger_\Delta(\nu) \hat{I}_\Delta(\nu')} = \Gb \delta(\nu - \nu')\ ,
\label{eqn:PhotocurrentSpectrumDifference}
\ee
and illustrated in Figure~\ref{fig:TwoDetectorNoiseSpectrum}.

\section{Spectrum of Shot Noise with a Variable Event Rate}
\label{sec:ShotNoise}

As discussed in section~\ref{sec:Introduction},
in the semiclassical picture one views photon bunching
as being caused by a stochastically varying photon arrival rate.
Here I calculate the spectrum of classical shot noise for
a time-varying event rate
and demonstrate that the shot noise remains white and uncorrelated,
as discussed qualitatively in section~\ref{sec:Introduction}.
I further demonstrate that the output noise spectrum of a single
detector calculated in the semiclassical picture can reproduce
the quantum-mechanical result given in section~\ref{sec:PhotocurrentSpectrumSingle}.
\tmpbf{The principal result is given by equation~(\ref{eqn:ShotNoiseSpectrumRandomRate}),
which reproduces an earlier result of \cite{Pic70};
the purpose of presenting the detailed derivation here is to introduce
the formalism in preparation for the calculation of
shot noise intensity fluctuations in section~\ref{sec:SemiclassicalSingleDetector}.}

A classical current containing shot noise,
e.g. the current across a tunnel barrier with a low transmission probability,
may be considered to be a sum of impulses,
\be
I(t) = \sum_{i} \delta(t - t_i)
\ee
where $\left\{t_i\right\}$ represent the times at which discrete
charges (e.g., electrons) jump across the barrier. 
To remain consistent with sections~\ref{sec:PhotocurrentSpectrumSingle}
and \ref{sec:PhotocurrentSpectrumMultipleDetectors},
I have omitted the usual factor of electron charge $e$,
so $I(t)$ has units of $\mathrm{s}^{-1}$ or $\mathrm{Hz}$.
Suppose further that the average current is time-dependent,
\be
\ave{I(t)} = \Gamma(t) = \Gb + \dG(t)\ ,
\ee
where $\Gb$ is the mean event rate and $\dG(t)$
represents variations in the rate and therefore has zero mean.
If the event rate $\Gamma(t)$ is constant, i.e. $\dG(t) = 0$,
we know that the current has a shot noise spectrum that is white
and has intensity $\Gb$. 
On the other hand, if the event rate varies with time, we
expect the shot noise intensity to also vary.
Thus, a time-resolved measurement of the shot noise
intensity should allow us to measure the corresponding time-dependent current.
This possibility is investigated here and further in section~\ref{sec:SemiclassicalSingleDetector}.

%

I follow the approach of \citet{Kel64}
for calculating classical shot-noise statistics;
a similar but mathematically more formal approach is given by
\citet{Pic70}.
Let $y_i$ be a random variable
that represents the number  of charges that flow during
the time interval $[t_i, t_i+\Delta t_i]$.
I assume that $y_i$ is independent of all other $y_j$ for $j \ne i$,
and furthermore that for sufficiently small time intervals $\Delta t_i$,
$y_i$ has a probability distribution given by
\begin{align}
P(y_i = 1) &= \Gamma(t_i) \Delta t_i
\nonumber \\
P(y_i = 0) &= 1 - \Gamma(t_i) \Delta t_i \ .
\nonumber
\end{align}
The number of charges that cross during the time interval $[0, T]$
is a random variable given by
\be
N_T = \int_0^T I(t) dt \approx \sum_{i=1}^M y_i 
\ee
where the time interval $[0, T]$
has been split into $M$ nonoverlapping subintervals $[t_i, t_i + \Delta t_i]$.
The distribution of $N_T$ is encoded by the
moment generating function calculated in Appendix~\ref{app:PoissonGeneratingFunction}:
\begin{align}
G_N(s) & = \ave{e^{s N_T} }
= \sum_{k=0}^\infty \frac{s^k}{k!}
\sum_{i_1 ... i_k = 1}^{M} \ave{y_{i_1} ... y_{i_k}}
\nonumber  \\
&= \exp \left[ \mu_T \left(e^s - 1 \right) \right] 
\label{eqn:GeneratingFunctionN}
\\
&= e^{-\mu_T} \sum_{k=0}^\infty \frac{\mu_T^k}{k!} \ e^{s k}\ .
\nonumber
\end{align}
As expected \citep{Man58}, 
this result  shows that $N_T$ follows a Poisson distribution with mean
\be
\mu_T = \int_0^T dt \Gamma(t)\ .
\ee
Thus, the current $I(t)$ is a Poisson process with a time-dependent rate $\Gamma(t)$.

The same formalism can be used to
work out the spectrum of shot noise for
a time-dependent current.
The time-limited Fourier transform of the current is defined as
\be
\hI_T(\nu)  = \int_{-T/2}^{+T/2} dt I(t) e^{-i 2 \pi \nu t}\ ,
\ee
which allows the power spectrum to be computed by evaluating the limit
\be
\lim_{T \rightarrow \infty} \ave{\hI_T(\nu) \hI^*_T(\nu')}\ .
\ee
Expressing the current in terms of the random variables $y_i$ gives
\be
\hI_T(\nu) \approx \sum_{i=1}^M y_i e^{-i 2 \pi \nu t_i}\ .
\ee
Therefore,
\be
\ave{\hI_T(\nu) \hI^*_T(\nu')}_y 
\approx
\sum_{i,j = 1}^M \ave{y_i y_j} e^{-i 2 \pi \nu t_i} e^{+i 2 \pi \nu' t_j} \ .
\ee
Now
\begin{align}
\ave{y_i y_j} 
&=
\ave{y_i} \ave{y_j} + \delta_{ij} \left( \ave{y_i^2} - \ave{y_i}^2 \right)
\nonumber \\
&=
\Gamma(t_i) \Gamma(t_j) \Delta t_i \Delta t_j 
\nonumber \\
&
+ \delta_{ij} \left[ \Gamma(t_i) \Delta t_i - \Gamma^2(t_i) \left(\Delta t_i\right)^2 \right]\ ,
\end{align}
making use of $y_i^2 = y_i$.
Inserting the first term into the sum and taking the continuum limit gives
\begin{align}
\lim_{T \rightarrow \infty} &
\int_{-T/2}^{T/2} dt_1 \, \Gamma(t_1) e^{- i 2 \pi \nu t_1}\,
\int_{-T/2}^{T/2} dt_2 \, \Gamma(t_1) e^{+i 2 \pi \nu t_1}\, 
\nonumber \\
& =
\left( \Gb \delta(\nu) + \hdG(\nu) \right) \left( \Gb \delta(\nu') + \hdG^*(\nu') \right) 
\end{align}
while the second term yields
\be
\Gb \delta(\nu - \nu') + \hdG(\nu - \nu')\ ;
\ee
note that the $(\Delta t_i)^2$ term vanishes in the continuum limit.
Thus, the Fourier components of shot noise are correlated
when the event rate varies deterministically with time, according to
\begin{align}
\ave{\hI(\nu) \hI^*(\nu')}_y & = 
\left( \Gb \delta(\nu) + \hdG(\nu) \right) \left( \Gb \delta(\nu') + \hdG^*(\nu') \right) 
\nonumber \\
& + \Gb \delta(\nu - \nu') + \hdG(\nu - \nu')\ . 
\label{eqn:ShotNoiseSpectrumDetermRate}
\end{align}

If the event rate has stochastic time-dependent rate variations $\dG(t)$, 
taken to be a stationary random process with power spectrum $\SG(\nu)$,
the shot noise spectrum may be obtained by
averaging equation~(\ref{eqn:ShotNoiseSpectrumDetermRate}) over $\dG$:
\begin{align}
\ave{\hI(\nu) \hI^*(\nu')}_{y,\dG} & = 
\left[ \Gb^2 \delta(\nu) + \Gb + \SG(\nu)  \right]
\nonumber \\
& \times \delta(\nu - \nu')\ ;
\label{eqn:ShotNoiseSpectrumRandomRate}
\end{align}
the $\delta(\nu - \nu')$ factor indicates that 
different Fourier components are uncorrelated and therefore
this compound Poisson process is stationary, as expected.
Specifically, the shot noise component 
$\Gb \delta(\nu - \nu')$ remains white and uncorrelated,
as promised in section~\ref{sec:Introduction}.
Equation~(\ref{eqn:ShotNoiseSpectrumRandomRate})
agrees with the result of \citet{Pic70} (their equation~2.27),
who claim agreement with an earlier result by Mandel.

The power spectrum of this classical compound Poisson process
may be made identical to the spectrum of the photocurrent 
calculated quantum-mechanically (equation~\ref{eqn:PhotocurrentSpectrumSingle}),
provided we make the identifications
\be
\Gb  = \int_{0}^{+\infty} d\nu_1 n(\nu_1)
\label{eqn:MeanRatePoisson}
\ee
and
\be
\SG(\nu) = \int_{0}^{+\infty} d\nu_1 n(\nu_1) n(\nu_1 + \nu)\ .
\label{eqn:RateFluctuationsPoisson}
\ee
If the occupation number $n(\nu)$ is constant across an optical bandwidth
$\Delta \nu$ and zero outside, the relative importance of the bunching
and Poisson terms at noise frequencies well below $\Delta \nu$
is governed by
\be
\frac{\SG(0)}{\Gb} = 
\frac{\int_{0}^{+\infty} d\nu_1 n^2(\nu_1) }{\int_{0}^{+\infty} d\nu_1 n(\nu_1)}
=
\frac{n^2 \Delta \nu}{n \Delta \nu} = n\ .
\ee

\section{Shot Noise Measurement: Semiclassical Analysis for a Single Detector}
\label{sec:SemiclassicalSingleDetector}

I now turn to the computation of the fluctuations in the
intensity of classical shot noise with a time-varying
event rate and demonstrate that the shot noise
intensity reflects variations in the event rate.
I apply these results
to the case of photon detection under the assumption
that bunching may be described by 
a stochastic photon arrival rate whose mean
and power spectrum are described
by equations~(\ref{eqn:MeanRatePoisson})
and (\ref{eqn:RateFluctuationsPoisson}), respectively.
As discussed in section~\ref{sec:Introduction}
and Appendix~\ref{app:QuantumVsSemiclassical},
this assumption is consistent with the full quantum theory
and is supported by experiment.
\tmpbf{The principal results presented in this section
(equations~\ref{eqn:SingleDetectorFluctuations}, \ref{eqn:ShotNoiseUncertainty},
\ref{eqn:ShotNoiseUncertaintyB}, and \ref{eqn:ShotNoiseUncertaintyC})
are new and demonstrate
that shot noise measurements using a single detector 
suffer the same $\sqrt{1+n}$ sensitivity degradation due
to photon bunching as would occur for direct photon counting}.

\begin{figure}[htb]
\begin{center}
\includegraphics[width=3.25in]{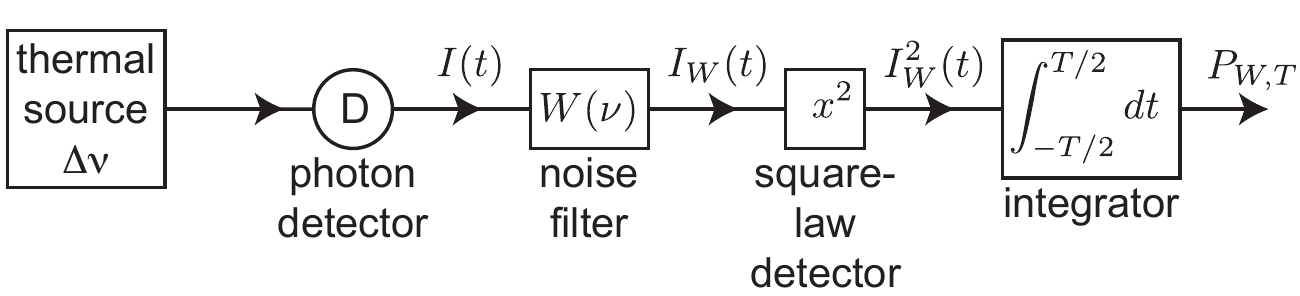}
\caption{Signal flow diagram for shot-noise detection using a single detector.
A noise filter $W(\nu)$ allows selection of the portion of the spectrum
where shot noise dominates 
(e.g., the hatched region in Figure~\ref{fig:SingleDetectorNoiseSpectrum})
prior to the measurement of the noise intensity using a square-law detector and integrator.
\label{fig:SingleDetSP}}
\end{center}
\end{figure}

Because the rate fluctuations $\SG(\nu)$ have a limited bandwidth
(equation~\ref{eqn:RateFluctuationsPoisson}),
only shot noise contributes to the current noise
(equation~\ref{eqn:ShotNoiseSpectrumRandomRate})
at high frequencies, 
and therefore a measurement of the intensity of the high-frequency noise
should give us information on the mean rate $\Gb$ and therefore
the mean current. 
Specifically, the output noise spectrum for a photon detector as
illustrated in Figure~\ref{fig:SingleDetectorNoiseSpectrum}
suggests use of the shot noise measurement setup
shown in Figure~\ref{fig:SingleDetSP},
which includes a noise filter $W(\nu)$ for isolating the shot noise dominated
portion of the spectrum.
We therefore consider applying a filter to the current,
\be
I_W(t) = \int_{-\infty}^T dt_1 W(t - t_1) I(t_1)\ ,
\ee
and then continuously integrating the noise power over a measurement interval $[-T/2, T/2]$,
\begin{align}
P_{W,T} &= \int_{-T/2}^{T/2} dt\, \left[ I_W(t) \right]^2
\nonumber \\
&= \int_{-T/2}^{T/2} dt\, \left[ \int_{-\infty}^t dt_1 W(t - t_1) I(t_1) \right]^2
\nonumber \\
&= \int_{-T/2}^{T/2} dt\, 
\left[ \int_{-\infty}^{+\infty} d\nu \, e^{i 2 \pi \nu t} W(\nu)) \hI(\nu) \right]^2
\nonumber \\
&= 
\int_{-\infty}^{+\infty} d\nu_1 \, d\nu_2
\hI(\nu_1) \hI^*(\nu_2) W(\nu_1) W^*(\nu_2) 
\nonumber \\
& \times
\int_{-T/2}^{T/2} dt\, e^{i 2 \pi \nu_1 t} e^{-i 2 \pi \nu_2 t}
\label{eqn:NoiseIntensity}\\
&\rightarrow
\int_{-\infty}^{+\infty} d\nu_1 \left|\hI(\nu_1) \right|^2 \left|W(\nu_1)\right|^2
\mathrm{\ \ \ as\ } T \rightarrow \infty.
\nonumber
\end{align}
The mean value of this measure of the high-frequency noise intensity is
\begin{align}
\ave{P_{W,T}} 
&=
\int_{-\infty}^{+\infty} d\nu_1 \, \int_{-\infty}^{+\infty} d\nu_2
\ave{\hI(\nu_1) \hI^*(\nu_2)} W(\nu_1) W^*(\nu_2) 
\nonumber \\
& \times
\int_{-T/2}^{T/2} dt\, e^{i 2 \pi \nu_1 t} e^{-i 2 \pi \nu_2 t}
\nonumber \\
&= 
\int_{-\infty}^{+\infty} d\nu_1 \, 
S_I(\nu_1) \left| W(\nu_1) \right|^2
\int_{-T/2}^{T/2} dt\, 
\nonumber \\
&=
\Gb T \, \int_{-\infty}^{+\infty} d\nu_1 \, \left| W(\nu_1) \right|^2
\nonumber \\
& + T \int_{-\infty}^{+\infty} d\nu_1 \, \left| W(\nu_1) \right|^2 \SG(\nu_1)\ .
\end{align}
The second term picks up the bunching noise component
but may be made negligible by choosing $W(\nu)$
to be zero at the lower frequencies where $\SG(\nu)$
has an appreciable value, as illustrated by
the hatched region in Figure~\ref{fig:SingleDetectorNoiseSpectrum}.
With this choice, if the noise
measurement bandwidth is defined as
\be
2 B = \int_{-\infty}^{+\infty} d\nu_1 \, \left| W(\nu_1) \right|^2\ ,
\label{eqn:Bdef}
\ee
where the factor of two accounts for negative frequencies, we have
\be
\ave{P_{W,T}} = 2 \Gb B T\ ,
\label{eqn:PWTave}
\ee
and therefore the measurement scheme 
shown in Figure~\ref{fig:SingleDetSP}
provides us with the desired information on $\Gb$.

However, we expect that the intensity of the
shot noise should be affected by the rate fluctuations $\dG(t)$.
We are therefore interested in the fluctuations
\be
\sigma_P^2 = \ave{P_{W,T}^2} - \ave{P_{W,T}}^2\ .
\ee
This quantity will require evaluation of the fourth moments of the form
\be
\ave{\hI(\nu_1) \hI(\nu_2)^* \hI(\nu_3) \hI(\nu_4)^*}\ .
\ee
Because of the presence of the high-pass filter $W(\nu)$,
we may safely assume that none of the frequencies are zero,
and therefore omit the DC terms.
Using the same approach as before, we write
\begin{align}
& \ave{\hI(\nu_1) \hI(\nu_2)^* \hI(\nu_3) \hI(\nu_4)^*}_y
\approx  \sum_{ijkl} \ave{y_i y_j y_k y_l} 
\nonumber \\
& \times 
e^{-i 2 \pi \nu_1 t_i}
e^{+i 2 \pi \nu_2 t_j}
e^{-i 2 \pi \nu_3 t_k}
e^{-i 2 \pi \nu_4 t_l}\ .
\label{eqn:FourthOrderPCmoment}
\end{align}
\tmpbf{Appendix~\ref{app:FourthOrderMoments} provides the details
of the evaluation of this quantity,
leading to an expression for $\sigma_P^2$
in the long measurement time limit $\Delta \nu T >> 1$
involving seven terms, labeled A1b+c, 
B2+B3+B4+B5, C2a+C3a, C1b, C2b+C3b, C4b+C5b+C6b+C7b, and D1.
Three of these terms drop out} if we design
our filter $W(\nu)$ so that it rejects noise due to the rate fluctuations,
i.e. 
\be
\int_{-\infty}^{+\infty} d\nu \, \left| W(\nu) \right|^2 \SG(\nu)
\rightarrow 0\ ,
\ee
as illustrated by the hatched region in Figure~\ref{fig:SingleDetectorNoiseSpectrum}.
The surviving terms (C2a+C3a, C1b, C2b+C3b, and D1)
contribute fractional fluctuations of
\begin{align}
& \frac{\sigma_P^2}{\ave{P_{W,T}}^2} = 
\frac{1}{T} \left\{ 
 \frac{ 2  \int d\nu |W(\nu)|^4  }{\left[ \int d\nu |W(\nu)|^2  \right]^2}
+  \frac{\SG(0)}{\Gb^2}  
\right.
\nonumber \\
& \left.
+ \frac{ 2   \int d\nu d\nu' |W(\nu)|^2   |W(\nu')|^2 \SG(\nu - \nu') }
{\Gb^2 \left[ \int d\nu |W(\nu)|^2  \right]^2}
+  \frac{1}{\Gb}
\right\}\ .
\label{eqn:SingleDetectorFluctuations}
\end{align}
The interpretation of these terms is simplified by choosing
a filter function $W(\nu)$ which is unity inside a measurement
bandwidth $B$ and zero outside, so that
\be
\int_{-\infty}^{+\infty} d\nu |W(\nu)|^2 = \int_{-\infty}^{+\infty} d\nu |W(\nu)|^4  = 2 B\ ,
\ee
including contributions from positive and negative frequencies.
If we define an effective bandwidth for the rate fluctuations,
\be
\Delta \nu = \frac{\int d\nu \SG(\nu)}{\SG(0)}
\ee
and evaluate the third term under
the assumption that a wide bandwidth is chosen in order
to optimize the shot-noise measurement,
$B >> \Delta \nu$,
the terms simplify to
\begin{align}
\frac{\sigma_P^2}{\ave{P_{W,T}}^2} = 
&   \frac{1}{BT}
+  \frac{\SG(0)}{\Gb^2 T}  
+ \frac{\SG(0) \Delta \nu}{\Gb^2 BT}
+  \frac{1}{\Gb T}  \ .
\label{eqn:ShotNoiseUncertainty}
\end{align}
The last term is due to the Poisson fluctuations
in the number of events over a time $T$ that one must
have even if the event rate is constant.
Meanwhile, the first term represents the noise
that results from measurement of a finite number of
independent samples associated with the
time-bandwidth product $BT$;
indeed, one sees that this term reproduces
Dicke's result \citep{Dic46} in the shot noise context.
Thus, it is helpful to use a large shot noise measurement
bandwidth, although the Poisson term dominates
when the bandwidth exceeds the mean event rate,
$B > \Gb$.
The second and third term represent the effect of
event rate fluctuations,
being proportional to the spectral density of
the fractional fluctuations 
$\SG(0) / \Gb^2$.
Here the spectral density $\SG(\nu)$ is evaluated
at zero frequency because the rate fluctuations
are being averaged over a long measurement time $T$.
The last three terms in equation~(\ref{eqn:ShotNoiseUncertainty})
may be rearranged to read
\begin{align}
\frac{\sigma_P^2}{\ave{P_{W,T}}^2} = 
&   \frac{1}{BT} + 
\frac{1}{\Gb T}  \left[ 1
+  \frac{\SG(0)}{\Gb}  
+ \frac{\SG(0)}{\Gb}  \frac{\Delta \nu}{B} \right]\ .
\label{eqn:ShotNoiseUncertaintyB}
\end{align}
In this form, the last term in the square brackets can be seen to
represent a correction to the rate fluctuation term
due to finite measurement bandwidth and is negligible
under the assumption $B >> \Delta \nu$.

If we make use of the identifications appropriate for
thermal photon noise given by equations
(\ref{eqn:MeanRatePoisson}) and
(\ref{eqn:RateFluctuationsPoisson}),
and furthermore assume that the occupation
number $n(\nu)$ is constant inside an optical
bandwidth $\Delta \nu$ and zero outside,
we have
\be
\Gb = n \Delta \nu
\label{eqn:PhotonRateGB}
\ee
and
\be
\SG(0) = n^2 \Delta \nu\ .
\label{eqn:PhotonRateFluctuationsZF}
\ee
The fractional fluctuation in the shot noise intensity is then given by
\begin{align}
\frac{\sigma_P}{\ave{P_{W,T}}} = 
&  \sqrt{ \frac{1}{BT} +  \frac{1 + n}{\Gb T} }\ .
\label{eqn:ShotNoiseUncertaintyC}
\end{align}
In the limit $B >> \Gb$,
we recover the usual result 
(equation~\ref{eqn:BunchingFractionalUncertainty})
that photon bunching 
gives a sensitivity penalty of $\sqrt{1+n}$ 
as compared to Poisson statistics.
This occurs despite the use of the white portion of the shot noise spectrum
to measure the photon rate.

\section{Semiclassical Analysis for Multiple Detectors}
\label{sec:SemiclassicalMultipleDetectors}

\begin{figure}[htb]
\begin{center}
\includegraphics[width=3.25in]{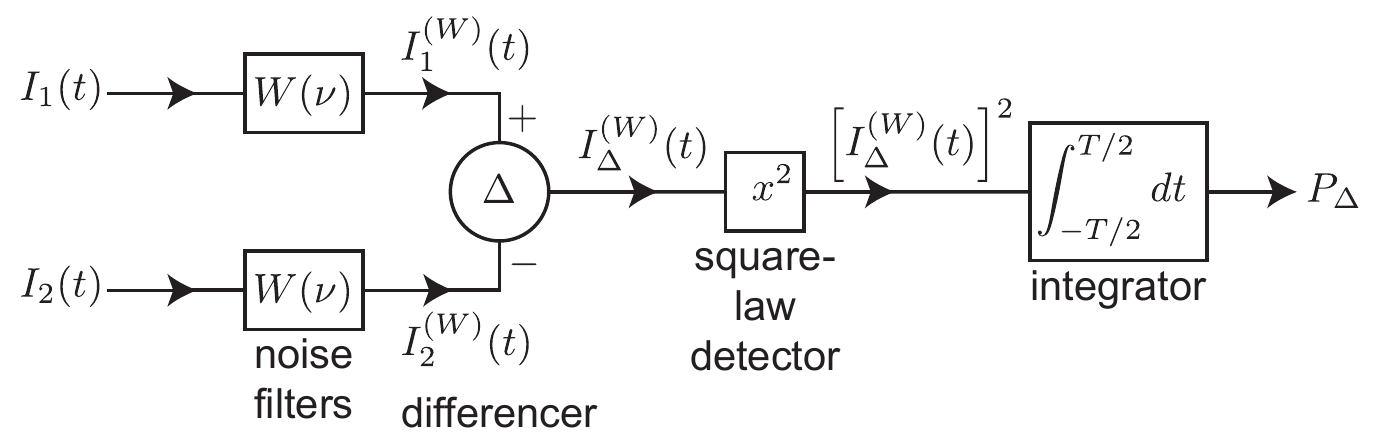}
\caption{Signal flow diagram for shot-noise detection using two detectors.
In principle, the ordering of the noise filtering and differencing operations may
be interchanged since both are linear.
\label{fig:TwoDetSP}}
\end{center}
\end{figure}

The extension of the semiclassical treatment in section~\ref{sec:SemiclassicalSingleDetector}
to the case of multiple detectors is straightforward
and allows us to analyze the sensitivity of shot noise measurement
schemes applied to the two-detector setup proposed by \citet{Lieu15},
shown in Figures~\ref{fig:TwoDetectors} and \ref{fig:CouplerDiagram}.
\tmpbf{The principal results (equations~\ref{eqn:SigmaPDelta} and \ref{eqn:MultipleDetectorNoise})
are new and agree with those in section~\ref{sec:SemiclassicalSingleDetector};
they demonstrate that shot noise measurements
applied to the two-detector scheme also cannot evade the $\sqrt{1+n}$
sensitivity degradation due to photon bunching.
We perform our analysis for a signal processing setup (Fig.~\ref{fig:TwoDetSP})
similar to those typically used for experimental 
measurements of shot noise \citep{Sch97},
though it differs in detail from the signal processing proposed by \citet{Lieu15}.
Nonetheless, our calculations are directly applicable to the regime that Lieu  et al. claim
leads to suppression of the bunching noise;
a detailed comparison of the calculations is given in section~\ref{sec:ComparisonLieu}}.

Suppose we have multiple currents exhibiting shot noise, 
\be
I_a(t) = \sum_i \delta(t - t_{i,a})
\ee
with time-dependent event rates
\be
\ave{I_a(t)} = \Gamma_a(t) = \Gb_a + \dG_a(t)\ .
\ee
Here the rate fluctuations are stationary stochastic
processes described by a cross-spectral correlation matrix,
\be
\ave{\hdG_a(\nu) \hdG^*_{b}(\nu')} = \CG_{ab}(\nu) \delta(\nu-\nu')\ .
\ee
Here $a$ and $b$ are discrete indices that label the currents.
As before, I discretize time and introduce random variables
$y_{a,i}$ to represent the number of events for current $a$ in the
time interval $[t_i, \Delta t_i]$.
The cross-spectral density between two currents is given by
\begin{align}
\ave{\hI_{a1}(\nu_1) \hI^*_{a2}(\nu_2)} & \approx
\sum_{i,j} \ave{y_{a1,i} y_{a2,j}} 
\nonumber \\
& \times e^{-2 \pi \nu_1 t_i} e^{+2 \pi \nu_2 t_j}\ .
\end{align}
The $y_{a,i}$ are all independent, so 
\begin{align}
\ave{y_{a1,i} y_{a2,j}} 
&= \ave{y_{a1,i}} \ave{ y_{a2,j}}
\nonumber \\
& + \delta_{ij} \delta_{a1,a2} \left( \ave{y^2_{a1,i}} - \ave{y_{a1,i}}^2 \right)\ .
\end{align}
The term $\ave{y_{a1,i}}^2$ is of higher order in $\Delta t_i$ and can be
neglected in the continuum limit:
\begin{align}
& \ave{\hI_{a1}(\nu_1) \hI_{a2}^*(\nu_2)}_y  = 
\nonumber \\
&
\left( \Gb_{a1} \delta(\nu_1) + \hdG_{a1}(\nu_1) \right) 
\left( \Gb_{a2} \delta(\nu_2) + \hdG_{a2}^*(\nu_2) \right)
\nonumber \\
& +
\delta_{a1,a2} \left[ \Gb_{a1} \delta(\nu_1 - \nu_2) + \hdG_{a1} (\nu_1 - \nu_2) \right]\ .
\label{eqn:ShotNoiseCrossSpectrumDetermRate}
\end{align}
Averaging over the random process $\dG_{a}(t)$ yields
\begin{align}
\ave{\hI_{a1}(\nu_1) \hI_{a2}^*(\nu_2)}_{y,\dG} 
& = 
\left[
\Gb_{a1}  \Gb_{a2} \delta(\nu_1)
+ \delta_{a1,a2} \Gb_{a1} 
\right.
\nonumber \\
& \left.
+ \CG_{a1,a2}(\nu_1)
\right] \delta(\nu_1 - \nu_2)\ ,
\label{eqn:ShotNoiseCrossSpectrumRandomRate}
\end{align}
which is simply a generalization of the single-detector result
given in equation~(\ref{eqn:ShotNoiseSpectrumRandomRate}).
If we compare this result to equation~(\ref{eqn:PhotocurrentSpectrumMultiple})
for the photocurrent correlations among detectors illuminated with
thermal radiation, we see that the expressions coincide if we
make the identifications
\be
\Gb_{a} = \bar{I}_a = \int_0^\infty d\nu B_{aa}(\nu)
\ee
and
\be
\CG_{ab}(\nu) = \int_0^\infty d\nu' B_{ab}(\nu') B_{ba}(\nu' + \nu)\ ,
\label{eqn:CrossSpectrumIdentification}
\ee
which are generalizations of
equations~(\ref{eqn:MeanRatePoisson}) and (\ref{eqn:RateFluctuationsPoisson}).

The shot-noise measurement scheme for a single detector
shown in Figure~\ref{fig:SingleDetSP}
may easily be adapted for use with two detectors as shown in Figure~\ref{fig:TwoDetSP};
this setup is designed to measure the shot noise intensity
in the difference of the two currents, $I_\Delta  = I_1 - I_2$,
as proposed by \citet{Lieu15}.
Although the filter $W(\nu)$ is no longer needed for rejection
of the bunching noise at low frequencies,
it is maintained in the setup since any real system has a finite bandwidth.
The output of the shot noise intensity measurement is given by
\begin{align}
 P_{\Delta} 
 & = \int_{-T/2}^{T/2} dt \left\{ \int_{-\infty}^t dt_1 W(t-t_1)
 \left[ I_1(t_1) - I_2(t_1) \right] \right\}^2
 \nonumber \\
& = 
 \int_{-\infty}^{+\infty} d\nu_1  d\nu_2 W(\nu_1) W^*(\nu_2)
\nonumber \\
& \times 
\left[ \hI_1(\nu_1) \hI_1^*(\nu_2)  + \hI_2(\nu_1) \hI_2^*(\nu_2)   - 2 \hI_1(\nu_1) \hI_2^*(\nu_2)  \right]
\nonumber \\
& \times  
\int_{-T/2}^{T/2} dt\, e^{i 2 \pi \nu_1 t} e^{-i 2 \pi \nu_2 t}\ .
\label{eqn:PDeltaDef}
\end{align}
and has an average value
\begin{align}
\ave{ P_{\Delta} } & = T \int_{-\infty}^{+\infty} d\nu |W(\nu)|^2
\left[ \Gb_1 + \CG_{11}(\nu) + \Gb_2 \right.
\nonumber \\
& \left. + \CG_{22}(\nu) - 2 \CG_{12}(\nu) \right]\ .
\label{eqn:PDeltaAve}
\end{align}
Equations~(\ref{eqn:PhotocurrentSpectrumBeamsplitterBunching})
and  (\ref{eqn:CrossSpectrumIdentification}) give
\begin{align}
\CG_{11}(\nu) & = \CG_{22}(\nu) = \CG_{12}(\nu) 
\nonumber \\
& = \frac{1}{4} \int_0^\infty d\nu' n_4(\nu') n_4(\nu'+\nu)
\end{align}
and therefore
\be
\ave{ P_{\Delta} } = 2 B T \left(\Gb_1 +  \Gb_2\right)
\label{eqn:PDeltaAveFinal}
\ee
is a measure of the total event rate regardless of the
choice of the filter passband $W(\nu)$.

Calculation of the sensitivity of this shot-noise intensity measurement
requires evaluation of fourth-order moments of the photocurrent,
\begin{align}
F_{abcd} 
&=
 \ave{\IW_{a}  \IW_{b} \IW_{c} \IW_{d}} 
\nonumber \\
& =
 \int_{-\infty}^{+\infty} d\nu_1  d\nu_2 d\nu_3 d\nu_4
\ave{\hI_{a}(\nu_1) \hI_{b}^*(\nu_2) \hI_{c}(\nu_3) \hI_{d}^*(\nu_4) }
\nonumber \\ 
& \times 
W(\nu_1) W^*(\nu_2) W(\nu_3) W^*(\nu_4)
\nonumber \\
& \times 
\int_{-T/2}^{T/2} dt\, e^{i 2 \pi \nu_1 t} e^{-i 2 \pi \nu_2 t}
\nonumber \\
& \times 
\int_{-T/2}^{T/2} dt'\, e^{i 2 \pi \nu_3 t'} e^{-i 2 \pi \nu_4 t'}\ .
\label{eqn:FourthOrderMomentsMultiple}
\end{align}
\tmpbf{This expression may be evaluated using the same approach
as used for the second moment; the details are given in 
Appendix~\ref{app:FourthOrderMomentsMultiple}.
The resulting variance of the shot noise intensity
is derived at the end of the Appendix}:
\begin{align}
\sigma^2_{P_\Delta} 
& =  \ave{P_\Delta^2} - \ave{P_\Delta}^2
\nonumber \\
& = 2 F_{1111} + 2 F_{1122} + 4 F_{1212} - 8 F_{1112}
\nonumber \\
&= 
+ 8T \Gb_1^2 \int d\nu |W(\nu)|^4 
+ 4T \CG_{11}(0) \left[ \int d\nu |W(\nu)|^2\right]^2
\nonumber \\
&
+ 8T  \int d\nu d\nu' |W(\nu)|^2   |W(\nu')|^2  \CG_{11}(\nu - \nu')
\nonumber \\
&
+ 2 T \Gb_1 \left[ \int d\nu |W(\nu)|^2 \right]^2\ .
\label{eqn:SigmaPDelta}
\end{align}
To compare to the previous single-detector case
(equation~\ref{eqn:ShotNoiseUncertainty}),
we make the substitutions
\be
\Gb_1 = \frac{1}{2} \Gb
\ee
and
\be
\CG_{11}(\nu) = \frac{1}{4} \SG(\nu)\ .
\ee
Using the mean value of $P_\Delta$ given by
equation~(\ref{eqn:PDeltaAve}),
we may express the fractional fluctuation in the noise intensity
of the difference current as
\begin{align}
& \frac{\sigma^2_{P_\Delta}}{\ave{P_\Delta}^2}
= \frac{1}{T} \left[ 
\frac{2 \int d\nu |W(\nu)|^4}{\left[\int d\nu |W(\nu)|^2\right]^2}
+ \frac{\SG(0)}{\Gb^2}
\right.
\nonumber \\
& \left.
+ \frac{2 \int d\nu d\nu' |W(\nu)|^2 |W(\nu')|^2 \SG(\nu - \nu')}
{\Gb^2 \left[\int d\nu |W(\nu)|^2\right]^2}
+ \frac{1}{\Gb}
\right]
\label{eqn:MultipleDetectorNoise}
\end{align}
which is exactly our previous result for a single detector
given by equation~(\ref{eqn:SingleDetectorFluctuations}).
In particular, the rate fluctuation term $\SG(0)$ leads to the
$\sqrt{1 + n}$ sensitivity degradation due to photon bunching.
However, 
it is no longer necessary to make the assumption that
the filter $W(\nu)$ rejects the low-frequency excess noise;
differencing the two detectors fed by a 50/50 beamsplitter
takes care of the rejection instead.

\section{Shot Noise Measurement: A Quantum Calculation}
\label{sec:ShotNoiseQuantum}

The semiclassical analyses given in 
sections~\ref{sec:SemiclassicalSingleDetector}
and \ref{sec:SemiclassicalMultipleDetectors}
for shot noise measurements are revisited in this section,
but now making use of  a fully quantum-mechanical treatment.
I focus first  on the single-detector case;
the generalization to  multiple detectors is straightforward and
is given at the end of this section.
As before, the photocurrent operator is given by equation~(\ref{eqn:PhotocurrentOperator})
and has Fourier components given by equation~(\ref{eqn:PhotocurrentSpectrumOperator}).
The shot noise intensity is measured in the same way:
a  filter $W(\nu)$ is applied to the photocurrent 
before a square-law detector and integrator are used to measure
the intensity, as illustrated in Figure~\ref{fig:SingleDetSP}.
This measurement scheme produces the quantity
$P_{W,T}$ as defined by equation~(\ref{eqn:NoiseIntensity}).
However, calculation of the statistics of $P_{W,T}$  now requires
quantum operator averages, which I perform in the usual way 
appropriate for thermal radiation, namely by combining photon creation
and destruction operators pairwise.
The quantum computation of the second-order moment
is detailed in equation~(\ref{eqn:PhotocurrentPowerSpectrumCalc}),
with a result that is identical to the semiclassical second-order moment,
\begin{align}
\ave{I(\nu_1) I^\dagger(\nu_2)} 
& = \left[ \Gb^2 \delta(\nu_1) + \Gb + \SG(\nu_1) \right] 
\nonumber \\
& \times \delta(\nu_1 - \nu_2)\ .
\end{align}
In fact the semiclassical spectrum was chosen to coincide with the quantum result
through the definitions of $\Gb$ and $\SG(\nu)$ given in
equations~(\ref{eqn:MeanRatePoisson}) and (\ref{eqn:RateFluctuationsPoisson}).
Thus, we conclude that $\ave{P_{W,T}} = \Gb B T$
for the quantum calculation just as for the semiclassical case
(equation \ref{eqn:NoiseIntensity}),
provided that we choose the noise filter $W(\nu)$
to avoid the excess low-frequency noise as discussed in section~\ref{sec:SemiclassicalSingleDetector}.

Evaluation of the fluctuations of $P_{W, T}$
requires a quantum computation of the fourth-order moment of 
Fourier components of the photocurrent,
which in turn requires eighth-order moments of the photon operators:
\begin{align}
F(\nu_1, & \, \nu_2, \nu_3, \nu_4)  =
\ave{I(\nu_1) I^\dagger(\nu_2) I(\nu_3) I^\dagger(\nu_4) } 
\nonumber \\
& =  \int d \nu'_1 d \nu'_2 d \nu'_3 d \nu'_4
\nonumber \\
& \times
\left<
b^\dagger(\nu'_1) b(\nu'_1 + \nu_1)
b^\dagger(\nu'_2 + \nu_2) b(\nu'_2)
\right.
\nonumber \\
& \times
\left.
b^\dagger(\nu'_3) b(\nu'_3 + \nu_3)
b^\dagger(\nu'_4 + \nu_4) b(\nu'_4)
\right> \ .
\label{eqn:EighthOrderPhotonMoments}
\end{align}
Combining operators pairwise produces $4! = 24$ terms.
However, as for the semiclassical calculation, many of these 
represent DC terms that are rejected by the filter $W(\nu)$
and therefore do not contribute to the shot noise intensity.
For example, if the first two operators are paired, we will have a factor
\begin{align}
\ave{11'} & = 
\int d\nu'_1 \ave{b^\dagger(\nu'_1) b(\nu'_1 + \nu_1)}
\nonumber \\
& = \int d\nu'_1 n(\nu'_1) \delta(\nu_1) 
\nonumber \\
& = \Gb \delta(\nu_1)
\end{align}
which vanishes except at DC, $\nu_1 = 0$, and may
therefore be ignored.
This shows that we may ignore any similar pairing,
e.g. $\ave{22'}$ in the obvious notation.
Any pairing may be represented by a permutation,
e.g. $\ave{11'} \ave{23'} \ave{34'} \ave{2'4}$ corresponds
to the permutation $(1) (234)$ expressed in cyclic notation.
All permutations that include a cycle of length 1, e.g. $(3)$,
will give DC terms that we may ignore. This leaves 9 permutations
left to consider:
\be
\begin{array}{ccc}
(12)(34) & (1234) & (1243) \\
(13)(24) & (1324) & (1342) \\
(14)(23) & (1423) & (1432)
\end{array}
\ee
\tmpbf{The detailed evaluation of some of these pairings is given in
Appendix~\ref{app:EighthOrderMoments}.
These pairings reproduce the terms found in the semiclassical
calculation outlined in section~\ref{sec:SemiclassicalSingleDetector}
and detailed in Appendix~\ref{app:FourthOrderMoments},
and also generate some extra terms that arise from the non-gaussianity of
the photon arrival rate fluctuations as described in Appendix~\ref{app:QuantumVsSemiclassical}
that are neglected in our semiclassical calculation.
In particular, the (1432) permutation includes the contributions
expressed by
equations~(\ref{eqn:QuantumPoisson}) and (\ref{eqn:QuantumBunchingTerm})
in Appendix~\ref{app:EighthOrderMoments}}:
\begin{align}
\ave{P_{W,T}^2}_{(1432)} 
& =
T \left[ \Gb + \SG(0) \right] 
\left[ \int d\nu \left| W(\nu) \right|^2 \right]^2 
\nonumber \\
& + ...\ ,
\end{align}
which are same as the semiclassical terms D1 and and C1b 
listed in Appendix~\ref{app:FourthOrderMoments} that
correspond to Poisson noise and bunching noise, respectively.
The latter term contributes $\SG(0) / \Gb^2$ to the
fractional fluctuations $\sigma_P^2 / \ave{P_{W,T}^2}$
(see equations~\ref{eqn:SingleDetectorFluctuations}, \ref{eqn:ShotNoiseUncertaintyC})
and thus represents the shot noise intensity fluctuations
due to photon bunching. It is this term that gives the same
$\sqrt{1+n}$ sensitivity degradation due to bunching 
as occurs for ordinary photon counting.

It is not difficult to translate these results to the case of multiple detectors.
\tmpbf{We again focus our attention on the (1432)
operator pairing in the corresponding quantum calculation,
which includes the contributions
(equations~\ref{eqn:QuantumPoissonMultiple} and \ref{eqn:QuantumBunchingTermMultiple})
\begin{align}
\left[F_{abcd} \right]_{(1432)} & = T \delta_{ab} \delta_{cd}
\left[ \Gb_a \delta_{ac} + \CG_{ac}(0) \right]
\left[ \int d\nu \left| W(\nu) \right|^2 \right]^2
\nonumber \\
& + ...
\end{align}
that correspond to the Poisson and bunching terms
D1 and C1b found in the semiclassical calculation, as
outlined in section~\ref{sec:SemiclassicalMultipleDetectors} and 
detailed in Appendix~\ref{app:FourthOrderMomentsMultiple}}.
The latter term contributes
\be
4T \CG_{11}(0) \left[ \int d\nu |W(\nu)|^2\right]^2
\ee
to the measurement variance $\sigma_{P_\Delta}^2$
(equation~\ref{eqn:SigmaPDelta})
for the two-detector setup shown in Figures~\ref{fig:TwoDetectors}
and \ref{fig:CouplerDiagram},
and leads to the $\sqrt{1+n}$ photon bunching degradation.

Thus, we conclude that a full quantum calculation reproduces the conclusions of
the semiclassical analyses for one or two detectors given in 
sections~\ref{sec:SemiclassicalSingleDetector} and \ref{sec:SemiclassicalMultipleDetectors},
namely that shot noise intensity measurements cannot evade
the bunching noise.

\section{Correlation of Shot Noise and Photon Counts}
\label{sec:Correlation}

We have two ways of measuring the photon flux:
direct photon counting using
the time integral of the photocurrent, $N_T$,  defined  in equation~(\ref{eqn:NTdef}), 
or through a shot noise intensity measurement
represented by $P_{W,T}$ and defined in equation~(\ref{eqn:NoiseIntensity}).
According to our semiclassical and quantum calculations,
both are affected by photon bunching;
therefore, these quantities must be correlated if our results are correct.
Conversely, if we can establish a correlation between these quantities,
the correlation may be used together with the well-known results for
bunching noise in direct photon counting to establish a lower bound for the
bunching noise that must also be present in the shot noise measurements.
In this section, I present a fully quantum treatment of these topics.

We are interested in evaluating the correlation
\begin{align}
\ave{P_{W,T} N_T} = & 
\int_{-\infty}^{+\infty} d\nu_1   d\nu_2  d\nu_3
\ave{I(\nu_1) I^\dagger(\nu_2) I(\nu_3) }
\nonumber \\
& \times
W(\nu_1) W^*(\nu_2) \int_{-T/2}^{T/2} dt_1\, e^{i 2 \pi (\nu_1 - \nu_2) t_1} 
\nonumber \\
& \times 
\int_{-T/2}^{T/2} dt_3\, e^{i 2 \pi \nu_3 t_3} \ .
\label{eqn:PNcorrelation}
\end{align}
We thus require the sixth-order moments of photon operators,
\begin{align}
& \ave{I(\nu_1) I^\dagger(\nu_2) I(\nu_3) } 
=  \int d \nu'_1 d \nu'_2 d \nu'_3
\nonumber \\
& \times \ave{
b^\dagger(\nu'_1) b(\nu'_1 + \nu_1)
b^\dagger(\nu'_2 + \nu_2) b(\nu'_2)
b^\dagger(\nu'_3) b(\nu'_3 + \nu_3)
}\ ,
\nonumber
\end{align}
to be evaluated as usual by computing the $3! = 6$ operator pairings.
As in section~\ref{sec:ShotNoiseQuantum}, 
the noise filters $W(\nu_1)$ and $W^*(\nu_2)$
allow us to ignore the DC terms in those variables;
however, we must now retain DC terms for $\nu_3$.
We can thus neglect permutations involving the cycles $(1)$
and $(2)$, which leaves only $(12)(3)$, $(123)$, and $(132)$.
We find:
\begin{align}
(12)(3) & = \left[ \Gb + \SG(\nu_1) \right] \Gb \delta(\nu_1 - \nu_2) \delta(\nu_3)
\nonumber \\
(123) & = \int d\nu'_1 \, n(\nu'_1) n(\nu_1+\nu_3+\nu'_1) 
\nonumber \\
& \times
\left[1 + n(\nu_1 + \nu'_1) \right] \delta(\nu_1 - \nu_2 + \nu_3)
\nonumber \\
(132) & = \int d\nu'_1 \, n(\nu'_1) 
\left[ 1 + n(\nu_1+ \nu'_1) \right] 
\nonumber \\
& \times
\left[1 + n(\nu_1 + \nu'_1 - \nu_2) \right] \delta(\nu_1 - \nu_2 + \nu_3)\ .
\nonumber
\end{align}
Performing the integrations indicated in equation~(\ref{eqn:PNcorrelation})
gives
\begin{align}
(12)(3) & = 
T^2 \int_{-\infty}^{+\infty} d\nu_1 \left|W(\nu_1)\right|^2
\Gb \left[ \Gb + \SG(\nu_1) \right] 
\nonumber \\
(123) & = 
T \int_{-\infty}^{+\infty} d\nu_1 \left|W(\nu_1)\right|^2
\nonumber \\
& \times 
\left[ \SG(\nu_1) + \int d\nu'_1 n(\nu'_1) n^2(\nu'_1 + \nu_1) \right]
\nonumber \\
(132) & = 
T \int_{-\infty}^{+\infty} d\nu_1 \left|W(\nu_1)\right|^2
\left[ \Gb + \SG(\nu_1) + \SG(0) 
\right.
\nonumber \\
& \left. 
+ \int d\nu'_1 n^2(\nu'_1) n(\nu'_1 + \nu_1) \right]\ .
\nonumber
\end{align}
The $(12)(3)$ term just gives the product of averages $\ave{P_{W,T}}\ave{N_T}$,
so
\begin{align}
& \ave{ P_{W,T} N_T}  - \ave{P_{W,T}}\ave{N_T} 
\nonumber \\
& = 
T \int_{-\infty}^{+\infty} d\nu_1 \left|W(\nu_1)\right|^2
\left\{ 
\Gb + \SG(0) + 2 \SG(\nu_1) 
\right.
\nonumber \\
& \left.
+ \int d\nu'_1 n(\nu'_1) \left[n(\nu'_1) + n(\nu'_1 + \nu_1) \right] n(\nu'_1 + \nu_1) 
\right\}\ .
\nonumber
\end{align}
Most of the terms vanish if we design the noise filter to reject the low-frequency
noise as illustrated by the hatched region in Figure~\ref{fig:SingleDetectorNoiseSpectrum};
the terms that survive are
\begin{align}
\ave{ P_{W,T} N_T} & - \ave{P_{W,T}}\ave{N_T} = 
2B T \left[ \Gb + \SG(0) \right]\ ,
\label{eqn:PNcorrResult}
\end{align}
using our standard definition of the shot noise 
measurement bandwidth $B$ (equation~\ref{eqn:Bdef}).
We therefore see that the shot noise intensity $P_{W,T}$
and photon counts $N_T$ are indeed correlated,
as we expect if both are affected by photon bunching.

The value of the correlation given by equation~(\ref{eqn:PNcorrResult})
allows us to set a lower bound on the variance of the shot noise intensity
$P_{W,T}$
given the well-established results for the variance of the 
photon counts $N_T$.
Indeed, if $X$ and $Y$ are two random variables
with zero mean and finite variance,
the Cauchy-Schwarz inequality holds:
\be
\ave{X Y}^2 \le \ave{X^2} \ave{Y^2}\ ,
\ee
which establishes a lower limit for the variance of $X$,
\be
\ave{X^2} \ge \frac{\ave{X Y}^2}{\ave{Y^2}}\ .
\ee
Now set $X = P_{W,T} - \ave{P_{W,T}}$ and $Y = N_T - \ave{N_T}$.
From equations~(\ref{eqn:sigmaNT}), (\ref{eqn:MeanRatePoisson}) 
and (\ref{eqn:RateFluctuationsPoisson}) we have
\be
\ave{Y^2} = T \left[ \Gb + \SG(0) \right]\ .
\ee
Using the known value of the correlation $\ave{XY}$ given by equation~(\ref{eqn:PNcorrResult}),
we have
\be
\sigma^2_P = \ave{X^2} 
\ge \frac{\ave{X Y}^2}{\ave{Y^2}}
= 4B^2 T \left[ \Gb + \SG(0) \right]\ .
\ee
Dividing by the square of the mean value $\ave{P_{W,T}} = 2 B T \Gb$
gives the fractional fluctuations
\be
\frac{\sigma^2_P}{\ave{P_{W,T}}^2}
\ge \frac{1}{\Gb T} + \frac{\SG(0)}{\Gb^2 T} 
= \frac{1 + n}{\Gb T}
\ee
where we have used equations~(\ref{eqn:PhotonRateGB})
and (\ref{eqn:PhotonRateFluctuationsZF}) in writing the second expression.
Thus, using a fully quantum-mechanical calculation,
we have demonstrated that the shot noise intensity measurement
must suffer at least the same $\sqrt{1+n}$ sensitivity degradation
due to photon bunching  as does standard photon counting.
Comparison to the result of the semiclassical calculation,
equation~(\ref{eqn:ShotNoiseUncertaintyC}),
shows that the correlation bound does not include
the $1/BT$ noise term associated with a finite bandwidth for the
shot noise measurement. This is to be expected:
the finite-bandwidth noise 
does not influence the direct photon counts, represented by $N_T$,
and is therefore absent in the correlation $\ave{P_{W,T} N_T}$.

The extension to the case of two detectors is straightforward
but will be omitted. However, it is easy to see that the results
above may be applied independently to each detector in
the two-detector setup shown in Figure~\ref{fig:TwoDetectors}.
Thus, the shot noise intensity for each detector must be correlated
with its photocurrent. Furthermore, the two photocurrents are correlated,
as demonstrated by \citet{Han56}. Therefore, the (high-frequency)
shot noise \emph{intensities} of the two detectors must also be correlated,
even though the shot noise itself is not: this distinction,
between moments of the form $\ave{I_1^2 I_2^2}$ vs.  $\ave{I_1 I_2}$,
is elucidated further in section~\ref{sec:Resolution}.

\section{Comparison to the results of Lieu et al.}
\label{sec:ComparisonLieu}

\begin{table*}[ht]
\begin{center}
\caption{\tmpbf{Comparison of symbols in this paper vs. those of \citet{Lieu15}}.}
\label{tbl:Symbols}
\begin{tabular}{cccc}
\hline
Quantity 				& This paper			& Lieu et al.				& Lieu et al. \\
					&					& (as used here)			& (original notation) \\
\hline
Number of samples		& (continuous)			& $N$					& $N$ \\
Sample time			& (continuous)			& $\Delta t$				& $T$ \\
Total measurement time	& $T$				& $T = N \Delta t$			& $N T$ \\
Optical bandwidth		& $\Delta \nu$			& $\Delta \nu = 1/\tau$		& $1/\tau$ \\
Shot noise bandwidth	& $B$				& $B_{\Delta t} = 1/2\Delta t$	& $1/2T$ \\
Photon arrival rate		&$\Gb = n \Delta \nu$	& $n_0 \Delta \nu$			& $n_0 / \tau$ \\
Photon rate fluctuations	& $\SG(0)$			& $\sqrt{\pi}n_0^2 \Delta \nu$	& $\sqrt{\pi}n_0^2 /\tau$ \\
\hline
\end{tabular}
\end{center}
\end{table*}

\tmpbf{In this section, I compare the results of the previous sections 
with those of \citet{Lieu15} for \emph{both} of the regimes
they examine, corresponding to 
long sample times $\Delta \nu \Delta t >> 1$ and
short sample times $\Gb \Delta t < 1$.
Here $\Delta t$ is the single-sample time defined by \citet{Lieu15};
to avoid confusion, I use $\Delta t$ 
instead of their chosen symbol, $T$,
and instead reserve $T = N \Delta t$ to signify the total time duration of the measurement
required for the acquisition and integration of $N$ samples.
The concept of sample time does not arise in
my calculations since I assume continuous time integration;
however, a connection can readily be made since the 
sample time $\Delta t$ defined by Lieu \textit{et al.}
sets the shot noise bandwidth $B_{\Delta t} = 1/2\Delta t$
associated with their measurement scheme.
I make use of this correspondence to compare the two calculations
for the same total measurement duration $T$, and find that
the results agree in the long sample time regime but disagree for short sample times.
Thus, my results directly contradict the claim of Lieu \textit{et al.} that bunching noise may be
avoided in the latter limit.
It is important to note that for both regimes, the total measurement duration
$T$ can be chosen to satisfy  $\Delta \nu T >> 1$, as I have assumed
for my calculations; indeed, long measurement durations
are essential for astronomical observations since the sensitivity improves as $1/\sqrt{T}$.
To aid in comparison of the results, and for ease of reference in the discussion below,
the relevant quantities and symbols used to represent them
in both papers are provided in Table~\ref{tbl:Symbols}}.

\citet{Lieu15} present a fully quantum calculation of the shot noise
fluctuations for the two-detector setup illustrated in Figure~\ref{fig:TwoDetectors}.
They use a very similar quantum formalism for photon detection
that differs only in minor and inconsequential detail.
For example, their definition of the operator representing the detector output
measures photon power instead of photon counts
as can be seen from their equation~(5).
Moments of photon operators are calculated in the standard way,
by combining operators pairwise,
as is appropriate for thermal radiation.
Lieu et al. consider the detector output averaged over some measurement time $\Delta t$
corresponding to the quantity
\be
I_{\Delta t} = \frac{1}{\Delta t} \int_0^{\Delta t} dt \, I(t)
\ee
in my notation. 
Lieu et al. focus on the
difference of the outputs of the two detectors in Figure~\ref{fig:TwoDetectors},
\be
I_\Delta(t) = I_1(t) - I_2(t)\ ,
\ee
and calculate both the second and fourth moments of
\be
I_{\Delta, \Delta t}  = \frac{1}{\Delta t} \int_0^{\Delta t} dt \, I_\Delta(t)\ .
\label{eqn:IDeltaDeltat}
\ee
Their fundamental conclusions
rely on evaluation of the mean and variance of the sum of $N$ consecutive measurements
of  $\left[ I_{\Delta, \Delta t} \right]^2$, obtained over a total time duration
of $T = N \Delta t$. This quantity may be expressed as
\be
P_{N,\mathrm{LKD}} = \sum_{k=1}^N \left[ I_{\Delta, \Delta t} (k) \right]^2
\label{eqn:PLKD}
\ee
where
\be
I_{\Delta, \Delta t}(k)  = \frac{1}{\Delta t} \int_{(k-1) \Delta t}^{k \Delta t} dt \, I_\Delta(t)\ 
\label{eqn:IDeltaDeltatK}
\ee
are the consecutive time-averaged samples of the photocurrent difference $I_\Delta(t)$.
In contrast, I first apply an arbitrary linear filter to the photocurrent,
\be
I_{\Delta}^{(W)}(t) = \int_{-\infty}^t dt' \, W(t-t') I_\Delta(t')
\ee
and then study the mean and variance of the shot noise intensity integrated over time,
\be
P_{\Delta} = \int_{-T/2}^{T/2} dt \, \left[I_{\Delta}^{(W)}(t)\right]^2 
\label{eqn:PdeltaDef}
\ee
as illustrated in Figure~\ref{fig:TwoDetSP}.

\begin{figure}[htb]
\begin{center}
\includegraphics[width=3.25in]{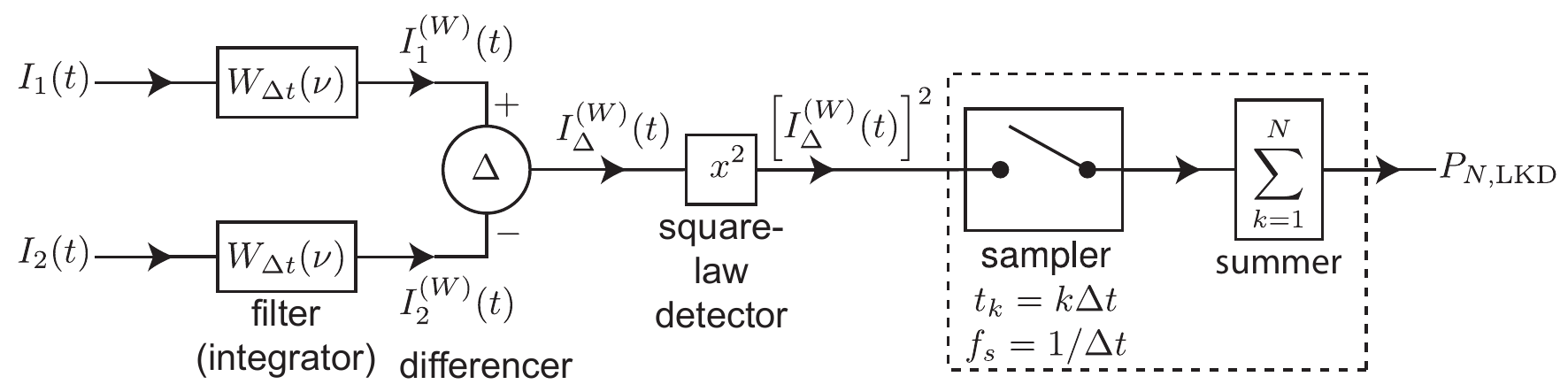}
\caption{Signal flow diagram for shot-noise detection
as proposed by \citet{Lieu15}.
The ordering of the filtering and differencing operations may
be interchanged.
This scheme differs from that shown in Figure~\ref{fig:TwoDetSP}
in two ways: 1) the choice of filter is fixed and 
corresponds to boxcar integrator with time duration $\Delta t$;
and 2) the (slow) time integration is not continuous but is instead
performed in a discrete fashion using a sampler and summer
(dashed box). 
The sampler operates at a rate $1/\Delta t$ and 
is synchronized to the filter.
\label{fig:LieuKibbleSP}}
\end{center}
\end{figure}

Although the definitions of $P_{N,\mathrm{LKD}}$ and $P_{\Delta}$
superficially appear to be different, 
these two quantities are closely related, as illustrated in Figure~\ref{fig:LieuKibbleSP}.
The averaging over $\Delta t$ performed by \citet{Lieu15} may be
represented by a particular (and inflexible)
choice for the linear filter, namely a time window function:
\be
W_{\Delta t}(t) = 
\left\{
\begin{array}{ll}
1/\Delta t, & 0 \le t \le \Delta t \nonumber \\
0, & \mathrm{otherwise}\ .
\end{array}
\right.
\ee
According to our definition (equation~\ref{eqn:Bdef}),
this filter has a bandwidth
\begin{align}
2 B_{\Delta t} & = \int_{-\infty}^{+\infty} d\nu \, \left| W_{\Delta t}(\nu) \right|^2 
\nonumber \\
& = \int_{-\infty}^{+\infty} dt \, W^2_{\Delta t}(t) 
\nonumber \\
&= \frac{1}{\Delta t}\ ;
\label{eqn:LKDWsquared}
\end{align}
we will also need
\be
\int_{-\infty}^{+\infty} d\nu \, \left| W_{\Delta t}(\nu) \right|^4 =  \frac{2}{3 \Delta t}\ .
\label{eqn:LKDWfourth}
\ee
Note that this filter does not reject DC or low-frequency noise,
but these are automatically rejected anyway 
by differencing the currents in a two-detector setup.
Another distinction is that \citet{Lieu15} perform the time integration
operation as a discrete sum rather than a continuous integration:
the output of the square-law detector is sampled at
times $t_k = k \Delta t$, and then summed,
as represented by the dashed box in Figure~\ref{fig:LieuKibbleSP}.
This choice does not significantly affect the results,
though the discrete sampling operation of Lieu et al.
may result in a minor degradation in performance due to noise aliasing.

Equation (29) in \citet{Lieu15} gives the second moment of one sample:
\be
\ave{I_{\Delta, \Delta t}^2} = \frac{n_0}{\Delta t \tau}\ ,
\ee
where we have omitted their factor of $\omega_0^2$ so that the operator
represents photon flux.  
According to their equation (7),
their symbol $\tau$ is related to the optical coherence time and is
inversely proportional to the optical bandwidth,
$\tau \sim 1/\Delta \nu$. Thus, translated to our notation,
\be
\ave{I_{\Delta, \Delta t}^2} \sim n_0 \Delta \nu 2 B_{\Delta t} = 2 \Gb B_{\Delta t}\ 
\ee
where $\Gb = n_0 \Delta \nu$ is the photon rate before the beamsplitter.
Meanwhile, the corresponding equation for our observable
(equation~\ref{eqn:PDeltaAveFinal}) reads
\be
\ave{ P_{\Delta} } = 2 B T \Gb\ .
\ee
These may be reconciled by using equation~(\ref{eqn:PdeltaDef}),
\be
\ave{\left[I_{\Delta}^{(W)}(t)\right]^2 } = \lim_{T \rightarrow 0} \frac{\ave{ P_{\Delta} }}{T}
= 2 \Gb B \ .
\ee
Thus, our results for the mean value of the shot noise intensity agree
with \citet{Lieu15} if we make the replacements $\tau \rightarrow 1/\Delta \nu$
and $B \rightarrow B_{\Delta t} = 1/2 \Delta t$.

We now turn to the variance of a single output sample of the \citet{Lieu15} setup, 
which they calculate using a clever evaluation of the eighth-order moments
of the photon operators in which most of the terms are discarded since
they cancel in the two-detector scheme. Specifically, they calculate
\be
\sigma_{1,\mathrm{LKD}}^2 = \ave{I_{\Delta, \Delta t}^4} -  \ave{I_{\Delta, \Delta t}^2}^2\ ,
\ee
and express the result as a fractional variance in their equation~(32)
when the number of samples $N=1$,
\be
\frac{\sigma_{1,\mathrm{LKD}}^2 }{\ave{I_{\Delta, \Delta t}^2}^2} 
= 2 + 3 \frac{\tau}{\Delta t} F\left(\frac{\Delta t}{\tau}\right) +  \frac{\tau}{\Delta t \,n_0}\ .
\ee
Translated into our notation, this reads
\be
\frac{\sigma_{1,\mathrm{LKD}}^2 }{\ave{I_{\Delta, \Delta t}^2}^2} 
= 2 +  \frac{3}{\Delta \nu \Delta t} F\left(\Delta \nu \Delta t \right)
+  \frac{1}{\Gb \Delta t}\ .
\label{eqn:LKDvariance}
\ee
Here $F(x)$ is a smooth function that allows  
both the $\Delta \nu \Delta t <<1 $ and $ \Delta \nu \Delta t  >>1 $ limits to be
examined, 
and is derived under the assumption of a Gaussian spectral profile for the thermal radiation.
\tmpbf{Comparison of the single-sample variance with our results requires
use of the latter limit} because we assume $\Delta \nu T >> 1$, where $T$ is the
duration of the measurement, for evaluation of the Fourier integrals.
In this limit, $F(x) \rightarrow \sqrt{\pi}$, and 
\be
\frac{\sigma_{1,\mathrm{LKD}}^2 }{\ave{I_{\Delta, \Delta t}^2}^2} 
= 2 +  \frac{3 \sqrt{\pi}}{\Delta \nu \Delta t} +  \frac{1}{\Gb \Delta t}\ .
\ee
We may safely assume that consecutive samples are uncorrelated
in the $\Delta \nu \Delta t >> 1$ limit, 
because the sample time $\Delta t$ is long compared to the optical coherence
time $\tau$. 
Thus, the fractional variance for a sum of $N$ samples 
(equation~\ref{eqn:PLKD}) would be
\be
\frac{\sigma_{N,\mathrm{LKD}}^2 }{\ave{I_{\Delta, \Delta t}^2}^2} 
= \frac{1}{N} \left[ 2 +  \frac{3 \sqrt{\pi}}{\Delta \nu \Delta t} +  \frac{1}{\Gb \Delta t} \right] \ .
\label{eqn:LKDvarianceNsamples}
\ee
Meanwhile, our result for the fractional variance (equation~\ref{eqn:MultipleDetectorNoise}) reads
\begin{align}
& \frac{\sigma^2_{P_\Delta}}{\ave{P_\Delta}^2}
= \frac{1}{T} \left[ 
\frac{2 \int d\nu |W(\nu)|^4}{\left[\int d\nu |W(\nu)|^2\right]^2}
+ \frac{\SG(0)}{\Gb^2}
\right.
\nonumber \\
& \left.
+ \frac{2 \int d\nu d\nu' |W(\nu)|^2 |W(\nu')|^2 \SG(\nu - \nu')}
{\Gb^2 \left[\int d\nu |W(\nu)|^2\right]^2}
+ \frac{1}{\Gb}
\right]\ .
\end{align}
The first term is readily evaluated using equations~(\ref{eqn:LKDWsquared}) and (\ref{eqn:LKDWfourth}).
The numerator in the third term must be evaluated in our chosen limit $\Delta \nu \Delta t >> 1$,
so $|\nu|, |\nu'| << \Delta \nu$ for the integrals; we obtain
\begin{align}
\frac{\sigma^2_{P_\Delta}}{\ave{P_\Delta}^2}
&= \frac{1}{T} \left[ \frac{4}{3} \Delta t
+ 3 \frac{\SG(0)}{\Gb^2} + \frac{1}{\Gb}
\right]\ .
\end{align}
For the Gaussian spectral profile used by \citet{Lieu15},
\begin{align}
\Gb = \int_{-\infty}^{+\infty} d\nu \, n(\nu) 
& = \frac{n_0}{\tau} \rightarrow n_0 \Delta \nu
\nonumber \\
\SG(0) = \int_{-\infty}^{+\infty} d\nu \, n^2(\nu) 
& = \frac{\sqrt{\pi} n^2_0}{\tau} \rightarrow \sqrt{\pi} n^2_0 \Delta \nu\ ,
\end{align}
so our result reads
\begin{align}
\frac{\sigma_P^2}{\ave{P_{\Delta}}^2} = 
&  \frac{4}{3} \frac{\Delta t}{T} +  \frac{3 \sqrt{\pi}}{\Delta \nu T}  +  \frac{1}{\Gb T}  \ .
\end{align}
Correspondence with \citet{Lieu15} is obtained by letting the
total measurement time $T$ coincide with $N \Delta t$, the time to obtain $N$
samples, yielding
\be
\frac{\sigma_P^2}{\ave{P_{\Delta}}^2} = \frac{1}{N} \left[ \frac{4}{3} 
+ \frac{3 \sqrt{\pi}}{\Delta \nu \Delta t}  +  \frac{1}{\Gb \Delta t} \right]  \ .
\ee
This expression reproduces the three terms of the Lieu et al. result stated
in equation~(\ref{eqn:LKDvarianceNsamples}),
which is derived from their equation~(32),
apart from a somewhat smaller numerical factor on our
first term which likely results from
our use of a continuous integration over the measurement time $T$
instead of a sum of discrete samples taken every $\Delta t$
as illustrated in Figure~\ref{fig:LieuKibbleSP}.
Note that the second term of the Lieu et al. result 
confirms our $S_\Gamma(0)$  term,
which is the signature of photon bunching in the shot noise intensity.
Thus, our results agree in the long sample time limit, $\Delta \nu \Delta t >> 1$.

In contrast, there is a major disagreement in the short sample time regime, 
$\Gb \Delta t < 1$,
which corresponds to the use of a wide bandwidth for 
measurement of the shot noise, $B_{\Delta t} > \Gb = n_0 \Delta \nu >> \Delta \nu$
as illustrated in
Figures~\ref{fig:SingleDetectorNoiseSpectrum} and \ref{fig:TwoDetectorNoiseSpectrum}.
Lieu et al. find $F(x) \approx 1/x$ for $x<<1$,
so their result (equation \ref{eqn:LKDvariance}) in this limit becomes
\be
\frac{\sigma_{1,\mathrm{LKD}}^2 }{\ave{I_{\Delta, \Delta t}^2}^2} 
\approx 5 +  \frac{1}{\Gb \Delta t} \approx \frac{1}{\Gb \Delta t}\ .
\ee
In other words, the goal is to choose $\Delta t$ small enough that
the Poisson term dominates the single-sample variance.
Their principal claim, namely that bunching noise can be avoided, 
rests on the statement that a sum
of $N$ such samples, as given by 
equation~(\ref{eqn:PLKD}) and illustrated in Figure~\ref{fig:LieuKibbleSP},
and acquired over a total measurement time
$T = N \Delta t$, would have a fractional variance 
\be
\frac{\sigma_{N,\mathrm{LKD}}^2 }{\ave{P_{N,\mathrm{LKD}}}}
= \frac{1}{N} \frac{\sigma_{1,\mathrm{LKD}}^2 }{\ave{I_{\Delta, \Delta t}^2}^2} 
= \frac{1}{N \Gb \Delta t} = \frac{1}{\Gb T}\ ,
\label{eqn:LieuShortSampleTimes}
\ee
as would be expected if the samples were statistically independent.
\tmpbf{Note that the limit $\Delta \nu T >> 1$ as required for our calculation
of the same two-detector scheme 
is reached simply by choosing $N >> 1 / \Delta \nu \Delta t$, so the comparison 
is immediate through use of
equations~(\ref{eqn:MultipleDetectorNoise}),  (\ref{eqn:SingleDetectorFluctuations})
and (\ref{eqn:ShotNoiseUncertaintyC})}:
\begin{align}
\frac{\sigma^2_{P_\Delta}}{\ave{P^2_{\Delta}}} = 
&  \frac{1}{BT} +  \frac{1 + n}{\Gb T} \ .
\label{eqn:PDeltaFractionalFluctuations}
\end{align}
\tmpbf{The second term dominates because we have assumed $B > \Gb$.
Thus, our calculation gives a sensitivity of $(1+n)/ \Gb T$ while Lieu et al. find 
$1 / \Gb T$;
our result includes the $(1 + n)$ bunching penalty, while Lieu et al.
claim it can be avoided.}

\section{Resolving the contradiction}
\label{sec:Resolution}

\tmpcom{(This section is new, and includes material previously in Section 9.)}

\tmpbf{Section~\ref{sec:ComparisonLieu} shows that the detection
sensitivities derived in this paper
agree with those of \citet{Lieu15} for long sample times
but disagree in the short sample time regime that is of central interest.
In obtaining their result for short sample times (equation~\ref{eqn:LieuShortSampleTimes}),
Lieu et al. assume statistical independence} 
but do not actually prove this by computing the correlations between samples,
$\ave{\left[ I_{\Delta, \Delta t} (k) \right]^2 \left[ I_{\Delta, \Delta t} (l) \right]^2}$.
They instead state, after their equation~(32):
``...data in non-overlapping time periods are uncorrelated, because the
correlation function $\ave{I_T^d(t) I_T^d(0)}$ is proportional to a delta function...''.
Translated to our notation, their statement relates to
\begin{align}
& \ave{I_{\Delta, \Delta t} (k) I_{\Delta, \Delta t} (l)}  
\nonumber \\
& =
\frac{1}{(\Delta t)^2} \int_{(k-1) \Delta t}^{k \Delta t} dt_1 \int_{(l-1) \Delta t}^{l \Delta t} dt_2 \,
\ave{I_\Delta(t_1) I_\Delta(t_2)}
\nonumber \\
& = \delta_{kl} \frac{\Gb}{\Delta t}\ .
\end{align}
This demonstrates that $I_{\Delta, \Delta t} (k)$ are uncorrelated
as Lieu et al. claim,
which is to be expected because the spectral density of $I_\Delta(t)$
is white as illustrated in Figure~\ref{fig:TwoDetectorNoiseSpectrum}.
Nonetheless, this does not mean that the \emph{squares} of these
random variables, $\left[ I_{\Delta, \Delta t} (k) \right]^2$, are uncorrelated.

Thus, we see that the fundamental claim of Lieu et al. is invalidated by a simple error,
the fallacy of the converse. Suppose we have two 
zero-mean random variables, $X$ and $Y$. If they are independent,
they must be uncorrelated, because $\ave{XY} = \ave{X}\ave{Y} = 0$.
However, the converse is not necessarily true.
If it were true, we could claim that when $\ave{XY} = 0$, we must
also have $\ave{X^2 Y^2} = \ave{X^2} \ave{Y^2}$,
which is the statement upon which the Lieu et al. result rests.
A simple counterexample suffices: suppose the joint distribution
of $X$ and $Y$ is given by
\be
f(x,y) = \frac{1}{2} \left[ \delta(x-y) + \delta(x+y) \right] g(x)\ ,
\ee
where $g(x)$ is a Gaussian distribution with zero mean
and variance $\sigma^2$.
This distribution  cannot be factorized into the form
$f(x,y) = f_x(x) f_y(y)$, so clearly $X$ and $Y$ cannot be independent.
We may readily compute
\begin{align}
\ave{X} & = \int dx dy \, f(x,y) x = \int dx g(x) x = 0
\nonumber \\
\ave{Y} & = \int dx dy \, f(x,y) y = \int dx \frac{1}{2} \left[ x - x \right] g(x) = 0
\nonumber \\
\ave{XY} & = \int dx dy \, f(x,y) xy = \frac{1}{2} \left[ x^2 - x^2  \right] g(x) = 0
\nonumber \\
\ave{X^2} & = \int dx dy \, f(x,y) x^2 = \int dx g(x) x^2 = \sigma^2
\nonumber \\
\ave{Y^2} & = \int dx dy \, f(x,y) y^2 = \frac{1}{2} \left[ x^2 + x^2  \right] g(x) = \sigma^2
\nonumber \\
\ave{X^2 Y^2} & = \int dx dy \, f(x,y) x^2 y^2 = \frac{1}{2} \left[ x^4 + x^4  \right] g(x) = 3 \sigma^4\ .
\nonumber
\end{align}
Thus $\ave{X^2 Y^2} \ne \ave{X^2} \ave{Y^2} $ even though
$\ave{XY} = 0$.

\tmpbf{It is quite easy to see that the samples $I^2_{\Delta, \Delta t} (k)$
must be correlated using a simple physical argument.
Consider the quantity defined in (equation~(\ref{eqn:IDeltaDeltatK}):
\be
I_{\Delta, \Delta t}(k)  = \frac{1}{\Delta t} \int_{(k-1)\Delta t}^{k \Delta t} dt \, \left[ I_1(t) - I_2(t) \right]\ .
\nonumber 
\ee
There are only three events that can occur with non-negligible probability
when $\Gb \Delta t << 1$, corresponding to a short sample time: 
a) detector 1 receives a photon; b) detector 2 receives a photon;
c) neither detector receives a photon. These events correspond
to values of 
$I_{\Delta, \Delta t}(k) = \left\{ +1/\Delta t, -1/\Delta t, 0 \right \}$,
respectively; therefore, $I^2_{\Delta, \Delta t}(k)$ takes on
the value of $1/(\Delta t)^2$ if \emph{either} detector receives a photon,
and zero otherwise.
Thus, in the limit $\Gb \Delta t << 1$,
the measurement scheme proposed by \citet{Lieu15}
(equation~\ref{eqn:PLKD}) can be expressed as
\be
P_{N,\mathrm{LKD}} = \sum_{k=1}^N \left[ I_{\Delta, \Delta t} (k) \right]^2
= \frac{N_T}{(\Delta t)^2}
\ee
where $N_T$ is the total number of photons received by both detectors
over the course of a measurement of duration $T = N \Delta t$.
Note that a single detector, replacing the two detectors and beamsplitter, would also have
received $N_T$ photons during such a measurement,
so the statistics of the Lieu et al. observable 
$P_{N,\mathrm{LKD}}$ must be the
same as those for $N_T$, corresponding to photon counting with a single detector.
The statistics of the latter are well known to be affected by bunching,
as stated in equation~(\ref{eqn:BunchingFractionalUncertainty}); thus
the Lieu et al. claim that the bunching noise can be avoided
is contradicted not only by the calculations presented in
this paper, but also by the extensive experimental
and theoretical work on photon bunching over the past six decades.
A more rigorous discussion is given in Appendix~\ref{app:LieuCorrelations},
which provides a detailed quantum-mechanical
calculation that demonstrates that
the samples $I^2_{\Delta, \Delta t} (k)$ are indeed correlated,
and that accounting for these correlations in the sensitivity calculation
leads again to the standard photon bunching penalty,
in agreement with the calculations 
for both one and two detectors presented in 
sections~\ref{sec:SemiclassicalSingleDetector},
\ref{sec:SemiclassicalMultipleDetectors},
\ref{sec:ShotNoiseQuantum},
and \ref{sec:Correlation}.
}

\vspace{1cm}

\section{Acknowledgements}
I thank Jim Moran and John Kovac at Harvard for bringing this
interesting problem to my attention, and Richard Lieu,
Tom Kibble, and Lingze Duan for extensive discussions.
This paper is dedicated to the memory of my father,
Jonas Stasys Zmuidzinas, 
who first introduced me to coherent-state integrals.

\appendix

\section{Generating function for a Poisson process with a time-variable rate}
\label{app:PoissonGeneratingFunction}

Here we evaluate the generating function introduced in
equation~(\ref{eqn:GeneratingFunctionN}),
\be
G_N(s) = \ave{e^{s N_T} }
= \sum_{k=0}^\infty \frac{s^k}{k!}
\sum_{i_1 ... i_k = 1}^{M} \ave{y_{i_1} ... y_{i_k}}\ .
\nonumber
\ee
If the indices $\left\{i_1 ... i_k \right\}$ are all distinct,
we may write
\be
\ave{y_{i_1} ... y_{i_k}} = \ave{y_{i_1}} ... \ave{y_{i_k}}
= \Gamma(t_{i_1}) ... \Gamma(t_{i_k}) \Delta t_{i_1} ... \Delta t_{i_k}\ ,
\ee
because the $y_i$ are independent.
When one or more indices repeat, we may use $y_i^m = y_i$
(for $m \ge 1$) to again obtain a product of distinct factors.
We are thus faced with the problem of partitioning the
set of indices $\left\{i_1 ... i_k \right\}$ into one or more 
groups, where the indices belonging to a group have the same value,
and indices belonging to different groups have distinct values.
The number of partitions of $k$ objects into $p$ groups is given by
 the Stirling number of the second kind, $S(k,p)$,
 which are nonzero for $p \le k$ \citep{Bla07}. 
 We therefore write
 \be
 \sum_{i_1 ... i_k = 1}^{N}   y_{i_1} ... y_{i_k}
 = \sum_{p=0}^k  S(k,p) {\sum_{i_1 ... i_p}}' y_{i_1} ... y_{i_p}
 \ee
where the prime on the second sum indicates that the indices
take on only distinct values. 
We may make the replacement
\be
{\sum_{i_1 ... i_p}}' y_{i_1} ... y_{i_p}
\rightarrow p! {\sum_{i_1 > i_2 ... > i_p}} y_{i_1} ... y_{i_p}\ .
\ee
by considering permutations of the indices.
Taking the average,
\begin{align}
{\sum_{i_1 > i_2 ... > i_p}} \ave{y_{i_1} ... y_{i_p}}
&= {\sum_{i_1 > i_2 ... > i_p}} \Gamma(t_{i_1}) ... \Gamma(t_{i_p})
\Delta t_{i_1} ... \Delta t_{i_p}
\nonumber \\
&\approx  \int_0^T dt_1 \int_0^{t_1} dt_2 ... \int_0^{t_{p-1}} dt_p 
\Gamma(t_1) ... \Gamma(t_{p})
\nonumber \\
&= \frac{1}{p!} \left[ \int_0^T dt \Gamma(t) \right]^p\ .
\end{align}
Use of the following identity for the  Stirling numbers
\be
\sum_{k, p = 0}^\infty S(k,p) \frac{x^k}{k!} y^p = \exp \left[ y \left(e^x - 1\right) \right]
\ee
allows us to evaluate the generating function,
\begin{align}
G_N(s) &=
\sum_{k, p=0}^\infty S(k,p) \frac{s^k}{k!}  \left[ \int_0^T dt \Gamma(t) \right]^p\
\nonumber \\
&= \exp \left[ \mu \left(e^s - 1 \right) \right] ,
\label{eqn:GeneratingFunctionPoissonCounts}
\end{align}
where
\be
\mu = \int_0^T dt \Gamma(t)\ .
\ee

\section{Detailed evaluation of shot noise fluctuations: single detector}
\label{app:FourthOrderMoments}

The required average  in equation~(\ref{eqn:FourthOrderPCmoment})
may be performed by considering the partitions
of the indices (see also Picinbono, Benjdaballah \& Pouget 1970, eqn. 2.16):
\begin{align}
\ave{y_i y_j y_k y_l}  
&= 
\ave{y_i}\ave{y_j}\ave{y_k}\ave{y_l}
\tag{A1} \\
& 
+ \delta_{ij} \ave{y_i} \ave{y_k}\ave{y_l}
+ \delta_{ik} \ave{y_i} \ave{y_j}\ave{y_l}
+ \delta_{il} \ave{y_i} \ave{y_j}\ave{y_k}
\tag{B1, B2, B3} \\
& 
+ \delta_{jk} \ave{y_i} \ave{y_j}\ave{y_l}
+ \delta_{jl} \ave{y_i} \ave{y_j}\ave{y_k}
+ \delta_{kl} \ave{y_i} \ave{y_j}\ave{y_k}
\tag{B4, B5, B6} \\
&
+ \delta_{ij} \delta_{kl} \ave{y_i} \ave{y_k}
+ \delta_{ik} \delta_{jl} \ave{y_i} \ave{y_j}
+ \delta_{il} \delta_{jk} \ave{y_i} \ave{y_j}
\tag{C1, C2, C3} \\
&
+ \delta_{ij} \delta_{ik} \ave{y_i} \ave{y_l}
+ \delta_{ij} \delta_{il} \ave{y_i} \ave{y_k}
+ \delta_{jk} \delta_{jl} \ave{y_i} \ave{y_j}
\tag{C4, C5, C6} \\
&
+ \delta_{ik} \delta_{il} \ave{y_i} \ave{y_j}
+ \delta_{ij} \delta_{ik} \delta_{il} \ave{y_i}\ .
\tag{C7, D1}
\end{align}
We have neglected to subtract the correction terms 
such as $\ave{y_i}^2 \ave{y_k} \ave{y_l}$ because,
as in section~\ref{sec:ShotNoise}, 
they contain an extra factor of $\Delta t_i$
and therefore will vanish in the continuum limit.
The number of terms of each partition class, here labeled
A, B, C, and D, is $(1, 6, 7, 1)$ and
follows the sequence of Stirling numbers $S(4, k)$
\citep{Bla07}, as expected.
Taking the continuum limit and evaluating the Fourier integrals gives
the following terms: 

\begin{align}
& \ave{\hI(\nu_1) \hI^*(\nu_2) \hI(\nu_3) \hI^*(\nu_4)}_y
\nonumber \\
&=
\left[ \Gb \delta(\nu_1) + \hdG(\nu_1) \right]
\left[ \Gb \delta(\nu_2) + \hdG^*(\nu_2) \right]
\left[ \Gb \delta(\nu_3) + \hdG(\nu_3) \right]
\left[ \Gb \delta(\nu_4) + \hdG^*(\nu_4) \right]
\tag{A1} \\
&+
\left[ \Gb \delta(\nu_1 - \nu_2) + \hdG(\nu_1 - \nu_2) \right]
\left[ \Gb \delta(\nu_3) + \hdG(\nu_3) \right]
\left[ \Gb \delta(\nu_4) + \hdG^*(\nu_4) \right]
\tag{B1} \\
&+
\left[ \Gb \delta(\nu_1 + \nu_3) + \hdG(\nu_1 + \nu_3) \right]
\left[ \Gb \delta(\nu_2) + \hdG^*(\nu_2) \right]
\left[ \Gb \delta(\nu_4) + \hdG^*(\nu_4) \right]
\tag{B2} \\
&+
\left[ \Gb \delta(\nu_1 - \nu_4) + \hdG(\nu_1 - \nu_4) \right]
\left[ \Gb \delta(\nu_2) + \hdG^*(\nu_2) \right]
\left[ \Gb \delta(\nu_3) + \hdG(\nu_3) \right]
\tag{B3} \\
&+
\left[ \Gb \delta(\nu_2 - \nu_3) + \hdG(\nu_2 - \nu_3) \right]
\left[ \Gb \delta(\nu_1) + \hdG(\nu_1) \right]
\left[ \Gb \delta(\nu_4) + \hdG^*(\nu_4) \right]
\tag{B4} \\
&+
\left[ \Gb \delta(\nu_2 + \nu_4) + \hdG(\nu_2 + \nu_4) \right]
\left[ \Gb \delta(\nu_1) + \hdG(\nu_1) \right]
\left[ \Gb \delta(\nu_3) + \hdG(\nu_3) \right]
\tag{B5} \\
&+
\left[ \Gb \delta(\nu_3 - \nu_4) + \hdG(\nu_3 - \nu_4) \right]
\left[ \Gb \delta(\nu_1) + \hdG(\nu_1) \right]
\left[ \Gb \delta(\nu_2) + \hdG^*(\nu_2) \right]
\tag{B6} \\
&+
\left[ \Gb \delta(\nu_1 - \nu_2) + \hdG(\nu_1 - \nu_2) \right]
\left[ \Gb \delta(\nu_3 - \nu_4) + \hdG(\nu_3 - \nu_4) \right]
\tag{C1} \\
&+
\left[ \Gb \delta(\nu_1 + \nu_3) + \hdG(\nu_1 + \nu_3) \right]
\left[ \Gb \delta(\nu_2 + \nu_4) + \hdG^*(\nu_2 + \nu_4) \right]
\tag{C2} \\
&+
\left[ \Gb \delta(\nu_1 - \nu_4) + \hdG(\nu_1 - \nu_4) \right]
\left[ \Gb \delta(\nu_2 - \nu_3) + \hdG^*(\nu_2 - \nu_3) \right]
\tag{C3} \\
&+
\left[ \Gb \delta(\nu_1 - \nu_2 + \nu_3) + \hdG(\nu_1 - \nu_2 + \nu_3) \right]
\left[ \Gb \delta(\nu_4) + \hdG^*(\nu_4) \right]
\tag{C4} \\
&+
\left[ \Gb \delta(\nu_1 - \nu_2 - \nu_4) + \hdG(\nu_1 - \nu_2 - \nu_4) \right]
\left[ \Gb \delta(\nu_3) + \hdG(\nu_3) \right]
\tag{C5} \\
&+
\left[ \Gb \delta(\nu_1 + \nu_3 - \nu_4) + \hdG(\nu_1 + \nu_3 - \nu_4) \right]
\left[ \Gb \delta(\nu_2) + \hdG^*(\nu_2) \right]
\tag{C6} \\
&+
\left[ \Gb \delta(-\nu_2 + \nu_3 - \nu_4) + \hdG^*(\nu_2 - \nu_3 + \nu_4) \right]
\left[ \Gb \delta(\nu_1) + \hdG(\nu_1) \right]
\tag{C7} \\
&+
\Gb \delta(\nu_1-\nu_2 + \nu_3 - \nu_4) + \hdG(\nu_1 - \nu_2 + \nu_3 - \nu_4)\ .
\tag{D1}
\end{align}

Averaging over the stationary process $\dG(t)$ 
now involves evaluation of its third-order and fourth-order
moments.
However, these higher-order moments are not fully specified
by the second moment, which is determined by the power spectrum
given in equation~(\ref{eqn:RateFluctuationsPoisson}),
because $\dG(t)$ is not guaranteed to be Gaussian.
Indeed, that $\dG(t)$ is not Gaussian is shown in
Appendix~\ref{app:QuantumVsSemiclassical}.
Nonetheless, $\dG(t)$ may often be approximately Gaussian,
and we proceed with this assumption recognizing
that it may introduce small, detailed differences with
the full quantum calculation.
However, as could be anticipated,
the term describing the sensitivity degradation
due to photon bunching (labeled C1b below) 
involves only a second-order moment of $\dG(t)$
and is therefore secure.

For a Gaussian $\dG(t)$, and omitting the DC terms, we find
\begin{align}
& \ave{\hI(\nu_1) \hI^*(\nu_2) \hI(\nu_3) \hI^*(\nu_4)}_{y, \delta \Gamma}
\nonumber \\
& 
= \SG(\nu_1) \SG(\nu_3) \delta(\nu_1 - \nu_2) \delta(\nu_3 - \nu_4)
+ \SG(\nu_1) \SG(\nu_2) \delta(\nu_1 - \nu_4) \delta(\nu_2 - \nu_3)
\tag{A1a, A1b}\\
& 
+ \SG(\nu_1) \SG(\nu_4) \delta(\nu_1 + \nu_3) \delta(\nu_2 + \nu_4)
+ \Gb \SG(\nu_3) \delta(\nu_1 - \nu_2) \delta(\nu_3 - \nu_4)
\tag{A1c, B1}\\
& 
+ \Gb \SG(\nu_2) \delta(\nu_1 + \nu_3) \delta(\nu_2 + \nu_4)
+ \Gb \SG(\nu_2) \delta(\nu_1 - \nu_4) \delta(\nu_2 - \nu_3)
\tag{B2, B3}\\
& 
+ \Gb \SG(\nu_1) \delta(\nu_1 - \nu_4) \delta(\nu_2 - \nu_3)
+ \Gb \SG(\nu_1) \delta(\nu_1 + \nu_3) \delta(\nu_2 + \nu_4)
\tag{B4, B5}\\
& 
+ \Gb \SG(\nu_1) \delta(\nu_1 - \nu_2) \delta(\nu_3 - \nu_4)
+ \Gb^2 \delta(\nu_1 - \nu_2) \delta(\nu_3 - \nu_4)
\tag{B6, C1a}\\
& 
+ \Gb^2 \delta(\nu_1 + \nu_3) \delta(\nu_2 + \nu_4)
+ \Gb^2 \delta(\nu_1 - \nu_4) \delta(\nu_2 - \nu_3)
\tag{C2a, C3a}\\
&
+ \left[
\SG(\nu_1 - \nu_2) + \SG(\nu_1 + \nu_3)  + \SG(\nu_1 - \nu_4) 
+ \SG(\nu_4) + \SG(\nu_3) \right.
\tag{C1b -- C5} \\
& 
+ \left. \SG(\nu_2) + \SG(\nu_1) +\Gb \right]
\delta(\nu_1 - \nu_2 + \nu_3 - \nu_4)
\tag{C6, C7, D1}
\end{align}
noting that the A1 term gives three contributions, A1a - A1c
due to the Gaussian pairwise evaluation of the fourth-order moment
of $\dG$, while the factors in
terms C1--C3 combine to give two contributions each, e.g. C1a and C1b.

We now evaluate the second moment of the shot noise
intensity measure in the limit of a long measurement
time, $\Delta \nu T >> 1$. The result is
\begin{align}
\ave{P_{W,T}^2} 
& = 
\int_{-\infty}^{+\infty} d \nu_1 d \nu_2 d \nu_3 d \nu_4 \,
\ave{\hI(\nu_1) \hI^*(\nu_2) \hI(\nu_3) \hI^*(\nu_4)}_{y, \delta \Gamma}
\nonumber \\
& 
\times 
W(\nu_1) W^*(\nu_2) W(\nu_3) W^*(\nu_4) 
\int_{-T/2}^{+T/2} dt e^{i 2 \pi (\nu_1 -\nu_2) t}
\int_{-T/2}^{+T/2} dt' e^{i 2 \pi (\nu_3 -\nu_4) t'}
\nonumber \\
& 
= T^2 \left[ \int d\nu |W(\nu)|^2 \SG(\nu) \right]^2
+ 2 T  \int d\nu |W(\nu)|^4 \SG^2(\nu) 
\tag{A1a, A1b + A1c} \\
& 
+ 2 T^2 \Gb \int d\nu |W(\nu)|^2  \int d\nu' |W(\nu')|^2 \SG(\nu') 
\tag{B1 + B6} \\
& 
+ 4 T \Gb \int d\nu |W(\nu)|^4 \SG(\nu) 
+ T^2 \Gb^2 \left[ \int d\nu |W(\nu)|^2  \right]^2
\tag{B2+B3+B4+B5, C1a} \\
& 
+ 2 T \Gb^2  \int d\nu |W(\nu)|^4 
+ T \SG(0)  \left[ \int d\nu |W(\nu)|^2  \right]^2
\tag{C2a+C3a, C1b} \\
&
+ 2 T  \int d\nu d\nu' |W(\nu)|^2   |W(\nu')|^2 \SG(\nu - \nu') 
\tag{C2b+C3b} \\
&
+ 4 T  \int d\nu |W(\nu)|^2   \int d\nu'  |W(\nu')|^2 \SG(\nu') 
\tag{C4b+C5b+C6b+C7b} \\
&
+ T \Gb \left[ \int d\nu |W(\nu)|^2 \right]^2\ .
\tag{D1}
\end{align}
The terms proportional to $T^2$ sum to give $\ave{P_{W,T}}^2$;
the remaining terms proportional to $T$
give the variance 
\begin{align}
\sigma_P^2 = & \ave{P^2_{W,T}} - \ave{P_{W,T}}^2
\nonumber \\
= & T \left\{ 
2   \int d\nu |W(\nu)|^4 \SG^2(\nu) 
+ 4  \Gb \int d\nu |W(\nu)|^4 \SG(\nu) 
\right.
\tag{A1b+c, B2+B3+B4+B5}\\
&
+ 2  \Gb^2 \int d\nu |W(\nu)|^4  
+  \SG(0)  \left[ \int d\nu |W(\nu)|^2  \right]^2
\tag{C2a+C3a, C1b}\\
&
+ 2   \int d\nu d\nu' |W(\nu)|^2   |W(\nu')|^2 \SG(\nu - \nu') 
\tag{C2b+C3b}\\
&
+ 4   \int d\nu |W(\nu)|^2   \int d\nu'  |W(\nu')|^2 \SG(\nu') 
\tag{C4b+C5b+C6b+C7b}\\
&
\left.
+  \Gb \left[ \int d\nu |W(\nu)|^2 \right]^2
\right\}\ .
\tag{D1}
\end{align}

\section{Detailed evaluation of shot noise fluctuations: multiple detectors}
\label{app:FourthOrderMomentsMultiple}
Following the approach described in Appendix~\ref{app:FourthOrderMoments},
and omitting the DC terms, we have:
\begin{align}
& \ave{\hI_{a1}(\nu_1) \hI_{a2}^*(\nu_2) \hI_{a3}(\nu_3) \hI_{a4}^*(\nu_4)}_y
\nonumber \\
&=
\hdG_{a1}(\nu_1)
\hdG_{a2}^*(\nu_2)
\hdG_{a3}(\nu_3)
\hdG_{a4}^*(\nu_4)
\tag{A1}\\
&+
\delta_{a1,a2} 
\left[ \Gb_{a1} \delta(\nu_1 - \nu_2) + \hdG_{a1} (\nu_1 - \nu_2) \right]
\hdG_{a3}(\nu_3) \hdG_{a4}^*(\nu_4)
\tag{B1} \\
&+
\delta_{a1,a3} 
\left[ \Gb_{a1} \delta(\nu_1 + \nu_3) + \hdG_{a1}(\nu_1 + \nu_3) \right]
\hdG_{a2}^*(\nu_2) \hdG_{a4}^*(\nu_4)
\tag{B2} \\
&+
\delta_{a1,a4} 
\left[ \Gb_{a1} \delta(\nu_1 - \nu_4) + \hdG_{a1}(\nu_1 - \nu_4) \right]
\hdG_{a2}^*(\nu_2) \hdG_{a3}(\nu_3)
\tag{B3} \\
&+
\delta_{a2,a3} 
\left[ \Gb_{a2} \delta(\nu_2 - \nu_3) + \hdG_{a2}(\nu_2 - \nu_3) \right]
\hdG_{a1}(\nu_1)
\hdG_{a4}^*(\nu_4)
\tag{B4} \\
&+
\delta_{a2,a4} 
\left[ \Gb_{a2} \delta(\nu_2 + \nu_4) + \hdG_{a2}(\nu_2 + \nu_4) \right]
\hdG_{a1}(\nu_1)
\hdG_{a3}(\nu_3)
\tag{B5} \\
&+
\delta_{a3,a4} 
\left[ \Gb_{a3} \delta(\nu_3 - \nu_4) + \hdG_{a3}(\nu_3 - \nu_4) \right]
\hdG_{a1}(\nu_1) \hdG_{a2}^*(\nu_2)
\tag{B6} \\
&+
\delta_{a1,a2} \delta_{a3,a4} 
\left[ \Gb_{a1} \delta(\nu_1 - \nu_2) + \hdG_{a1}(\nu_1 - \nu_2) \right]
\left[ \Gb_{a3} \delta(\nu_3 - \nu_4) + \hdG_{a3}(\nu_3 - \nu_4) \right]
\tag{C1} \\
&+
\delta_{a1,a3} \delta_{a2,a4} 
\left[ \Gb_{a1} \delta(\nu_1 + \nu_3) + \hdG_{a1}(\nu_1 + \nu_3) \right]
\left[ \Gb_{a2} \delta(\nu_2 + \nu_4) + \hdG_{a2}^*(\nu_2 + \nu_4) \right]
\tag{C2} \\
&+
\delta_{a1,a4} \delta_{a2,a3} 
\left[ \Gb_{a1} \delta(\nu_1 - \nu_4) + \hdG_{a1}(\nu_1 - \nu_4) \right]
\left[ \Gb_{a2} \delta(\nu_2 - \nu_3) + \hdG_{a2}^*(\nu_2 - \nu_3) \right]
\tag{C3} \\
&+
\delta_{a1,a2} \delta_{a1,a3} 
\left[ \Gb_{a1} \delta(\nu_1 - \nu_2 + \nu_3) + \hdG_{a1}(\nu_1 - \nu_2 + \nu_3) \right]
\hdG_{a4}^*(\nu_4)
\tag{C4} \\
&+
\delta_{a1,a2} \delta_{a1,a4} 
\left[ \Gb_{a1} \delta(\nu_1 - \nu_2 - \nu_4) + \hdG_{a1}(\nu_1 - \nu_2 - \nu_4) \right]
\hdG_{a3}(\nu_3)
\tag{C5} \\
&+
\delta_{a1,a3} \delta_{a1,a4} 
\left[ \Gb_{a1} \delta(\nu_1 + \nu_3 - \nu_4) + \hdG_{a1}(\nu_1 + \nu_3 - \nu_4) \right]
\hdG_{a2}^*(\nu_2)
\tag{C6} \\
&+
\delta_{a2,a3} \delta_{a2,a4} 
\left[ \Gb_{a2} \delta(-\nu_2 + \nu_3 - \nu_4) + \hdG_{a2}^*(\nu_2 - \nu_3 + \nu_4) \right]
\hdG_{a1}(\nu_1)
\tag{C7} \\
&+
\delta_{a1,a2} \delta_{a1,a3} \delta_{a1,a4} 
\Gb_{a1} \delta(\nu_1-\nu_2 + \nu_3 - \nu_4) + \hdG_{a1}(\nu_1 - \nu_2 + \nu_3 - \nu_4)\ .
\tag{D1}
\end{align}
Inserting this result into equation~(\ref{eqn:FourthOrderMomentsMultiple})
and evaluating some  integrals, we find
\begin{align}
F_{abcd} 
& = 
T^2 \int d\nu |W(\nu)|^2 \CG_{ab} (\nu) \int d\nu' |W(\nu')|^2 \CG_{cd} (\nu') 
\tag{A1a} \\
&
+ T  \int d\nu |W(\nu)|^4 
\left[\CG_{ac}(\nu)  \CG_{db}(\nu) + \CG_{ad}(\nu)  \CG_{cb}(\nu) \right]
\tag{A1b + A1c} \\
& 
+ T^2  \int d\nu |W(\nu)|^2  \int d\nu' |W(\nu')|^2 
[\delta_{ab} \Gb_a \CG_{cd}(\nu') + \delta_{cd} \Gb_c \CG_{ab}(\nu') ]
\tag{B1 + B6} \\
& 
+ T \int d\nu |W(\nu)|^4 
\left[ \
\delta_{ac} \Gb_a \CG_{db}(\nu) 
+ \delta_{ad} \Gb_a \CG_{cb}(\nu) \right.
\nonumber \\ 
& \left.
+ \delta_{bc} \Gb_b \CG_{ad}(\nu) 
+ \delta_{bd} \Gb_b \CG_{ca}(\nu) 
\right]
\tag{B2+B3+B4+B5} \\
&
+ T^2 \delta_{ab} \delta_{cd} \Gb_a \Gb_c \left[ \int d\nu |W(\nu)|^2  \right]^2
\tag{C1a} \\
& 
+ T \left[ \delta_{ac} \delta_{bd} \Gb_a \Gb_b +
\delta_{ad} \delta_{bc} \Gb_a \Gb_c \right]
  \int d\nu |W(\nu)|^4 
\tag{C2a+C3a} \\
&
+ T \delta_{ab} \delta_{cd} \CG_{ac}(0)  \left[ \int d\nu |W(\nu)|^2  \right]^2
\tag{C1b} \\
&
+ T \int d\nu d\nu' |W(\nu)|^2   |W(\nu')|^2 
\left[ \delta_{ac} \delta_{bd} + \delta_{ad} \delta_{bc}  \right] \CG_{ab}(\nu - \nu')
\tag{C2b+C3b} \\
&
+ T  \int d\nu |W(\nu)|^2   \int d\nu'  |W(\nu')|^2 
\left[ \delta_{ab} \delta_{ac} \CG_{ad}(\nu') 
+ \delta_{ab} \delta_{ad} \CG_{ac}(\nu')
\right.
\nonumber \\
&
\left.
+ \delta_{ac} \delta_{ad} \CG_{ab}(\nu')
+ \delta_{bc} \delta_{bd} \CG_{ab}(\nu') \right]
\tag{C4b+C5b+C6b+C7b} \\
&
+ T \delta_{ab} \delta_{ac} \delta_{ad}
\Gb_a \left[ \int d\nu |W(\nu)|^2 \right]^2\ .
\tag{D1}
\end{align}
The fluctuations of the
noise intensity of the difference current are obtained by considering
\begin{align}
\ave{P_\Delta^2} 
&= F_{1111} + F_{2222} + 4 F_{1212}
+ 2 F_{1122} - 4 F_{1112} - 4 F_{2212}
\nonumber \\
&= 2 F_{1111} + 2 F_{1122} + 4 F_{1212} - 8 F_{1112}\ ,
\end{align}
as can be seen by computing $P^2_\Delta$ (equation~\ref{eqn:PDeltaDef})
and making use of the symmetry of the 50/50 beamsplitter.
The $T^2$ terms are eliminated by subtracting the square of the mean,
\begin{align}
\sigma^2_{P_\Delta}
&=
\ave{P_\Delta^2} - \ave{P_\Delta}^2
= 2 F_{1111} + 2 F_{1122} + 4 F_{1212} - 8 F_{1112}
\nonumber \\
& = 
4 T  \int d\nu |W(\nu)|^4 [\CG_{11}(\nu)]^2
+ 8T \Gb_1 \int d\nu |W(\nu)|^4 \CG_{11}(\nu)
\nonumber \\
&
+ 4T \Gb_1^2 \int d\nu |W(\nu)|^4 
+ 2T \CG_{11}(0) \left[ \int d\nu |W(\nu)|^2\right]^2
\nonumber \\
&
+ 4T  \int d\nu d\nu' |W(\nu)|^2   |W(\nu')|^2  \CG_{11}(\nu - \nu')
\nonumber \\
&
+ 8T  \int d\nu d\nu' |W(\nu)|^2   |W(\nu')|^2  \CG_{11}(\nu')
\nonumber \\
&
+ 2 T \Gb_1 \left[ \int d\nu |W(\nu)|^2 \right]^2
\tag{$2 F_{1111}$} \\
&
+ 4 T  \int d\nu |W(\nu)|^4 [\CG_{11}(\nu)]^2
\nonumber \\
&
+ 2T \CG_{11}(0) \left[ \int d\nu |W(\nu)|^2\right]^2
\tag{$2 F_{1122}$} \\
&
+ 8 T  \int d\nu |W(\nu)|^4 [\CG_{11}(\nu)]^2
+ 8T \Gb_1 \int d\nu |W(\nu)|^4 \CG_{11}(\nu)
\nonumber \\
&
+ 4T \Gb_1^2 \int d\nu |W(\nu)|^4 
\nonumber \\
&
+ 4T  \int d\nu d\nu' |W(\nu)|^2   |W(\nu')|^2  \CG_{11}(\nu - \nu')
\tag{$4 F_{1212}$} \\
&
- 16 T  \int d\nu |W(\nu)|^4 [\CG_{11}]^2(\nu)
- 16T \Gb_1 \int d\nu |W(\nu)|^4 \CG_{11}(\nu)
\nonumber \\
&
- 8T  \int d\nu d\nu' |W(\nu)|^2   |W(\nu')|^2  \CG_{11}(\nu')
\tag{$- 8 F_{1112}$} .
\end{align}
The resulting sum is
\begin{align}
\sigma^2_{P_\Delta}
& = 
+ 8T \Gb_1^2 \int d\nu |W(\nu)|^4 
+ 4T \CG_{11}(0) \left[ \int d\nu |W(\nu)|^2\right]^2
\nonumber \\
&
+ 8T  \int d\nu d\nu' |W(\nu)|^2   |W(\nu')|^2  \CG_{11}(\nu - \nu')
\nonumber \\
&
+ 2 T \Gb_1 \left[ \int d\nu |W(\nu)|^2 \right]^2\ .
\label{eqn:SigmaPDeltaApp}
\end{align}

\section{Quantum calculation: evaluation of eighth-order moments}
\label{app:EighthOrderMoments}

In this appendix, we evaluate eighth-order moments of the photon
operators that are needed for a quantum-mechanical calculation
of the sensitivity of a shot-noise measurement scheme,
as outlined in section~\ref{sec:ShotNoiseQuantum} 
and equation~(\ref{eqn:EighthOrderPhotonMoments}).
As described in that section, there are 9 operator permutations 
that give nonvanishing contributions
out of the 4!=24 possibilities.
We will not evaluate all of these terms but instead choose a few
that are instructive, including the term that is responsible for
the sensitivity degradation due to photon bunching.

We start with the $2^2$ permutations, (12)(34), (13)(24), and (14)(23):
\begin{align}
(12)(34) & = 
\ave{ b^\dagger(\nu'_1) b(\nu'_2)}
\ave{b(\nu'_1 + \nu_1) b^\dagger(\nu'_2 + \nu_2) }
\ave{b^\dagger(\nu'_3) b(\nu'_4)}
\ave{b(\nu'_3 + \nu_3) b^\dagger(\nu'_4 + \nu_4) }
\nonumber
\end{align}
Performing the indicated averages and integrations gives
\begin{align}
(12)(34) & =  \int d \nu'_1 d \nu'_2 d \nu'_3 d \nu'_4 \left\{
n(\nu'_1) \delta(\nu'_1 - \nu'_2) 
[n(\nu'_1+\nu_1) + 1 ] \delta(\nu'_1 + \nu_1 - \nu'_2 - \nu_2)
\right.
\nonumber \\
& \left.
\times n(\nu'_3) \delta(\nu'_3 - \nu'_4) 
[n(\nu'_3+\nu_3) + 1 ] \delta(\nu'_3 + \nu_3 - \nu'_4 - \nu_4)
\right\}
\nonumber \\
& =
\int d \nu'_1 d \nu'_3
n(\nu'_1) [ n(\nu'_1 + \nu_1) + 1] \delta(\nu_1 - \nu_2)
n(\nu'_3) [ n(\nu'_3 + \nu_3) + 1] \delta(\nu_3 - \nu_4)
\nonumber \\
& =
[ \Gb + \SG(\nu_1) ] [ \Gb + \SG(\nu_3) ] \delta(\nu_1 - \nu_2) \delta(\nu_3 - \nu_4)\ .
\label{eqn:perm(12)(34)}
\end{align}
Comparison with the semiclassical calculation detailed in
Appendix~\ref{app:FourthOrderMoments}
shows that we have
reproduced the terms A1a, B1, B6, and C1a;
these become proportional to $T^2$ after the integrations over time
and are related to the mean value of $P_{W,T}$ rather than
its fluctuations.
Next, we evaluate
\begin{align}
(13)(24) & = 
\ave{ b^\dagger(\nu'_1) b(\nu'_3 + \nu_3)}
\ave{b(\nu'_1 + \nu_1) b^\dagger(\nu'_3) }
\ave{b^\dagger(\nu'_2 + \nu_2) b(\nu'_4)}
\ave{b(\nu'_2) b^\dagger(\nu'_4 + \nu_4) }\ .
\nonumber
\end{align}
Upon averaging and integrating,
\begin{align}
(13)(24) 
& =  \int d \nu'_1 d \nu'_2 d \nu'_3 d \nu'_4 \left\{
n(\nu'_1) \delta(\nu'_1 - \nu'_3 - \nu_3) 
[n(\nu'_1+\nu_1) + 1 ] \delta(\nu'_3 - \nu'_1 - \nu_1)
\right.
\nonumber \\
& \left.
\times n(\nu'_4) \delta(\nu'_4 - \nu'_2 - \nu_2) 
[n(\nu'_4+\nu_4) + 1 ] \delta(\nu'_2  - \nu'_4 - \nu_4)
\right\}
\nonumber \\
& =  \int d \nu'_1 d \nu'_4 \left\{
n(\nu'_1)  [n(\nu'_1+\nu_1) + 1 ] \delta(\nu_1 + \nu_3)
n(\nu'_4) [n(\nu'_4+\nu_4) + 1 ] \delta(\nu_2 + \nu_4)
\right\}
\nonumber \\
&=
[ \Gb + \SG(\nu_1) ] [ \Gb + \SG(\nu_2) ] \delta(\nu_1 + \nu_3) \delta(\nu_2 + \nu_4)\ .
\label{eqn:perm(13)(24)}
\end{align}
Comparison with the semiclassical calculation shows that we have
reproduced the terms A1c, B2, B5, and C2a;
these become proportional to $T$ after the integrations over time
and are therefore related to the fluctuations of $P_{W,T}$.
Similarly, 
\be
(14)(23) =
[ \Gb + \SG(\nu_1) ] [ \Gb + \SG(\nu_2) ] \delta(\nu_1 - \nu_4) \delta(\nu_2 - \nu_3)
\label{eqn:perm(14)(23)}
\ee
corresponds to the semiclassical terms 
A1b, B3, B4, and C3a, which again are 
fluctuation terms since they are proportional to $T$.

We now turn to the $4^1$ terms: (1234), (1243), (1324), (1342), (1423), (1432).
The terms that we have derived so far using the $2^2$ permutations
reproduce the semiclassical results 
of section~\ref{sec:SemiclassicalSingleDetector} and
represent $1/BT$ noise,
or terms that vanish if we choose a noise filter
$W(\nu)$ that rejects the bunching noise component at low frequency
illustrated in Figure~\ref{fig:SingleDetectorNoiseSpectrum}.
The $4^1$ permutations are more interesting.
We start with
\begin{align}
(1234) & = 
\ave{ b^\dagger(\nu'_1) b(\nu'_2)}
\ave{ b^\dagger(\nu'_2 + \nu_2) b(\nu'_3 + \nu_3) }
\ave{b^\dagger(\nu'_3) b(\nu'_4)}
\ave{ b(\nu'_1 + \nu_1) b^\dagger(\nu'_4 + \nu_4) }\ .
\nonumber
\end{align}
Averaging and integrating,
\begin{align}
(1234) & = \int d \nu'_1 d \nu'_2 d \nu'_3 d \nu'_4 \left\{
n(\nu'_1) \delta(\nu'_1 - \nu'_2) n(\nu'_2 + \nu_2) \delta(\nu'_2 + \nu_2 - \nu'_3 - \nu_3)
\right.
\nonumber \\
& \times \left.
n(\nu'_3) \delta(\nu'_3 - \nu'_4) [n(\nu'_4 + \nu_4) +1] 
\delta(\nu'_1 + \nu_1 - \nu'_4 - \nu_4)
\right\}
\nonumber \\
& =
\int d \nu'_1 d \nu'_3   \left\{
n(\nu'_1) n(\nu'_1 + \nu_2) 
n(\nu'_3) [n(\nu'_3 + \nu_4) +1] 
\right.
\nonumber \\
& \times \left. 
\delta(\nu'_1 - \nu'_3  + \nu_2 - \nu_3)
\delta(\nu'_1 - \nu'_3 + \nu_1 - \nu_4)
\right\}
\nonumber \\
& = 
\int d \nu'_1 
n(\nu'_1) n(\nu'_1 + \nu_2)  n(\nu'_1 + \nu_1 - \nu_4) [n(\nu'_1 + \nu_1) +1] 
\nonumber \\
& \times \delta(\nu_1 - \nu_2 + \nu_3 - \nu_4)\ .
\nonumber
\end{align}
This represents a contribution to the shot noise fluctuations given by
\begin{align}
\ave{P_{W,T}^2}_{(1234)} 
& = 
\int_{-\infty}^{+\infty} d \nu_1 d \nu_2 d \nu_3 d \nu_4 \,
W(\nu_1) W^*(\nu_2) W(\nu_3) W^*(\nu_4) 
\nonumber \\
& \times F'_{(1234)}(\nu_1, \nu_2, \nu_3, \nu_4) \delta(\nu_1 - \nu_2 + \nu_3 - \nu_4)
\nonumber \\
& \times
\int_{-T/2}^{+T/2} dt e^{i 2 \pi (\nu_1 -\nu_2) t}
\int_{-T/2}^{+T/2} dt' e^{i 2 \pi (\nu_3 -\nu_4) t'}
\nonumber \\
& = T
\int_{-\infty}^{+\infty} d \nu_1 d \nu_3 |W(\nu_1)|^2 |W(\nu_3)|^2
\nonumber \\
& \times
\int d \nu'_1 
n(\nu'_1) n(\nu'_1 + \nu_1)  n(\nu'_1 + \nu_1 - \nu_3) [n(\nu'_1 + \nu_1) +1] 
\nonumber
\end{align}
This term did not appear in the semiclassical analysis, and
presumably represents non-gaussianity of the
photon arrival rate fluctuations which are expected in
the quantum calculation as shown in Appendix~\ref{app:QuantumVsSemiclassical}
but are neglected in section~\ref{sec:SemiclassicalSingleDetector}.
However, for this term to be appreciable, the 
noise frequencies $\nu_1$ and $\nu_3$
must be comparable to or smaller than the optical bandwidth $\Delta \nu$;
this term does not contribute if we choose a cutoff for
$W(\nu)$ that is well above $\Delta \nu$ 
as shown in Figure~\ref{fig:SingleDetectorNoiseSpectrum}.
Next is the pairing
\begin{align}
(1243) & = 
\ave{ b^\dagger(\nu'_1) b(\nu'_2)}
\ave{ b^\dagger(\nu'_2 + \nu_2) b(\nu'_4)}
\ave{b(\nu'_1 + \nu_1) b^\dagger(\nu'_3) }
\ave{b(\nu'_3 + \nu_3) b^\dagger(\nu'_4 + \nu_4) }\ .
\nonumber
\end{align}
Averaging and integrating,
\begin{align}
(1243) & = \int d \nu'_1 d \nu'_2 d \nu'_3 d \nu'_4 \left\{
n(\nu'_1) \delta(\nu'_1 - \nu'_2) n(\nu'_2 + \nu_2) \delta(\nu'_4 - \nu'_2 - \nu_2)
\right.
\nonumber \\
& \times \left.
[n(\nu'_1 + \nu_1) + 1] \delta(\nu'_3 - \nu'_1 - \nu_1) 
[n(\nu'_3 + \nu_3) +1]  \delta(\nu'_4 + \nu_4 - \nu'_3 - \nu_3)
\right\}
\nonumber \\
& = 
\int d \nu'_1 
n(\nu'_1) n(\nu'_1 + \nu_2)  [1 + n(\nu'_1 + \nu_1)] [1 + n(\nu'_1 + \nu_2 + \nu_4)]
\nonumber \\
& \times \delta(\nu_1 - \nu_2 + \nu_3 - \nu_4)\ .
\nonumber
\end{align}
This term contributes a shot noise fluctuation given by
\begin{align}
\ave{P_{W,T}^2}_{(1243)} 
& =
T
\int_{-\infty}^{+\infty} d \nu_1 d \nu_3 |W(\nu_1)|^2 |W(\nu_3)|^2
\nonumber \\
& \times
\int d \nu'_1 
n(\nu'_1) n(\nu'_1 + \nu_1)  [1 + n(\nu'_1 + \nu_1)] [1 + n(\nu'_1 + \nu_1 + \nu_3)]\ .
\nonumber
\end{align}
Again, this term is small if we chose the high-pass filter cutoff frequency well
above the optical bandwidth $\Delta \nu$. Note that there is a contribution
\begin{align}
&
T
\int_{-\infty}^{+\infty} d \nu_1 d \nu_3 |W(\nu_1)|^2 |W(\nu_3)|^2
\int d \nu'_1 
n(\nu'_1) n(\nu'_1 + \nu_1)  
\nonumber \\
& = T \int_{-\infty}^{+\infty} d \nu_1 |W(\nu_1)|^2 \SG(\nu_1) 
\int_{-\infty}^{+\infty} d \nu_3 |W(\nu_3)|^2
\end{align}
that reproduces the semiclassical term C7.
We skip ahead and look at
\begin{align}
(1432) & = 
\ave{ b^\dagger(\nu'_1) b(\nu'_4)}
\ave{b(\nu'_1 + \nu_1) b^\dagger(\nu'_2 + \nu_2) }
\ave{b(\nu'_2) b^\dagger(\nu'_3) }
\ave{ b(\nu'_3 + \nu_3) b^\dagger(\nu'_4 + \nu_4) }\ .
\nonumber
\end{align}
Averaging and integrating,
\begin{align}
(1432) & = \int d \nu'_1 d \nu'_2 d \nu'_3 d \nu'_4 \left\{
n(\nu'_1) \delta(\nu'_1 - \nu'_4) [1 + n(\nu'_1 + \nu_1)] \delta(\nu'_1 + \nu_1 - \nu'_2 - \nu_2)
\right.
\nonumber \\
& \times \left.
[1 + n(\nu'_2)] \delta(\nu'_2 - \nu'_3)
[1 + n(\nu'_3 + \nu_3)]  \delta(\nu'_4 + \nu_4 - \nu'_3 - \nu_3)
\right\}
\nonumber \\
& = 
\int d \nu'_1 
n(\nu'_1) [1 + n(\nu'_1 + \nu_1)  [1 + n(\nu'_1 + \nu_1 - \nu_2)] 
[1 + n(\nu'_1 + \nu_4)]
\nonumber \\
& \times \delta(\nu_1 - \nu_2 + \nu_3 - \nu_4)\ .
\nonumber
\end{align}
The product expands to eight terms. The first term is
\be
\int d \nu'_1 n(\nu'_1) \delta(\nu_1 - \nu_2 + \nu_3 - \nu_4)
= \Gb \delta(\nu_1 - \nu_2 + \nu_3 - \nu_4)
\label{eqn:QuantumPoisson}
\ee
and reproduces the semiclassical term responsible for Poisson noise, D1.
Another term is
\be
\int d \nu'_1 n(\nu'_1) n(\nu'_1 + \nu_1 - \nu_2) \delta(\nu_1 - \nu_2 + \nu_3 - \nu_4)
= \SG(\nu_1 - \nu_2) \delta(\nu_1 - \nu_2 + \nu_3 - \nu_4)
\label{eqn:QuantumBunchingTerm}
\ee
and reproduces the semiclassical term C1b in 
Appendix~\ref{app:FourthOrderMoments}.

For the case of multiple detectors, the photon operators are decorated with
a subscript to indicate the detector that they correspond to. 
Thus, we are interested in
\begin{align}
(1432) & = 
\ave{b_a^\dagger(\nu'_1) b_d(\nu'_4)}
\ave{b_a(\nu'_1 + \nu_1) b_b^\dagger(\nu'_2 + \nu_2) }
\ave{b_b(\nu'_2) b_c^\dagger(\nu'_3) }
\ave{ b_c(\nu'_3 + \nu_3) b_d^\dagger(\nu'_4 + \nu_4) }\ .
\nonumber
\end{align}
Averaging and integrating,
\begin{align}
(1432) & = \int d \nu'_1 d \nu'_2 d \nu'_3 d \nu'_4 \left\{
B_{ad}(\nu'_1) \delta(\nu'_1 - \nu'_4) 
[\delta_{ab} + B_{ba}(\nu'_1 + \nu_1)] \delta(\nu'_1 + \nu_1 - \nu'_2 - \nu_2)
\right.
\nonumber \\
& \times \left.
[\delta_{bc} + B_{cb}(\nu'_2)] \delta(\nu'_2 - \nu'_3)
[\delta_{cd} + B_{dc}(\nu'_3 + \nu_3)]  \delta(\nu'_4 + \nu_4 - \nu'_3 - \nu_3)
\right\}
\nonumber \\
& = 
\int d \nu'_1 
B_{ad}(\nu'_1) 
[\delta_{ab} + B_{ba}(\nu'_1 + \nu_1)]
[\delta_{bc} + B_{cb}(\nu'_1 + \nu_1 - \nu_2)]
[\delta_{cd} + B_{dc}(\nu'_1 + \nu_4)] 
\nonumber \\
& \times \delta(\nu_1 - \nu_2 + \nu_3 - \nu_4)\ .
\nonumber
\end{align}
Again there are eight terms. The first term is
\begin{align}
\delta_{ab} \delta_{bc} \delta_{cd}
\int d \nu'_1 B_{ad}(\nu'_1)  \delta(\nu_1 - \nu_2 + \nu_3 - \nu_4)
= \delta_{ab} \delta_{ac} \delta_{ad}
\Gb_a \delta(\nu_1 - \nu_2 + \nu_3 - \nu_4)
\label{eqn:QuantumPoissonMultiple}
\end{align}
and reproduces the semiclassical D1 term in
Appendix~\ref{app:FourthOrderMomentsMultiple}
that is responsible for Poisson noise.
The interesting term is
\begin{align}
\delta_{ab} \delta_{cd}
\int d \nu'_1 &
B_{ad}(\nu'_1) B_{cb}(\nu'_1 + \nu_1 - \nu_2) \delta(\nu_1 - \nu_2 + \nu_3 - \nu_4)
\nonumber \\
& = 
\delta_{ab} \delta_{cd}
\CG_{ac}(\nu_1 - \nu_2) \delta(\nu_1 - \nu_2 + \nu_3 - \nu_4)
\label{eqn:QuantumBunchingTermMultiple}
\end{align}
and reproduces the semiclassical term C1b 
in Appendix~\ref{app:FourthOrderMomentsMultiple}.

\section{Correlation of the Lieu et al. samples: a quantum calculation}
\label{app:LieuCorrelations}

\tmpcom{(This appendix is new.)}

This appendix presents a quantum-mechanical calculation that shows
that the samples $I^2_{\Delta, \Delta t} (k)$ 
introduced in equations~(\ref{eqn:PLKD}, \ref{eqn:IDeltaDeltatK})
and assumed by \citet{Lieu15} to be independent
are in fact correlated;
furthermore, these correlations are shown to lead to the standard
photon bunching penalty.
Let $I_a(t)$ be the output of detector $a$; here $a = 1$ or $2$. The corresponding quantum
operator is
\be
I_a(t) = \int d\nu_1 d\nu'_1 b_a^\dagger(\nu_1) b_a(\nu'_1) e^{- i 2 \pi (\nu_1 - \nu'_1) t}\ .
\ee
We define the integral over the time interval $[k\Delta t, (k+1)\Delta t]$ as
\be
I_{a,\Delta t}(k) = \int_{k\Delta t}^{(k+1)\Delta t} dt \, I_a(t)\ .
\ee

The Lieu et al. detection scheme involves summing the
squares of the differences $\IdT(k) = I_{1,\Delta t}(k) - I_{2,\Delta t}(k)$
of the two time-averaged and sampled 
outputs of a beamsplitter-fed detector pair (Figures~\ref{fig:TwoDetectors} and \ref{fig:LieuKibbleSP}):
\be
S = \sum_{k=0}^{N-1} \left[ \IdT(k) \right]^2 = \left( \Delta t \right)^2 P_{N,\mathrm{LKD}} 
\label{eqn:SsumDef}
\ee
where $ P_{N,\mathrm{LKD}}$ is defined in equation~(\ref{eqn:PLKD}).
The mean value of a single sample is given by
\begin{align}
\ave{\left[ \IdT(k) \right]^2} = G_{11}(k) - G_{12}(k) - G_{21}(k) + G_{22}(k)\ ,
\end{align}
where
\begin{align}
G_{ab}(k) &= \ave{I_{a,\Delta t}(k) I_{b,\Delta t}(k)}
\nonumber \\
&=  
\int_{k\Delta t}^{(k+1)\Delta t} dt_1  dt_2
 \int d\nu_1 d\nu'_1  d\nu_2 d\nu'_2
e^{- i 2 \pi (\nu_1 - \nu'_1) t_1} e^{- i 2 \pi (\nu_2 - \nu'_2) t_2}
\nonumber \\
& \times \ave{ b_a^\dagger(\nu_1) b_a(\nu'_1) b_b^\dagger(\nu_2) b_b(\nu'_2)}
\nonumber \\
& =
\int_{k\Delta t}^{(k+1)\Delta t} dt_1  dt_2
 \int d\nu_1 d\nu'_1  d\nu_2 d\nu'_2
e^{- i 2 \pi (\nu_1 - \nu'_1) t_1} e^{- i 2 \pi (\nu_2 - \nu'_2) t_2}
\nonumber \\
& \times \left\{
B_{aa}(\nu_1) \delta(\nu_1 - \nu_1') B_{bb}(\nu_2) \delta(\nu_2 - \nu_2') \right.
\nonumber \\
&
+ \left. B_{ab}(\nu_1) \delta(\nu_1 - \nu_2')
\left[ \delta_{ab} + B_{ba}(\nu_2) \right] \delta(\nu_2 - \nu_1')
\right\}
\nonumber \\
&\approx \Gb_a \Gb_b (\Delta t)^2
+ \delta_{ab} \Gb_a \Delta t
+ \left| \int d\nu B_{ab}(\nu) \right|^2 \left(\Delta t \right)^2\ ;
\end{align}
we have made use of $\Delta \nu \Delta t << 1$ in approximating the third term.
Note that the second term dominates in the short sample time regime,  $\Gb \Delta t << 1$.
Here $\Gb_a = \int d\nu B_{aa}(\nu)$ is the photon rate for detector $a$;
the quantity $B_{ab}(\nu)$ is introduced in section~\ref{sec:PhotocurrentSpectrumMultipleDetectors}
through equations~(\ref{eqn:BmatrixDef}), 
(\ref{eqn:BbeamsplitterDiag}), and (\ref{eqn:BbeamsplitterOffDiag}).
Therefore,
\begin{align}
\ave{\left[ \IdT(k) \right]^2} \approx \left(\Gb_1 + \Gb_2 \right) \Delta t = \Gb \Delta t\ .
\end{align}
This is exactly what we expect given the discussion in section~\ref{sec:Resolution}:
for small $\Delta t$, the value of $\left[ \IdT(k) \right]^2$ is unity if either detector
receives a photon, and zero otherwise, and the probability of either receiving a photon is 
$\Gb \Delta t$.
Thus, we conclude that mean value of the sum is
\be
\ave{S} = N \Gb \Delta t = \Gb T
\label{eqn:SsumAve}
\ee
where $T = N \Delta t$ is the total time duration of the measurement.

To calculate the fluctuations of the sum $S$, we first define the quantity
\begin{align}
F_{abcd}(k) 
&= 
\ave{I_{a,\Delta t}(k) I_{b,\Delta t}(k) I_{c,\Delta t}(0) I_{d,\Delta t}(0)}
\nonumber \\
& =
\int_{k\Delta t}^{(k+1)\Delta t} dt_1
\int_{k\Delta t}^{(k+1)\Delta t} dt_2 
\int_{0}^{\Delta t} dt_3 
\int_{0}^{\Delta t} dt_4
\ave{
I_a(t_1) 
I_b(t_2)
I_c(t_3)
I_d(t_4)
}\ .
\nonumber
\end{align}
We wish to evaluate the correlation
\be
C_{\Delta t}(k) = \ave{\left[ \IdT(k) \right]^2 \left[\IdT(0)\right]^2}\ .
\ee
We may easily express $C_{\Delta t}(k)$ in terms of $F_{abcd}(k)$:
\begin{align}
C_{\Delta t}(k) & =
F_{1111}(k)  - F_{2111}(k)  - F_{1211}(k)  + F_{2211}(k)  
\nonumber \\
&
- F_{1121}(k)  + F_{2121}(k)  + F_{1221}(k)  - F_{2221}(k)  
\nonumber \\
&
- F_{1112}(k)  + F_{2112}(k)  + F_{1212}(k)  - F_{2212}(k)  
\nonumber \\
&
+ F_{1122}(k)  - F_{2122}(k)  - F_{1222}(k)  + F_{2222}(k)  \ .
\label{eqn:Fexpansion}
\end{align}
As usual, evaluation of $F_{abcd}(k)$ involves an eighth-order moment
of photon operators,
\begin{align}
F_{abcd}(k) 
& =
\int_{k\Delta t}^{(k+1)\Delta t} dt_1
\int_{k\Delta t}^{(k+1)\Delta t} dt_2 
\int_{0}^{\Delta t} dt_3 
\int_{0}^{\Delta t} dt_4
\nonumber \\
&
\times \int_0^\infty d\nu_1 d\nu'_1 d\nu_2 d\nu'_2 d\nu_3 d\nu'_3 d\nu_4 d\nu'_4 
\nonumber \\
& \times 
e^{-i 2 \pi (\nu_1 - \nu'_1) t_1}
e^{-i 2 \pi (\nu_2 - \nu'_2) t_2}
e^{-i 2 \pi (\nu_3 - \nu'_3) t_3}
e^{-i 2 \pi (\nu_4 - \nu'_4) t_4}
\nonumber \\
& \times
\ave{
b_a^\dagger(\nu_1) b_a(\nu'_1)
b_b^\dagger(\nu_2) b_b(\nu'_2)
b_c^\dagger(\nu_3) b_c(\nu'_3)
b_d^\dagger(\nu_4) b_d(\nu'_4)
}\ .
\nonumber
\end{align}
Pairwise combination of the operators gives 4! = 24 terms.
It is not difficult to show that the (12)(34) permutation gives
\begin{align}
F_{abcd}^{(12)(34)}(k) & = G_{ab}(k) G_{cd}(0)\ ,
\end{align}
and inserting this result into equation~(\ref{eqn:Fexpansion}) gives
\begin{align}
C_{\Delta t}^{(12)(34)}(k) & =
\left[ G_{11}(0) - G_{12}(0) - G_{21}(0) + G_{22}(0)\right]^2
\nonumber \\
& = \ave{\left[ \IdT(k) \right]^2}^2\ .
\end{align}
This term will therefore contribute $\ave{S}^2$ when calculating $\ave{S^2}$,
which will subtract out when we calculate the variance of $S$.
As in Appendix~\ref{app:EighthOrderMoments}, the (12)(34) permutation
contributes to the mean value rather than to the fluctuations.

As before, the operator pairing corresponding to the (1432) permutation
is responsible for the Poisson and bunching noise:
\begin{align}
F_{abcd}^{(1432)}(k) 
& =
\int_{k\Delta t}^{(k+1)\Delta t} dt_1
\int_{k\Delta t}^{(k+1)\Delta t} dt_2 
\int_{0}^{\Delta t} dt_3 
\int_{0}^{\Delta t} dt_4
\nonumber \\
&
\times \int_0^\infty d\nu_1 d\nu'_1 d\nu_2 d\nu'_2 d\nu_3 d\nu'_3 d\nu_4 d\nu'_4 
\nonumber \\
& \times 
e^{-i 2 \pi (\nu_1 - \nu'_1) t_1}
e^{-i 2 \pi (\nu_2 - \nu'_2) t_2}
e^{-i 2 \pi (\nu_3 - \nu'_3) t_3}
e^{-i 2 \pi (\nu_4 - \nu'_4) t_4}
\nonumber \\
& \times
\ave{ b_a^\dagger(\nu_1) b_d(\nu'_4) }
\ave{ b_a(\nu'_1) b_b^\dagger(\nu_2) }
\ave{ b_b(\nu'_2) b_c^\dagger(\nu_3) }
\ave{ b_c(\nu'_3) b_d^\dagger(\nu_4) }
\ 
\nonumber \\
& = 
\int_{k\Delta t}^{(k+1)\Delta t} dt_1
\int_{k\Delta t}^{(k+1)\Delta t} dt_2 
\int_{0}^{\Delta t} dt_3 
\int_{0}^{\Delta t} dt_4
 \int_0^\infty d\nu_1 d\nu_2 d\nu_3  d\nu_4 
\nonumber \\
& \times 
e^{-i 2 \pi (\nu_1 - \nu_2) t_1}
e^{-i 2 \pi (\nu_2 - \nu_3) t_2}
e^{-i 2 \pi (\nu_3 - \nu_4) t_3}
e^{-i 2 \pi (\nu_4 - \nu_1) t_4}
\nonumber \\
& \times
B_{ad}(\nu_1) 
\left[ \delta_{ab} + B_{ba}(\nu_2) \right]
\left[ \delta_{bc} + B_{cb}(\nu_3) \right]
\left[ \delta_{cd} + B_{dc}(\nu_4) \right]\ .
\label{eqn:perm4321}
\end{align}
This expression leads to eight terms;
among these is the term that gives rise to the C1b
contribution in the semiclassical and quantum calculations
in Appendices~\ref{app:FourthOrderMoments} and \ref{app:EighthOrderMoments}:
\begin{align}
F_{abcd}^{(1432), \mathrm{C1b}}(k) 
& =
\int_{k\Delta t}^{(k+1)\Delta t} dt_1
\int_{k\Delta t}^{(k+1)\Delta t} dt_2 
\int_{0}^{\Delta t} dt_3 
\int_{0}^{\Delta t} dt_4
 \int_0^\infty d\nu_1 d\nu_2 d\nu_3  d\nu_4 
\nonumber \\
& \times 
e^{-i 2 \pi \nu_1 (t_1 - t_4)}
e^{-i 2 \pi \nu_2 (t_2 - t_1)}
e^{-i 2 \pi \nu_3 (t_3 - t_2)}
e^{-i 2 \pi \nu_4 (t_4 - t_3)}
\nonumber \\
& \times
\delta_{ab} 
\delta_{cd}
B_{ac}(\nu_1)  B_{ca}(\nu_3) \ .
\label{eqn:perm4321part}
\end{align}
The $\nu_2$ integral gives $\delta(t_2 - t_1)$ while the $\nu_4$ integral gives $\delta(t_4 - t_3)$;
therefore
\begin{align}
F_{abcd}^{(1432), \mathrm{C1b}}(k) 
& =
\int_{k\Delta t}^{(k+1)\Delta t} dt_1
\int_{0}^{\Delta t} dt_3 
 \int_0^\infty d\nu_1 d\nu_3 
\nonumber \\
& \times 
e^{-i 2 \pi \nu_1 (t_1 - t_3)}
e^{-i 2 \pi \nu_3 (t_3 - t_1)} \,
\delta_{ab} 
\delta_{cd}
B_{ac}(\nu_1)  B_{ca}(\nu_3) \ .
\label{eqn:perm4321part2}
\end{align}
Note that 
\be
B_{ac}(\nu_1) B_{ca}(\nu_3) = \frac{1}{4} n(\nu_1) n(\nu_3)
\ee
regardless of the choice of indices.
Furthermore,
\begin{align}
 \int_0^\infty d\nu_1 d\nu_3 \, n(\nu_1) n(\nu_3)
 e^{-i 2 \pi \nu_1 \tau } e^{+i 2 \pi \nu_3 \tau}
 & = 
 \int_0^\infty d\nu d\nu' \, n(\nu) n(\nu + \nu')
 e^{i 2 \pi \nu' \tau }
 \nonumber \\
 & = 
 \int_0^\infty d\nu' \, \SG(\nu')   e^{i 2 \pi \nu' \tau }
 = A_\Gamma(\tau)\ ,
\nonumber
 \end{align}
where  $A_\Gamma(\tau)$ is the Fourier transform of $\SG(\nu)$
and represents the time autocorrelation function of the photon rate fluctuations.
Note that $A_\Gamma(\tau)$ decays on a timescale $\tau \sim 1/\Delta \nu$,
and that $A(0) = \Gb^2$.
Thus
\begin{align}
F_{abcd}^{(1432), \mathrm{C1b}}(k) 
& =
\frac{1}{4}
\delta_{ab} 
\delta_{cd}
\int_{k\Delta t}^{(k+1)\Delta t} dt_1
\int_{0}^{\Delta t} dt_3 
A_\Gamma(t_1 - t_3)
\nonumber \\
& \approx \frac{1}{4} \delta_{ab} \delta_{cd} (\Delta t)^2 A_\Gamma(k\Delta t)
\end{align}
where the approximation holds because $\Delta t \Delta \nu << 1$.
The (1432) pairing also contributes a term 
that corresponds to Poisson
noise, labeled D1 in the semiclassical calculation:
\begin{align}
F_{abcd}^{(1432), \mathrm{D1}}(k) 
& =
\frac{1}{2}
\delta_{ab} 
\delta_{cd}
\delta_{ac}
\delta_{k,0}
\Delta t \Gb\ .
\end{align}

Of the sixteen terms in equation~(\ref{eqn:Fexpansion}),
the only nonzero contributions for the C1b piece of the (1432) permutation
come from $F_{1111}$, $F_{1122}$, $F_{2211}$, and $F_{2222}$,
due to the $\delta_{ab} \delta_{cd}$ factor; and all have the same sign.
For the D1 piece, the additional $\delta_{ac}$ factor means that
only $F_{1111}$ and $F_{2222}$ contribute.
These two pieces give a
contribution to $C_{\Delta t}(k)$ given by
\begin{align}
C_{\Delta t}^{(1432), \mathrm{C1b+D1}}(k)  
& \approx \Gb \Delta t \delta_{k,0} + (\Delta t)^2 A_\Gamma(k\Delta t)\ .
\label{eqn:CT1432part}
\end{align}
The second term in this expression shows that the quantities $\left[ \IdT(k)\right]^2$ 
are indeed correlated, in contradiction to the assumption of \citet{Lieu15}.
The corresponding contribution to the variance of 
$S$ (equation~\ref{eqn:SsumDef}) is:
\begin{align}
\sigma_S^2 &= \ave{S^2} - \ave{S}^2 
\nonumber \\
& = \left[ \sum_{k, l=0}^{N-1} C_{\Delta t}(k-l) \right] - \ave{S}^2
\nonumber \\
& = \sum_{k, l=0}^{N-1} \left\{ \Gb \Delta t \delta_{kl} + (\Delta t)^2 A_\Gamma[(k-l) \Delta t]\ \right\} + ...
\nonumber \\
& \approx N \Gb \Delta t + N (\Delta t)^2 \frac{n^2 \Delta \nu}{\Delta t} + ...
\nonumber \\
& = \Gb T (1 + n) + ...
\end{align}
where I made use of
\begin{align}
\sum_{l} A_\Gamma[(k-l) \Delta t] 
& \approx \frac{1}{\Delta t} \int_{-\infty}^{+\infty} A_\Gamma(\tau) d\tau
 = \frac{1}{\Delta t} \int_0^\infty n^2(\nu) d\nu = \frac{n^2 \Delta \nu}{\Delta t}\ ,
\end{align}
$T = N \Delta t$, and $\Gb = n \Delta \nu$.
Using equation~(\ref{eqn:SsumAve}), we find
\be
\frac{\sigma_{N,\mathrm{LKD}}^2 }{\ave{P_{N,\mathrm{LKD}}}} = 
\frac{\sigma_S^2}{\ave{S}^2} \approx \frac{1 + n}{\Gb T} + ...
\ee
This expression agrees with the other results presented in this paper 
(equation~\ref{eqn:PDeltaFractionalFluctuations})
but contradicts the fundamental result of \citet{Lieu15}
(equation~\ref{eqn:LieuShortSampleTimes}).


\section{Equivalence of Quantum and Semiclassical Approaches}
\label{app:QuantumVsSemiclassical}

In this Appendix, I
use a straightforward extension of the arguments developed by \citet{Sud63}
to show that the full quantum-mechanical calculation of photon bunching
is equivalent to a semiclassical calculation that makes use
of a compound Poisson random process with a stochastically varying 
count rate.
The equivalence is shown
by comparing the generating functionals, defined as
\be
G[s] = \ave{\exp\left[ \int_{-\infty}^{+\infty} dt\, s(t) I(t) \right]}\ ,
\ee
The semiclassical and quantum-mechanical versions will be denoted by 
$\Gsemi[s]$ and $\Gquant[s]$, respectively.
These generating functions fully encode the statistics of the photocurrent $I(t)$;
the statistics must be the same if $\Gsemi[s] = \Gquant[s]$.

A Poisson process with a deterministic time-varying rate $\Gamma(t)$
obeys equation~\ref{eqn:GeneratingFunctionPoissonCounts}):
\be
\ave{\exp\left[ s \int_{t_k}^{t_{k+1}} I(t) dt\right] }_y = 
\exp\left\{ \left[\int_{t_k}^{t_{k+1}} \Gamma(t) dt \right] \left(e^s - 1 \right) \right\}\ .
\ee
If we make the time interval small enough, we may approximate
\be
\ave{\exp\left[ s \int_{t_k}^{t_{k+1}} I(t) dt\right] }_y \approx
\exp\left\{ \Gamma(t_k) \Delta t_k \left(e^s - 1 \right) \right\}\ .
\ee
If the intervals $[t_k, t_{k+1}]$ span the region over which
$s(t)$ is nonzero, we may write
\be
\int_{-\infty}^{+\infty} dt\, s(t) I(t) \approx \sum_{k} s(t_k) \int_{t_k}^{t_{k+1}} I(t) dt\ .
\ee
Making use of the independence of the subinterval counts $\left\{y\right\}$,
we have
\begin{align}
G[s] 
& \approx   
\ave{\exp\left[ \sum_{k} s(t_k) \int_{t_k}^{t_{k+1}} I(t) dt \right]}_y
\nonumber \\
& = 
\prod_k \ave{\exp\left[ s(t_k) \int_{t_k}^{t_{k+1}} I(t) dt \right]}_y
\nonumber \\
& \approx 
\prod_k \exp\left\{ \Gamma(t_k) \Delta t_k \left(e^{s(t_k)} - 1 \right) \right\}
\nonumber \\
& = 
\exp\left\{ \sum_k \Gamma(t_k) \Delta t_k \left(e^{s(t_k)} - 1 \right) \right\}
\nonumber
\end{align}
and by taking the continuum limit we find
\be
G[s] = 
\exp\left\{ \int_{-\infty}^{+\infty} dt \, \Gamma(t)  \left(e^{s(t)} - 1 \right) \right\}\ .
\label{eqn:GeneratingFunctionSemiclassicalDeterministic}
\ee
If we now allow the rate $\Gamma(t)$ to be stochastic instead of deterministic,
we must also perform an average over $\Gamma(t)$. We 
obtain a formal expression for the semiclassical generating function
by writing this average as a functional integral
\be
\Gsemi[s] = \int [d\Gamma(t)]  f[\Gamma(t)]
\exp\left\{ \int_{-\infty}^{+\infty} dt \, \Gamma(t)  \left(e^{s(t)} - 1 \right) \right\}\ ,
\label{eqn:GeneratingFunctionSemiclassicalRandom}
\ee
where $f[\Gamma(t)]$ represents the probability density functional for the
rate process $\Gamma(t)$ \citep{Ued89}
and $[d\Gamma(t)]$ is the functional integration measure.

We now show that the quantum generating function may also be
written in this manner and obtain an expression for the 
resulting probability density $f[\Gamma(t)]$.
The quantum-mechanical averages require traces over the thermal density matrix given by
equation~(\ref{eqn:ThermalDensityMatrix}):
\begin{align}
\Gquant[s] &= \ave{\exp\left[ \int_{-\infty}^{+\infty} dt\, s(t) I(t) \right]}
\nonumber \\
&=
\tr \left\{\exp\left[ \int_{-\infty}^{+\infty} dt\, s(t) b^\dagger(t) b(t) \right] \rho  \right\}\ .
\nonumber \\
&=
\tr \left\{\exp\left[ \int_0^{\infty} d\nu_1 d\nu_2 
\hat{s}(\nu_1 - \nu_2) b^\dagger(\nu_1) b(\nu_2) \right] \rho  \right\}\ .
\end{align}
where $\hat{s}(\nu)$ is the Fourier transform of $s(t)$,
\be
\hat{s}(\nu) = \int_{-\infty}^{+\infty} dt\, s(t) e^{i 2 \pi \nu t}\ .
\ee
In the following discussion,
we will find it useful to 
switch between operators labeled by a continuous frequency index
and a discrete approximation using
\be
\int_0^{\infty} d\nu_1 d\nu_2 \,
\hat{s}(\nu_1 - \nu_2) b^\dagger(\nu_1) b(\nu_2)
\leftrightarrow
\sum_{ij} S_{ij} b^\dagger_i b_j
\ee
where
\be
b_i = \frac{1}{\sqrt{\Delta \nu_i}} \int_{\nu_i}^{\nu_i + \Delta \nu_i} b(\nu)
\ee
and similarly for $b^\dagger_i$, and therefore $[b_i, b_j^\dagger] = \delta_{ij}$,
while
\be
 S_{ij} = \hat{s}(\nu_i - \nu_j) \sqrt{\Delta \nu_i \Delta \nu_j} 
 \ee
is a Hermitian matrix by virtue of $\hat{s}(-\nu) = \hat{s}^*(\nu)$.

The coherent state representation is convenient for calculating 
$\Gquant[s]$.
The coherent states \citep{Gla63} are given by
\be
\ket{z} = \exp\left( \sum_i z_i b^\dagger_i \right) \ket{0}
\ee
and satisfy the normalization
\be
\braket{z}{z'} = \exp\left( \sum_i z^*_i z_i \right) = e^{z^\dagger z}\ ,
\ee
where $z$ represents the column vector with components $\{ z_i \}$
and $z^\dagger$ is its Hermitian conjugate, a row vector
with components $\{ z^*_i \}$.
The coherent states satisfy the overcompleteness relation
\be
1 = \int d\mu(z) e^{-z^\dagger z} \ket{z} \bra{z}
\ee
where the integration measure is
\be
d \mu(z) = \prod_i \frac{d(\re{z_i}) d(\im{z_i})}{\pi}\ .
\ee
The thermal density matrix has a diagonal coherent-state representation
\be
\rho = \det \left( N^{-1} \right)
\int d\mu(z) e^{-z^\dagger z}  
e^{-z^\dagger N^{-1} z}
\ket{z} \bra{z}
\ee 
where $N$ is a diagonal matrix of mode occupation numbers with elements
\be
N_{ij} = n_i \delta_{ij} = \frac{1}{e^{x_i} - 1} \delta_{ij}\ .
\ee
Thermal averages may be computed using this representation,
\begin{align}
\tr(A \rho) 
&=  
\det \left( N^{-1} \right)
\int d\mu(z) e^{-z^\dagger z} e^{-z^\dagger N^{-1} z} \mat{z}{A}{z}\ .
\nonumber
\end{align}
The operator we are interested in has the form
\be
A = \exp \left( \sum_{ij} S_{ij} b^\dagger_i b_j \right) = \exp \left( b^\dagger S b \right)
\ee
where we use the vector notation for the photon operators in which
$b$ represents a column vector with elements $\{b_i \}$ and 
$b^\dagger$ represents its Hermitian conjugate.
Coherent-state matrix elements may be evaluated using the normal
ordering theorem \citep{Bla07}
\be
\exp\left[b^\dagger S b \right] = : \exp\left[b^\dagger \left(e^S - 1 \right) b \right] :
\ee
which gives a compact result for the quantum-mechanical generating function,
\begin{align}
\Gquant[s] &= 
\det \left( N^{-1} \right)
\int d\mu(z) e^{-z^\dagger z} e^{-z^\dagger N^{-1} z} 
\mat{z}{  : \exp\left[b^\dagger \left(e^S - 1 \right) b \right] : }{z}
\nonumber \\
& = 
\det \left( N^{-1} \right)
\int d\mu(z) e^{-z^\dagger z} e^{-z^\dagger N^{-1} z} 
\exp\left[z^\dagger \left(e^S - 1 \right) z \right] \braket{z}{z}
\nonumber \\
& = 
\det \left( N^{-1} \right)
\int d\mu(z)
\exp\left[ - z^\dagger\left( N^{-1} - e^S + 1  \right) z \right]
\nonumber \\
& =
\frac{\det \left( N^{-1} \right)}{\det \left( N^{-1} - e^S + 1 \right)}
\nonumber \\
& =
\frac{1}{\det \left[ 1 - N \left( e^S - 1 \right) \right]} 
\nonumber \\
& =
\exp \left\{ - \tr \ln \left[ 1 - N \left( e^S - 1 \right) \right] \right \}\ .
\label{eqn:QuantumGeneratingFuntionSimple}
\end{align}
Here we have made use of the complex Gaussian integral \citep{Neg88}
\be
\int d\mu(z) \exp\left[ - z^\dagger B z - c^\dagger z - z^\dagger d \right]
= \frac{\exp\left[ c^\dagger B^{-1} d \right]}{\det B}  \ .
\ee
Equation~(\ref{eqn:QuantumGeneratingFuntionSimple})
resembles other results for thermal radiation,
e.g. those of \citet{Bee98}.

Although equation~(\ref{eqn:QuantumGeneratingFuntionSimple})
is a relatively simple expression for the quantum-mechanical
generating function, 
it is not easy to compare this result to our semiclassical generating
function given by equation~(\ref{eqn:GeneratingFunctionSemiclassicalRandom}).
If we hold off the $z$-integration, we have
\begin{align}
\Gquant[s] &= 
\det \left( N^{-1} \right)
\int d\mu(z)
\exp\left[ - z^\dagger N^{-1} z - z^\dagger \left( e^S -1  \right) z \right]\ .
\nonumber
\end{align}
In the continuum limit, the second term in the argument of the exponential is
\begin{align}
z^\dagger \left(e^S - 1\right)  z 
& =
\int d\nu d\nu' z^*(\nu) \left(e^S - 1\right)_{\nu, \nu'} z(\nu')\ .
\nonumber
\end{align}
Now
\begin{align}
\left(e^S - 1\right)_{\nu, \nu'}
& =
\hat{s}(\nu-\nu') + \frac{1}{2!} \int d\nu_1 \hat{s}(\nu-\nu_1) \hat{s}(\nu_1-\nu') 
+ ...
\nonumber \\
&=
\int dt_1 s(t_1) \left[ e^{i 2 \pi (\nu-\nu') t_1} \right]
\nonumber \\
& +
\frac{1}{2!} \int dt_1 dt_2 s(t_1) s(t_2) 
\int d\nu_1 e^{i 2 \pi (\nu-\nu_1)t_1} e^{i 2 \pi (\nu_1-\nu') t_2 } 
+ ...
\nonumber \\
& =
\int dt_1 \left[ s(t_1) + \frac{1}{2!} s^2(t_1) + ... \right] e^{i 2 \pi (\nu-\nu') t_1}
\nonumber \\
& =
\int dt \left[ e^{s(t)} - 1 \right] e^{i 2 \pi (\nu-\nu') t}\ .
\end{align}
Thus we obtain
\be
z^\dagger \left(e^S - 1\right)   z 
=
\int dt \Gamma(t | z) \left(  e^{s(t)} - 1 \right)
\ee
where
\be
\Gamma(t | z) = \int d\nu d\nu' e^{i 2 \pi (\nu-\nu') t}
z^*(\nu) z(\nu')\ .
\label{eqn:QMRateProcess}
\ee

We thus conclude that
the quantum-mechanical generating function may be written
in a form identical to that of a compound Poisson process 
as expressed in
equation~(\ref{eqn:GeneratingFunctionSemiclassicalRandom}),
\begin{align}
\Gquant[s] 
& = 
\exp \left[  - \tr \ln N \right]  
\int d\mu(z) 
\exp \left( - z^\dagger N^{-1} z \right) 
\exp \left[ \int dt \Gamma(t | z) \left(  e^{s(t)} - 1 \right) \right]
\nonumber \\
& = 
\int [d \Gamma(t)]  f[\Gamma(t) | N]
\exp\left\{ \int_{-\infty}^{+\infty} dt \, \Gamma(t)  \left(e^{s(t)} - 1 \right) \right\}
\label{eqn:GeneratingFunctionQuantum} \\
& = \Gsemi[s]\ ,
\nonumber
\end{align}
provided that the probability density functional for the stochastic rate
process is given by
\be
f[\Gamma(t) | N] = 
\exp \left[  - \tr \ln N \right]  
\int d\mu(z) 
\exp \left( - z^\dagger N^{-1} z \right) 
\delta \left[ \Gamma(t) - \Gamma(t|z) \right] 
\ee
and where $\Gamma(t|z)$ is given by equation~(\ref{eqn:QMRateProcess}).
Note that while $z(\nu)$ has a Gaussian distribution with variance $n(\nu)$,
$\Gamma(t|z)$ is a quadratic form of $z(\nu)$ and therefore is
not strictly Gaussian.

It is not difficult to demonstrate that
\be
\exp \left[  - \tr \ln N \right]  
\int d\mu(z) 
\exp \left( - z^\dagger N^{-1} z \right) 
z(\nu_1) z^*(\nu_2) = n(\nu_1) \delta(\nu_1 - \nu_2)
\ee
while
\begin{align}
\exp \left[  - \tr \ln N \right]  
&
\int d\mu(z) 
\exp \left( - z^\dagger N^{-1} z \right) 
z(\nu_1) z^*(\nu_2) z(\nu_2) z^*(\nu_4) 
\nonumber \\
&= 
n(\nu_1) n(\nu_3) \delta(\nu_1 - \nu_2) \delta(\nu_3 - \nu_4)
+
n(\nu_1) n(\nu_3) \delta(\nu_1 - \nu_4) \delta(\nu_3 - \nu_2)\ .
\nonumber
\end{align}
Thus, the mean of the equivalent stochastic rate process is
\begin{align}
\ave{\Gamma(t_1)} 
& = 
\int \, d\mu[\Gamma(t) ] \, f[\Gamma(t) | n ] \, \Gamma(t_1)
\nonumber \\
& =
\exp \left[  - \tr \ln N \right]  
\int d\mu(z) 
\exp \left( - z^\dagger N^{-1} z \right) 
\Gamma(t_1 | z)
\nonumber \\
& = 
\int d\nu d\nu' e^{i 2 \pi (\nu-\nu') t_1}
n(\nu) \delta(\nu - \nu')
\nonumber \\
& =
\int d\nu n(\nu) = \Gb\ ,
\end{align}
which is the expected result.
Meanwhile, the second moment is
\begin{align}
\ave{\Gamma(t_1) \Gamma(t_2)} 
& = 
\int d\nu_1 d\nu_2 d\nu_3 d\nu_4 
\, e^{i 2 \pi (\nu_1-\nu_2) t_1}
\, e^{i 2 \pi (\nu_3-\nu_4) t_2}
\nonumber \\
&
\times\left[
n(\nu_1) \delta(\nu_1 - \nu_2)
n(\nu_3) \delta(\nu_3 - \nu_4)
+
n(\nu_1) \delta(\nu_1 - \nu_4)
n(\nu_3) \delta(\nu_3 - \nu_2)
\right]
\nonumber \\
& =
\int d\nu_1 d\nu_3 
\left[
n(\nu_1) n(\nu_3) 
+ \, e^{i 2 \pi (\nu_1-\nu_3) (t_1 - t_2)}
n(\nu_1) 
n(\nu_3)
\right]
\nonumber \\
& = \Gb^2 + \int d\nu \SG(\nu) e^{i 2 \pi \nu (t_1 - t_2)}\ .
\label{eqn:QuantumToSemiclassical}
\end{align}
These results coincide with equations~(\ref{eqn:MeanRatePoisson})
and (\ref{eqn:RateFluctuationsPoisson}).

\end{document}